\documentclass[prx,aps,twocolumn,floatfix,showpacs,superscriptaddress]{revtex4}
\usepackage{graphicx,graphics,psfrag,amsmath,calc}
\usepackage{epsfig}
\usepackage{color}
\usepackage{comment}
\begin{document}

\title{Color superfluidity of neutral ultra-cold fermions \\
in the presence of color-flip and color-orbit fields}

\author{Doga Murat Kurkcuoglu}

\affiliation{
Department of Physics, Georgia Southern University, Statesboro, Georgia 30460, USA
}
\affiliation{
School of Physics, Georgia Institute of Technology, Atlanta, 
Georgia 30332, USA
}

\author{C. A. R. S{\'a} de Melo}
\affiliation{
School of Physics, Georgia Institute of Technology, Atlanta, 
Georgia 30332, USA
}

\date{\today}

\begin{abstract}
We describe how color superfluidity is modified 
in the presence of color-flip and color-orbit fields in the context of 
ultra-cold atoms, and discuss connections between this problem
and that of color superconductivity 
in quantum chromodynamics. We study the case
of s-wave contact interactions between different colors, and we identify 
several superfluid phases, with five being nodal and one being fully gapped.
When our system is described in a mixed color basis, the superfluid
order parameter tensor is characterized by six independent components 
with explicit momentum dependence induced by color-orbit coupling.
The nodal superfluid phases are topological in nature, and the low temperature
phase diagram of color-flip field versus interaction parameter exhibits
a pentacritical point, where all five nodal color superfluid phases converge.
These results are in sharp contrast to the case of zero color-flip and 
color-orbit fields, where the system has perfect U(3) symmetry
and possesses a superfluid phase that is characterized by 
fully gapped quasiparticle excitations with a single complex order 
parameter with no momentum dependence and by inert unpaired fermions 
representing a non-superfluid component.
In the latter case, just a crossover between a Bardeen-Cooper-Schrieffer and
a Bose-Einstein-Condensation superfluid occurs.
Furthermore, we analyse the order parameter tensor in a total pseudo-spin basis, 
investigate its momentum dependence in the singlet, triplet and 
quintet sectors, and compare the results with the simpler case of spin-1/2
fermions in the presence of spin-flip and spin-orbit fields, where only
singlet and triplet channels arise.
Finally, we analyse in detail spectroscopic properties of color 
superfluids in the presence of color-flip and color-orbit fields, 
such as the quasiparticle excitation spectrum, momentum distribution, and 
density of states to help characterize all the encountered 
topological quantum phases, which can be realized in fermionic isotopes 
of Lithium, Potassium and Ytterbium atoms with three internal states trapped. 
\end{abstract}
\maketitle

%
%

Ultra-cold atoms have become preferred systems to study experimentally,
because they can be used as quantum simulators of various phenomena 
accross different areas of physics. Today it is possible to engineer 
Hamiltonians in the laboratory that describe models that have been 
investigated in the context of condensed matter physics. 
For instance, two very succesful examples
of these experimental quantum simulations are studies of the 
superfluid-insulator transition~\cite{bloch-2005} and 
the evolution from Bardeen-Cooper-Schrieffer (BCS) to Bose-Einstein condensation 
(BEC) superfluidity~\cite{jin-2003, hulet-2003, ketterle-2003, jin-2004, 
salomon-2004, grimm-2004, thomas-2004}, which were performed in 
recent years. 

The success of cold atoms as quantum simulators is largely due to the 
flexibility that these systems have. It is now routinely 
possible to change atomic species, dimensionality, density and interactions 
in clouds of ultra-cold atoms, while in the case of optical lattices, it
is possible, in addition, to change the lattice structure.
For instance, recent experimental advances lead to the trapping of
three hyperfine states in Fermi gases, such as in $^6$Li
which have tunable interactions via an external
magnetic field, which become SU(3) symmetric in the limit of
high magnetic fields~\cite{jochim-2008,ohara-2009}. 
Even larger component systems have now been
produced in the laboratory, such as in the case of 
fermion isotope of Ytterbium, $^{173}$Yb, 
where six internal states exist with essentially
SU(6) symmetry, that can be reduced to SU(3) 
by selectively trapping only three internal 
states~\cite{takahashi-2010a}. Current temperatures
that can be achieved in these systems are approximately
$T = 0.3 T_F$, where $T_F$ is the Fermi temperature 
set by the total density of fermions. Additional 
experiments with the goal of reducing further the 
temperature are underway~\cite{fallani-2017}.
Thus, links to quantum chromodynamics (QCD)
are possible~\cite{sademelo-2008}, 
where dense cold matter with SU(3) color symmetry
is created by the forces that confine quarks inside baryons and mesons 
through the exchange of SU(3) gauge bosons known as gluons.
In particular, connections to color superconductivity in 
the absence of color-flip and color-orbit fields were made by several
authors for continuum~\cite{zhuang-2006,demler-2007,baym-2010}
and lattice~\cite{hofstetter-2007} systems.  

The relation between three-component ultra-cold fermions and 
color superconductivity in QCD is not only of academic interest, 
but also of experimental interest, as the gap between theoretical proposals 
and experimental realization closes due to technical advances. 
The type of color superfluidity in neutral ultra-cold fermions 
is expected to be related, but somewhat different from the possibilities 
encountered in QCD, as quarks are electrically charged, 
have different masses, colors and flavors, and thus color superconductivity
of quarks is generically 
different~\cite{barrois-1977, love-1984, velkovsky-1998, wilczek-1998}. 
For instance, it is quite difficult to realize in neutral ultra-cold 
fermions analagous phases to color flavor locking (CFL) 
superconductivity~\cite{wilczek-1998, wijewardhana-1999, velkovsky-2000, 
schafer-2000a, schwetz-2000}, where three flavors are degenerate, because
the number of internal degrees of freedom is very large. Even analogous
phases to two-flavor superconductivity 
(2SC)~\cite{barrois-1977, love-1984, velkovsky-1998, wilczek-1998}, 
where a preferred color pairing channel is selected are
difficult to realize in the context of ultra-cold fermions. 
However, analogous phases to one-flavor superconductivity with
three colors~\cite{hosek-1998, schafer-2000b, hosek-2000} 
are easier to realize in cold-atom systems, because one has to deal 
with only three internal states of the constitutive fermions.

Moreover, it has been suggested that color superconductivity of quark matter
may occur naturally in compact neutron 
stars~\cite{schaab-1997, blaschke-2000, page-2000}. 
Since the higher temperature regions of the QCD phase
diagram are now being studied in heavy ion collisions, it is important
to use our knowledge of neutron star phenomena to understand if color
superconductivity indeed appears in the high density region of the 
QCD phase diagram. There have been even further suggestions that 
inhomogeneous color superconductivity phases emerge in neutron stars and
are responsible for the glitches in rotational period of these compact
stars~\cite{alford-2000}. Therefore, there has been observational interest in 
determining astrophysical consequences of color superconductivity.

Experiments involving color superconductivity in quark matter are not easy
to come along, but table-top experimental set-ups involving 
ultra-cold fermions with three internal states can serve
as simulators of color superfluidity similar to color superconductivity
in QCD. In addition, experiments with ultra-cold fermions with three internal
states can also stand alone to serve as pointers for new directions and
new phases of color superfluids that have no counterpart in QCD. 
A recent experimental development with a fermionic isotope of Ytterbium
($^{173}$Yb) demonstrated that three internal states of the atom can
be coupled to artificially created spin-orbit fields~\cite{fallani-2016}
via a Raman scheme with two counter-propagating lasers used earlier 
by the NIST group in the context of a bosonic isotope
of Rubidium ($^{87}$Rb) with two internal states~\cite{spielman-2011}.
The interactions of $^{87}$Rb cannot be
adjusted, but it was possible to study the low temperature 
phase diagram of $^{87}$Rb in the presence of spin-orbit coupling 
and Zeeman fields both experimentally~\cite{spielman-2011}
and theoretically~\cite{wu-2011,ho-2011,stringari-2012} for 
fixed interactions. Similar Raman 
schemes were used succesfully in fermionic isotope of Potassium ($^{40}$K),
where Fano-Feshbach resonances exist and interactions can 
be tuned in the presence of 
spin-orbit coupling for two internal states~\cite{spielman-2013,jiang-2014}.
These experimental efforts on $^{40}$K were developed concomitantly 
with various theoretical proposals~\cite{chapman-2011, 
shenoy-2011,zhang-2011,zhai-2011,pu-2011,han-2012,seo-2012,devreese-2014} 
of spin-orbit coupled fermions with two internal states, 
where interactions can be changed.
Although experiments involving fermions in the Raman scheme are still 
performed at high temperatures $(T \approx 0.3 E_F)$, new methods of 
reducing the temperature and of creating artificial spin-orbit or 
color-orbit fields in the laboratory are underway~\cite{spielman-2017}
using radio-frequency chip technology~\cite{goldman-2010}.

Thus, in this paper, we describe theoretically the possibility of 
color superfluidity in the presence of color-orbit and color-flip fields
for trapped fermionic isotopes of Lithium ($^{6}$Li), Potassium ($^{40}$K)
or Ytterbium ($^{173}$Yb) with three internal states, 
to which we assign the color indices Red $(R)$, Green $(G)$ and Blue $(B)$. 
The remainder of the paper is organized as follows.
In section~\ref{sec:hamiltonian}, we discuss in detail the independent 
particle Hamiltonian and its spectrum, as well as the interaction terms 
between fermions of different colors. We also introduce the mixed-color 
representation that diagonalizes the independent particle Hamiltonian, 
which is used later in 
section~\ref{sec:hamitonian-mixed-color-basis} to clarify the origin of 
various color superfluid phases. 
In section~\ref{sec:saddle-point-approximation}, we discuss the emergence of color 
superfluidity in the presence of color-orbit and color-flip fields,
within the saddle point approximation at low temperatures. We solve 
the self-consistent equations for the order parameter tensor and 
particle number, and obtain the low temperature phase diagram in the space of
color-flip versus interaction parameter, for fixed color-orbit coupling. 
Due to the presence of color-orbit coupling and color-flip fields, a set 
of five color superfluid phases emerge, with characteristic nodal structures
in their excitation spectrum. These phases have a topological structure 
similar to Lifshitz transitions in metals under 
high pressure~\cite{lifshitz-1960}, but arise only due to the simultaneous
presence of color-orbit coupling, color-flip fields and interactions.
Where five nodal color superfluid phases merge, we identifiy a quintuple 
point that is also pentacritical, given that the transition between superfluid
phases is continuous. We also show that the low temperature transitions 
between normal and color superfluid phases are continuous when color-orbit 
coupling is present and are discontinuous, when color-orbit coupling is absent.
In section~\ref{sec:hamitonian-mixed-color-basis}, we describe 
the Hamiltonian in the mixed-color basis to make evident the emergence
of the momentum dependence of the order parameter tensor and to shine 
light on the physical origin of the nodal structure of the quasiparticle
and quasihole excitation spectrum. Furthermore, we also write the order 
parameter tensor in a total pseudo-spin basis to show that 
pairing can occur in singlet, triplet and quintet sectors and to make 
comparisons with the case of spin-1/2 fermions, where only the singlet
and triplet channels exist. 
In section~\ref{sec:spectroscopic-properties}, we analyse the quasiparticle
and quasihole excitation spectra in the mixed-color basis and show how 
the nodal structure and gaps emerge in the elementary excitation spectrum. 
Furthermore, we compute the momentum distributions for different colored 
fermions in various quantum phases to show how this easily measurable 
quantity can be used to identify different normal and superfluid phases.
We also compute the density of states of colored fermions and show how
they change as different normal and superfluid phases are visited in
the phase diagram. In section~\ref{sec:final-remarks}, we discuss the 
applicability and limitations of the current work and comment on the role of
Efimov states at lower particle densities, the possible emergence of 
non-uniform color superfluidity over a narrow region of the phase diagram, and
the effects of fluctations near the critical temperature between normal and
color superfluid states. Lastly, in section~\ref{sec:conclusions}, we summarize 
our conclusions.

%
%

%
\section{Hamiltonian}
\label{sec:hamiltonian}

In order to describe interacting three-component fermions labeled
by color states Red ($R$), Green ($G$) and Blue ($B$), under the influence of 
color-orbit and color-flip fields, we begin with the most general 
independent particle Hamiltonian resulting from  
suitably designed radio-frequency chip or Raman beams 
in the rotating frame
\begin{eqnarray}
\label{eqn:independent-particle-hamiltonian-matrix-first}
{\bf H}_0({\bf k})
= 
\left(
\begin{array}{ccc}
\varepsilon_R({\bf k}) 	& \Omega_{RG}	         &   \Omega_{RB}	  \\
\Omega_{RG}^* 		& \varepsilon_G({\bf k}) &   \Omega_{GB}	  \\
\Omega_{GB}^* 		& \Omega_{GB}^*	         & \varepsilon_B({\bf k}) \\
\end{array} 
\right),
\end{eqnarray}
where 
$
\varepsilon_c({\bf k}) = ({\bf k}-{\bf k}_c)^2/(2m) + \eta_c
$
represents the energy of internal state with color
$
c = 
\{
R,G,B
\}
$
after a color dependent net momentum transfer
${\bf k}_c$ provided by the Raman beams or by the radio-frequency chip. 
Here, $\eta_c$ is a color-dependent reference energy of the atom in 
color state $c$. 
The matrix elements $\Omega_{c c^{\prime}}$ represent a color-flip tensor
(Rabi frequencies) that couple different atomic color states $c$ and $c^{\prime}$. 
We note in passing the use of units where Planck's and Boltzmann's constants
are equal to one, that is,  $\hbar = k_B = 1$.

We will be using throughout the manuscript the terminology 
{\it independent particle} instead of {\it single particle} when referring
to Hamitonians, energies or other properties that describe a collection 
of a large number of particles that do not interact with each other. 
We follow the Feynman school tradition~\cite{feynman-1963}, where 
the {\it single particle} nomenclature is reserved to describe just 
an individual particle rather than a collection of non-interacting particles.  

Instead of analysing the most general theoretical case
shown in Eq.~(\ref{eqn:independent-particle-hamiltonian-matrix-first}), 
we explore the simplest non-trivial experimental configuration, where 
the component $\Omega_{RB}$ of the color-flip tensor is zero, 
indicating that there is no coupling between states $R$ and $B$, 
that is, $\Omega_{RB} = 0$. In addition, we consider that the matrix elements
that couple states $R$ to $G$ or $G$ to $B$ are real and 
equal, that is, 
$\Omega_{RG} = \Omega_{RG}^* = \Omega_{GB} = \Omega_{GB}^* = \Omega$.
Furthermore, we choose a symmetric situation, where momentum 
transfers occur only to color states R and B, such that 
${\bf k}_R = k_T {\hat {\bf x}}$,
${\bf k}_G = 0$,
and ${\bf k}_B = -k_T {\hat {\bf x}}$,
where $k_T$ is the magnitude of the momentum transferred 
to the atom by photons. 
Lastly, we can define an overall energy reference via the sum 
$
\sum_c \eta_c = \eta,
$
leading to internal energies 
$
\eta_R = -\delta,
$
$
\eta_B = \eta
$ 
and
$
\eta_G = +\delta,
$
where $\delta$ represents the detuning of the photon frequencies for transitions
between color states. Thus, next, we discuss the simplest experimentally 
relevant independent particle Hamiltonian for color states with 
color-dependent momentum transfer and color-flip terms.

\subsection{Independent particle Hamiltonian}
\label{sec:independent-particle-hamiltonian}

For the simplest experimental realization discussed above,
the independent particle Hamiltonian for three color states described
in Eq.~(\ref{eqn:independent-particle-hamiltonian-matrix-first}) becomes
\begin{equation}
\label{eqn:independent-particle-hamiltonian-matrix}
{\bf H}_{\rm IP}({\bf k}) 
= 
\varepsilon ({\bf k}){\bf 1} 
- h_x({\bf k}){\bf J}_x 
- h_z({\bf k}){\bf J}_z 
+ b_z {\bf J}_z^2
\end{equation}
where ${\bf J}_{\ell}$, with $\ell = \{ x, y, x \}$, 
are spin-one angular momentum matrices, 
$
\varepsilon ({\bf k}) = {\bf k}^2/(2m) + \eta 
$ 
is a reference kinetic energy which is identical 
for all color states,
$
h_x({\bf k}) = -\sqrt{2} \Omega
$ 
plays the role of a color-flip field (like a spin-flip 
Zeeman field for spins), 
and 
$
h_z({\bf k}) = 2k_T k_x /(2m) + \delta
$
plays the role momentum dependent Zeeman field along $z$-axis.
Notice that 
$
h_z({\bf k}) = 2k_T k_x /(2m) + \delta
$
has two components. The first one $2k_T k_x /(2m)$ represents color-orbit 
coupling controlled by the momentum transfer magnitude $k_T$, and the
second one represents a color-shift field controlled by the 
detuning $\delta$ (like a Zeeman shift for spins).
Notice that $h_z({\bf k})$ is transverse to the momentum transfer direction 
($x$-axis).  Lastly, the term $b_z = k_T^2/(2m) - \eta$ is a 
quadratic color-shift (quadratic Zeeman shift) associated with the momentum 
transfer along the $x$ direction.

To make further connections to QCD, we note that the independent particle 
Hamiltonian described in Eq.~(\ref{eqn:independent-particle-hamiltonian-matrix})
in general does not commute with the Gell-Mann matrices 
$\boldsymbol{\lambda}_j$, which 
are the eight generators of SU(3). To see this explicitly,
it sufficient to recall that the angular momentum matrices ${\bf J}_{\ell}$ can be
written in terms of $\boldsymbol{\lambda}_j$ as
${\bf J}_x = ({\boldsymbol \lambda}_1 + \boldsymbol{\lambda}_6)/\sqrt{2}$ 
along the x-direction;
${\bf J}_y = (\boldsymbol{\lambda}_2 + \boldsymbol{\lambda}_7)/\sqrt{2}$ 
along the y-direction; and
${\bf J}_z = ({\boldsymbol\lambda}_3 + \sqrt{3}\boldsymbol{\lambda}_8)/2$ 
along the z-direction and to 
show that the commutator $[{\bf H}_{\rm IP}, \boldsymbol {\lambda}_j] \ne 0$.
The Hamiltonian above becomes SU(3) invariant only when 
the coefficients $h_x ({\bf k}) = h_z ({\bf k}) = b_z = 0$, rendering
${\bf H}_{\rm IP} ({\bf k})$ diagonal and proportional to the unit matrix 
${\bf 1}$, that is, all color states become degenerate.

A very similar independent particle Hamiltonian was created in 
the laboratory for spin-one bosonic $^{87}$Rb atoms~\cite{spielman-2016}, 
where magnetic phases were investigated. Here, however, the independent
particle Hamiltonian corresponds to colored fermions, with potential
candidates being $^{6}$Li, $^{40}$K and $^{173}$Yb. Thus, the independent 
particle Hamiltonian matrix of 
Eq.~(\ref{eqn:independent-particle-hamiltonian-matrix}) takes the 
explicit matrix form
\begin{eqnarray}
\label{eqn:independent-hamiltonian-matrix}
{\bf H}_{\rm IP}({\bf k}) 
= 
\left(
\begin{array}{c c c}
\varepsilon_{\rm R} ({\bf k}) & -h_x ({\bf k})/\sqrt{2} & 0 \\
-h_x ({\bf k})/\sqrt{2} & \varepsilon_{\rm G} ({\bf k}) & 
-h_x ({\bf k})/\sqrt{2} \\
0 & -h_x ({\bf k})/\sqrt{2} & \varepsilon_{\rm B} ({\bf k}) \\  
\end{array}
\right),
\end{eqnarray}
where the function  
$
\varepsilon_{\rm R} ({\bf k}) 
=
\varepsilon ({\bf k}) - h_z ({\bf k}) + b_z
$
represents the diagonal matrix element for the Red (R) fermion, 
the function 
$
\varepsilon_{\rm G} ({\bf k}) 
=
\varepsilon ({\bf k}) 
$
represents the diagonal matrix element 
for the Green (G) fermion, 
and the function 
$
\varepsilon_{\rm B} ({\bf k}) 
=
\varepsilon ({\bf k}) + h_z ({\bf k}) + b_z
$
represents the diagonal matrix element 
for the Blue (B) fermion, 

In second-quantized notation, the independent particle Hamiltonian is
\begin{equation}
\label{eqn:independent-particle-hamiltonian-operator}
{\hat H}_{\rm IP}
= 
\sum_{\bf k}
{\bf F}^{\dagger} ({\bf k})
{\bf H}_{\rm IP}({\bf k})
{\bf F} ({\bf k})
\end{equation}
where the spinor operator is 
$
{\bf F}^{\dagger} ({\bf k}) 
= 
\left[
f_R^{\dagger}({\bf k}),
f_G^{\dagger}({\bf k}),
f_B^{\dagger}({\bf k})
\right]
$,
with
$
f_c^{\dagger}({\bf k})
$
creating a fermion with momentum 
${\bf k} - {\bf k}_c$ in internal color state $c = \{ R, G, B \}$. 
In order to bring the independent particle Hamiltonian matrix into
a diagonal form, we introduce next a mixed-color representation.

\subsection{Mixed color representation}
\label{sec:mixed-color-representation}

The Hamiltonian matrix ${\bf H}_{\rm IP} ({\bf k})$ is diagonalized via 
the rotation 
$
{\bf \Phi} ({\bf k}) 
= 
{\bf R}({\bf k}) 
{\bf F} ({\bf k}),
$
where the rotation matrix ${\bf R}({\bf k})$ 
satisfies the unitarity condition 
${\bf R}^\dagger ({\bf k}) {\bf R} ({\bf k}) = {\bf 1}$.
The spinor ${\bf \Phi} ({\bf k})$ represents the basis of 
independent particle eigenstates, whose elements are expressed as linear 
combinations of the elements of spinor ${\bf F} ({\bf k})$ in 
the original color basis via the rotation matrix
\begin{eqnarray}
\label{eqn:unitary-matrix}
{\bf R}({\bf k}) 
= 
\left(
\begin{array}{c c c}
R_{\Uparrow R} ({\bf k}) & R_{\Uparrow G} ({\bf k}) & R_{\Uparrow B} ({\bf k}) \\
R_{0 R} ({\bf k}) & R_{0 G} ({\bf k}) & R_{0 B} ({\bf k}) \\
R_{\Downarrow R} ({\bf k}) & R_{\Downarrow G} ({\bf k}) & R_{\Downarrow B} ({\bf k})  
\end{array}
\right),
\end{eqnarray}
where the normalization condition
$
\sum_{c}
\vert R_{\alpha c} ({\bf k}) \vert^2 
= 
1
$
for each row guarantees the unitarity of ${\bf R} ({\bf k})$.

In this case, the independent particle Hamiltonian becomes 
\begin{equation}
{\hat H}_{\rm IP} 
=
\sum_{\bf k}
{\bf \Phi}^\dagger ({\bf k}) {\bf H}_{\rm M} ({\bf k}) {\bf \Phi} ({\bf k}),
\end{equation}
where the spinor representing the diagonal basis is
$
{\bf \Phi}^{\dagger} ({\bf k}) 
= 
\left[
\phi^{\dagger}_{\Uparrow} ({\bf k}),
\phi^{\dagger}_{0} ({\bf k}),
\phi^{\dagger}_{\Downarrow} ({\bf k})
\right],
$ 
with $\phi^{\dagger}_{\alpha} ({\bf k})$ being the creation 
operator of a fermion with eigenenergy $\mathcal{E}_{\alpha} ({\bf k})$
and mixed-color  label $\alpha = \{ \Uparrow, 0, \Downarrow \}$.
The Hamiltonian matrix in diagonal form is
\begin{equation}
{\bf H}_{\rm M}({\bf k}) 
= 
{\bf R}({\bf k})
{\bf H}_{\rm IP}({\bf k}) 
{\bf R}^{\dagger} ({\bf k})
\end{equation}
with matrix elements 
$
\left[
{\bf H}_{\rm M}
\right]_{\alpha \beta} ({\bf k}) 
=
{\mathcal E}_{\alpha} ({\bf k}) \delta_{\alpha \beta},
$ 
where $\mathcal{E}_{\alpha} ({\bf k})$ 
are the eigenvalues of the matrix ${\bf H}_{\rm IP} ({\bf k})$ shown 
in Eqs.~(\ref{eqn:independent-particle-hamiltonian-matrix})
and~(\ref{eqn:independent-hamiltonian-matrix}).
Finally, the independent particle Hamiltonian in the mixed-color basis 
is simply written as 
\begin{equation}
{\hat H}_{\rm IP} 
=
\sum_{\bf k}
{\cal E}_{\alpha} ({\bf k}) 
\phi_{\alpha}^\dagger ({\bf k}) \phi_{\alpha} ({\bf k}),
\end{equation}
where the mixed-color operator $\phi_{\alpha} ({\bf k})$ is 
written as a linear combination of the color operators
$f_{c} ({\bf k})$ via the matrix elements $R_{\alpha c} ({\bf k})$, 
that is, 
$
\phi_{\alpha} ({\bf k})
=
\sum_{c}
R_{\alpha c} ({\bf k}) f_c ({\bf k}).
$
This is the general structure of eigenenergies and eigenstates of 
a system of independent colored particles in the presence of color-flip
and color-orbit fields. Next, we discuss a couple of simple limits and 
a specific example of this eigensystem. 

\subsection{Independent particle spectrum}
\label{sec:independent-particle-spectrum}

We discuss first the independent particle eigenenergies  
in the limit where the quadratic color-shift term is zero, 
that is, $b_z = 0$ or $\eta = k_T^2/(2m)$, but the Rabi coupling $\Omega \ne 0$ 
and the detuning $\delta \ne 0$. In this situation,
the eigenvalues of ${\hat H}_{\rm IP}$ have the simple form 
\begin{equation}
{\cal E}_{\alpha} ({\bf k})
= 
\varepsilon ({\bf k}) 
- m_{\alpha} 
\vert h_{\rm eff} ({\bf k}) \vert,
\end{equation}
where the effective {\it magnetic} field magnitude is 
$
\vert h_{\rm eff} ({\bf k}) \vert
= 
\sqrt{h_x^2 ({\bf k}) + h_z^2 ({\bf k})},
$
and $m_{\Uparrow} = +1$, $m_0 = 0$, and $m_{\Downarrow} = -1$. 
In this case, the independent particle Hamiltonian ${\hat H}_{\rm IP}$
describes simply a pseudo-spin 1 system in the presence of 
the effective external field 
$
{\bf h}_{\rm eff}({\bf k}) = 
\left[ 
h_x ({\bf k}), 0, h_z ({\bf k})
\right].  
$

\begin{figure} [hbt]
\centering
\epsfig{file=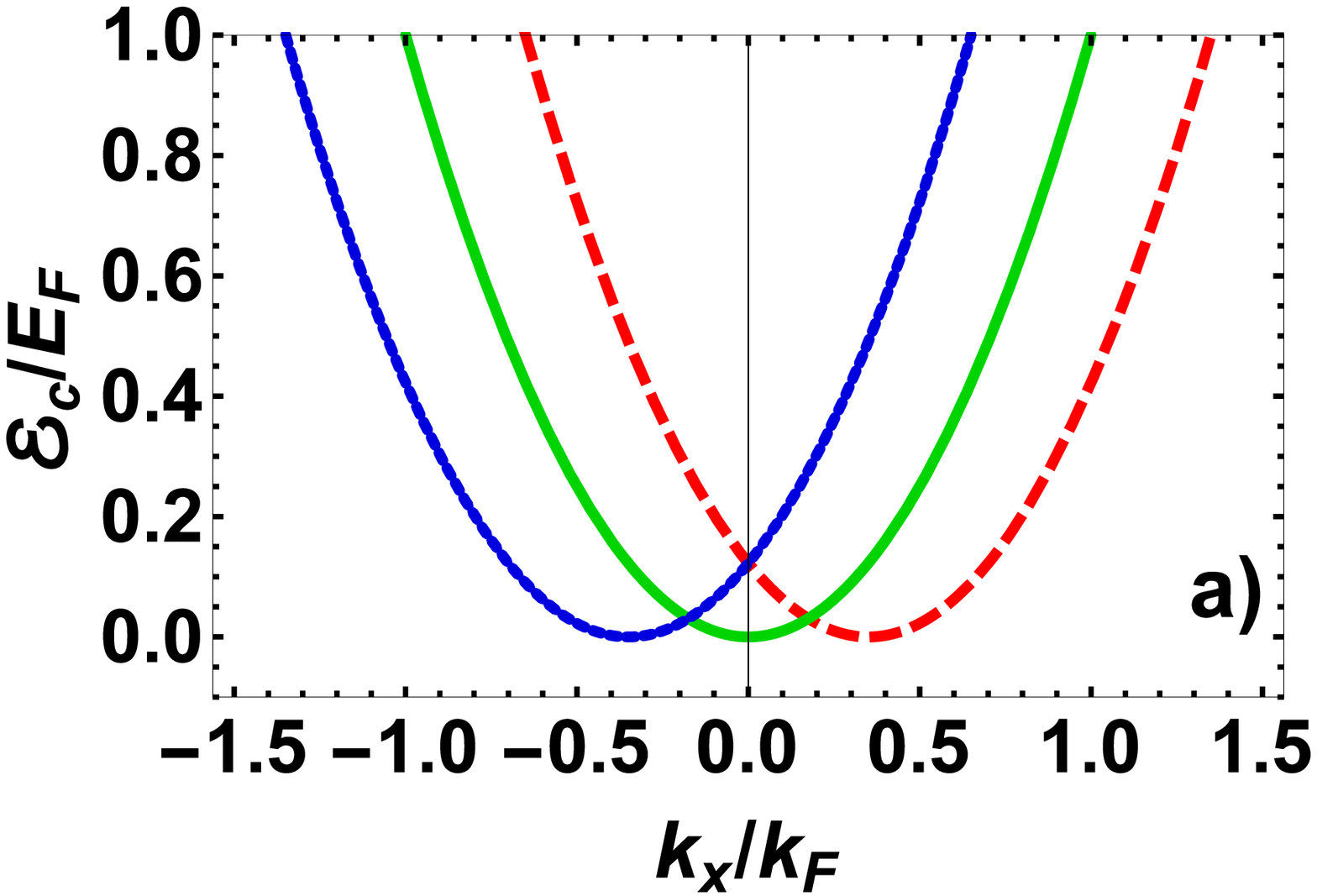,width=0.48 \linewidth}
\epsfig{file=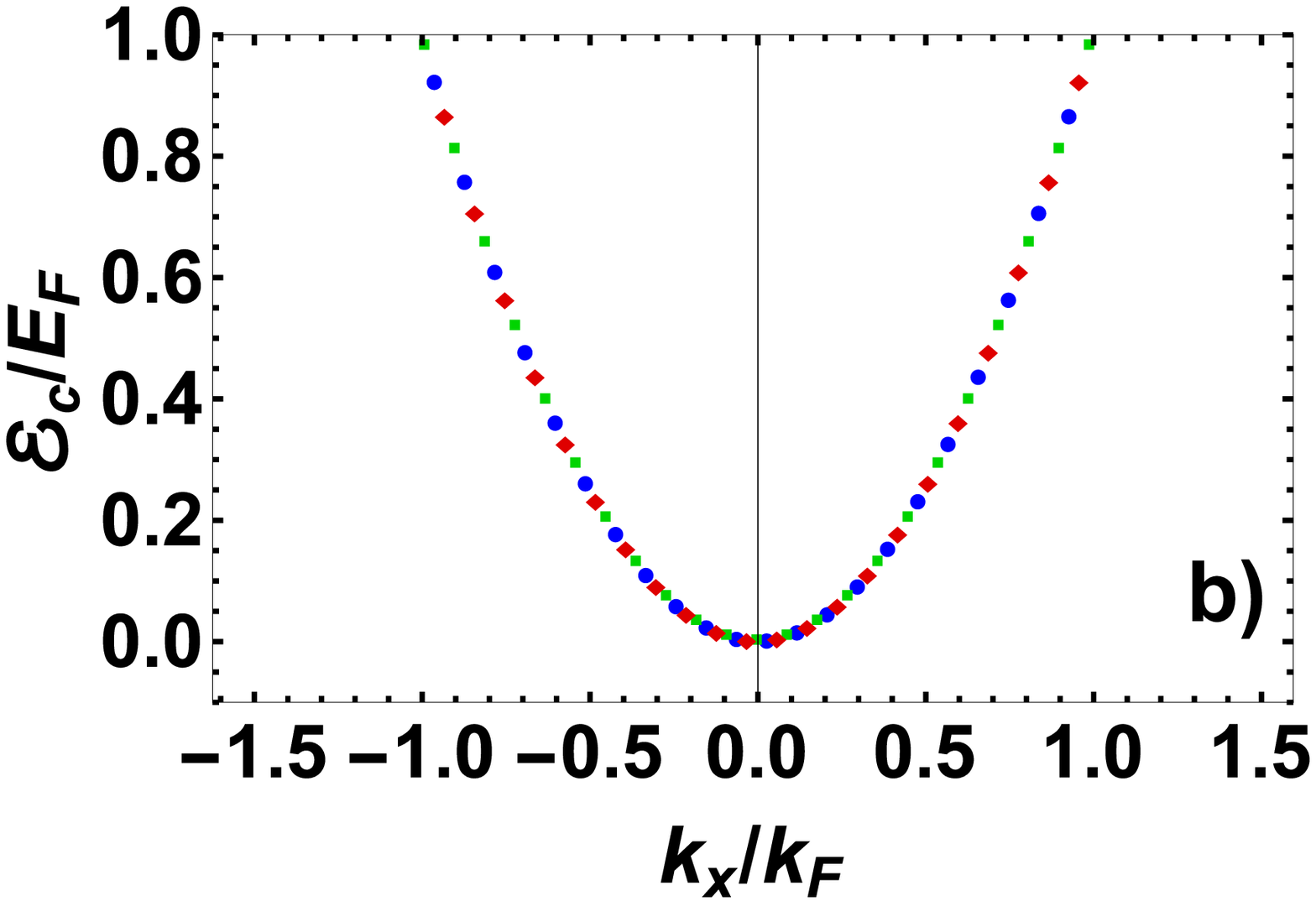,width=0.48 \linewidth}
\caption{
\label{fig:one}  
(Color online) 
Energy dispersions $\varepsilon_c ({\bf k})$ 
for Red ($R$), Green ($G$) and Blue ($B$) states
versus momentum along the $k_x$ direction, for parameters
$b_z = k_T^2/(2m)$ $(\eta = 0)$ and $\delta = 0$. The momentum transfer
in a) is $k_T = 0.35 k_F$ and in b) is $k_T = 0$. 
The dashed red curve describes the $R$ states,
solid green curve describes the $G$ states, and the dotted blue 
curve describes the $B$ states. 
Notice that 
the Red dispersion is shifted to the right, 
the Green dispersion has no shift, and
the Blue dispersion is shifted to the left when $k_T \ne 0$, 
but that all dispersions are identical when $k_T = 0$. 
}
\end{figure}

A second simple limit of the more general color problem 
with $b_z \ne 0$, $\eta \ne 0$, $\Omega \ne 0$ and $\delta \ne 0$
discussed in Sec.~\ref{sec:independent-particle-hamiltonian}
corresponds to the case where
$b_z = k_T^2/(2m)$, $\eta = 0$, $\Omega = 0$ and $\delta = 0$. 
In this situation the kinetic energies of the Red, Green and Blue states 
are respectively 
$
\varepsilon_R ({\bf k}) = \varepsilon ({\bf k} - {\bf k}_T),
$
$
\varepsilon_G ({\bf k}) = \varepsilon ({\bf k}),
$
$
\varepsilon_B ({\bf k}) = \varepsilon ({\bf k} + {\bf k}_T ),
$
indicating that the Blue states are shifted towards negative momenta along
the x direction, while the Red states are shifted towards positive momenta along
the x direction, since ${\bf k}_T = k_T \hat{\bf x}.$ In the limit
where there is no color-orbit coupling, that is $k_T = 0$, it is clear 
that the three kinetic 
energies are identical $\varepsilon_R ({\bf k}) = \varepsilon_G ({\bf k}) =
\varepsilon_B ({\bf k}) = \varepsilon ({\bf k}).$ These two situations are 
illustrated in Fig.~\ref{fig:one}.
%

From now on, in addition to setting $\eta = 0$ with $b_z = k_T^2/(2m)$, 
we will also set the detuning $\delta$ to zero, 
leading to $h_z ({\bf k}) = 2 k_T k_x/(2m)$, but we keep the
color-flip term $h_x ({\bf k} = - \sqrt{2}\Omega$ with zero or non-zero
Rabi frequencies $\Omega$. This case is chosen 
to simplify the number of parameters involved, given that the problem 
of color superfluidity in the presence 
of color-orbit coupling and color-flip fields is sufficiently novel, 
and thus there is no need to make matters more complex than they need to be. 
In passing, we mention that 
the case of finite detuning $(\delta \ne 0)$ is also very rich given 
that parity is not conserved, thus affecting the normal state and 
superfluid phases that emerge, as well 
as their topological properties.

Throughout the text, the energy scale that we use is the Fermi energy 
$
E_F = k_F^2/(2m),
$
where 
$
k_F = (2 \pi^2 n)^{1/3},
$
is the Fermi momentum defined from 
total density of fermions $n = 3 k_F^3/(6 \pi^2)$, where the
factor of $3$ reflects the three colors $\{ R, G, B \}$. The 
same kinetic energies $\varepsilon ({\bf k}) = {\bf k}^2/(2m)$ are used here 
for all internal color states, since our momentum $(k_F)$ and energy 
$(E_F)$ units are fixed
by setting the parameters $\eta, k_T, \Omega$ and $\delta$ to zero. 

\begin{figure} [hbt]
\centering
\epsfig{file=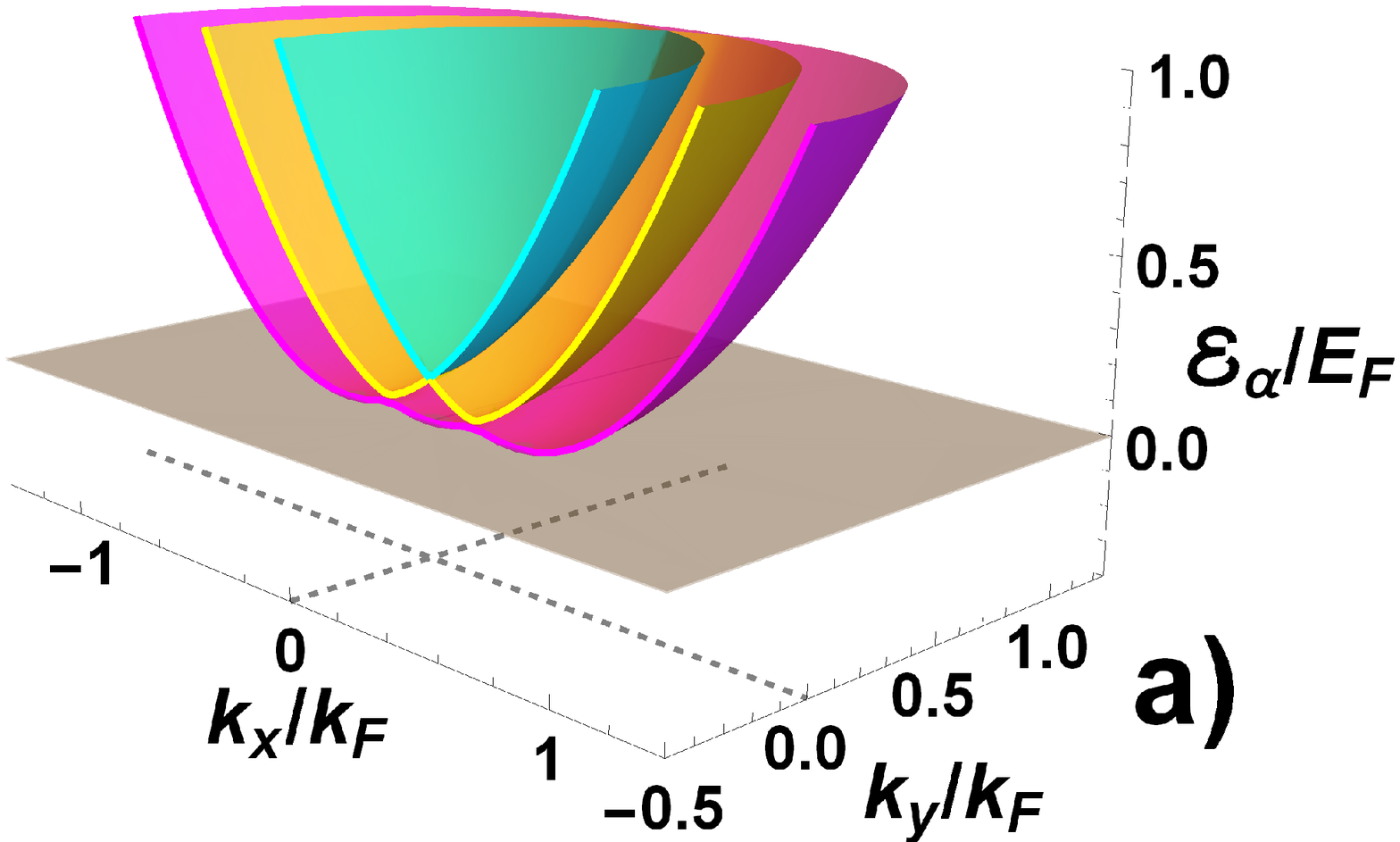,width=0.48 \linewidth}
\epsfig{file=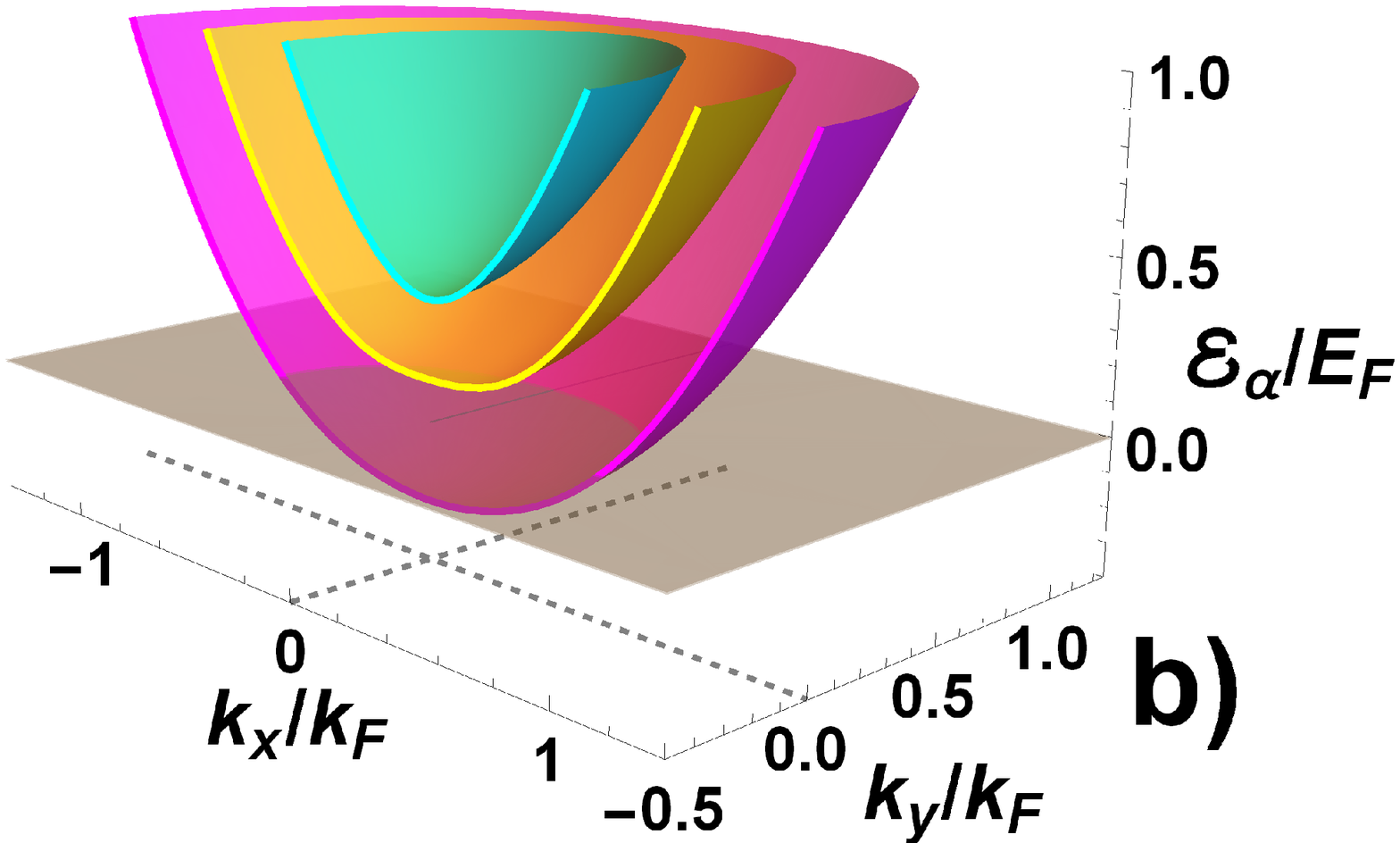,width=0.48 \linewidth}
\caption{
\label{fig:two}  
(Color online) 
Three dimensional plots of the mixed-color
eigenvalues $\mathcal{E}_{\alpha} ({\bf k})$ 
versus $(k_x, k_y)$ with $k_z = 0$, 
when the quadratic color-shift is $b_z = k_T^2/(2m)$ $(\eta = 0)$ 
and the color-orbit coupling momentum transfer is $k_T = 0.35k_F$.  
In a) the color-flip field is $\Omega = 0.01 E_F$,
where the lower band (magenta) has three minima,
the middle band (yellow) has two minima, and the
upped band (cyan) has one minimum. 
In b) the color-flip field is $\Omega = 0.2E_F$,
where all three bands (lower, middle, upper) have a single minimum.
}
\end{figure}

In Fig.~{\ref{fig:two}}, we show plots of 
eigenvalues $\mathcal{E}_{\alpha} ({\bf k})$ versus 
momentum $(k_x, k_y)$ with $k_z = 0$, 
for fixed momentum transfer $k_T = 0.35 k_F$,
zero detuning $\delta = 0$, and quadratic color-shift
$b_z = k_T^2/(2m)$ $(\eta = 0)$. 
In Fig.~{\ref{fig:two}}a, we show the case 
where the color flip term  is small $( \Omega = 0.01E_F )$
where single, double and triple minima are shown in the 
upper, middle and lower eigenergies, respectively. 
In Fig.~{\ref{fig:two}}b, we show the case 
where the Rabi frequency (color-flip term) 
is sufficiently large $\Omega = 0.2E_F$,
such that the upper, middle and lower eigenergies have only a single minimum.
The presence of color-orbit $(k_T \ne 0)$ and color-flip 
$(\Omega \ne 0)$ fields lifts the degeneracy of the color states 
$\{ R, G, B \}$, which are assumed to have the same dispersion 
$\varepsilon ({\bf k}) = {\bf k}^2/(2m)$ when $k_T = 0$, 
$\Omega = 0$, and $\eta = 0$. For a fixed color-orbit coupling
$k_T = \gamma k_F$, we can estimate when the three minima in the lower 
mixed-color band disappear by comparing the energy of the crossing 
point between the Red and Green or Green and Blue energy dispersions. 
The crossing points occur at momenta 
$k_x = \pm k_T/2 = \pm (\gamma/2) k_F$, so when 
$\Omega \sim (k_T/2)^2/(2m) = (\gamma^2/4) E_F$, the three minima of the 
lowest mixed-color band coalesce into one. The two minima in the middle 
mixed-color band become a single one, when the color-flip field $\Omega$ is 
of the order of the energy diference between the crossing points between 
Red-Blue bands and Red-Green or Green-Blue bands, that is, when
$\Omega \sim \left[ k_T^2/(2m) - (k_T/2)^2/(2m) \right] = (3\gamma^2/4)E_F$. 
For the specific case of $k_T = 0.35 k_F$, the three minima of
the lowest mixed-color band disappear when $\Omega \sim 0.03 E_F$ 
and the two minima of the middle mixed-color band disappear when 
$\Omega \sim 0.09 E_F$, therefore in Fig.~\ref{fig:two}b, where
$\Omega = 0.2 E_F$, each one of the three mixed-color bands have 
a single minimum.

%

Now that we analysed the independent particle Hamiltonian for three-color
fermions in the presence of color-orbit and 
color-flip fields, we are ready to discuss next the effects of 
interactions.

\subsection{Interaction Hamiltonian}
\label{sec:interaction-hamiltonian}

We begin our discussion of the effects of interactions between different
color states in cold atoms by recalling that gluons are the mediators of 
quark-quark interactions in the color superconductivity problem encountered 
in quantum chromodynamics (QCD). Therefore, due to their dynamical nature, 
the interactions between quarks have a finite range
contribution, and cannot be assumed to be zero-ranged. However, the situation 
encountered in cold atoms is simpler than in QCD, because the atomic fermions 
interact essentially via zero-ranged forces.

The interactions between different color states of cold fermions 
are essentially zero-ranged and attractive  
$-g_{cc^{\prime}} \delta ({\bf r} - {\bf r}^{\prime})$ of 
strength $g_{c c^{\prime}} > 0$, between internal states with different
colors only, that is, $c \ne c^{\prime}$. 
The use of contact (zero-ranged) interactions really  
means that the interaction range $R_{cc^\prime}$
is much smaller than the interparticle distance $k_F^{-1}$, 
which is indeed the situation encountered experimentally in ultra-cold fermions,
since these atoms are neutral.  

Experimentally, interactions between atoms in different color (internal) states
occur predominantly in the s-wave channel at low temperatures.
Thus, we consider only interactions between the Red-Green 
$(g_{RG})$, Red-Blue $(g_{RB})$ and Green-Blue $(g_{GB})$ states 
to be non-zero, while all the Red-Red, Green-Green
and Blue-Blue interactions are negligible, that is, $g_{RR} = g_{GG} = g_{BB} = 0$.
However, within the set of s-wave interactions, we could
still have different interaction parameters, that is, $g_{RG} \ne g_{RB} \ne g_{GB}$.
The zero-ranged nature of the interactions between colored fermions
allows us to describe our system in terms of 
s-wave scattering lengths $a_{s, cc^\prime}$ between different colors, as we
shall see later.

In momentum coordinates, the interaction part of the Hamiltonian has
the structure 
\begin{eqnarray}
\label{eqn:atom-atom-interaction-hamiltonian}
{\hat H}_{\rm INT} 
= 
-
\frac{1}{V} 
\sum \limits_{{\bf Q},\left\{c \neq c^{\prime}\right\}} 
g_{cc^{\prime}} 
a_{cc^{\prime}}^{\dagger}({\bf Q})
a_{cc^{\prime}}({\bf Q}),
\end{eqnarray}
where the volume of space is $V$, the paired-colors creation operators are 
$
a_{cc^{\prime}}^{\dagger}({\bf Q}) 
= 
\sum_{\bf k} 
f_c^{\dagger}({\bf k} + {\bf Q}/2)
f_{c^{\prime}}^{\dagger}(-{\bf k} - {\bf Q}/2),
$ 
with their center of mass momentum being ${\bf Q}$. 
Here, the operators $f_c^{\dagger}({\bf K})$ represent the creation
of a fermion with color $c = \{R, G, B\}$ and momentum ${\bf K}$.
As we shall see soon, the expectation values of 
the operator $a_{cc^{\prime}}^{\dagger}({\bf Q})$ in a superfluid state 
describe the emergence of Cooper pairs in the BCS regime, and of 
tightly-bound pairs in BEC limit at low temperatures. We note in passing 
that $g_{c c^{\prime}}$ has dimensions of energy times volume. 

Having analysed the interactions between different colored fermions, we will 
discuss next the full Hamiltonian, including the addition of the chemical
potential to describe thermodynamic states with fixed average number of 
particles.

\subsection{Full Hamiltonian}
\label{sec:full-hamiltonian}

The full Hamiltonian for a color superfluid with color-orbit coupling,
color-flip fields and contact interactions is
\begin{equation}
\label{eqn:full-hamiltonian}
{\hat H}
=
{\hat H}_{\rm IP}
+
{\hat H}_{\rm INT}
-
\mu {\hat N},
\end{equation}
with
$
{\hat N} 
= 
\sum_{c,{\bf k}}
f_c^{\dagger}({\bf k})
f_c({\bf k})
$ 
representing the total number of colored fermions. 

Having written the full Hamiltonian for the colored fermion problem with 
attractive interactions, we will discuss next the emergence of various 
color superfluid ground states, the nodal structure of their quasiparticle 
excitations and the low temperature phase diagram 
in the color-flip versus interaction parameter space for 
fixed color-orbit coupling.

\section{Saddle-point approximation}
\label{sec:saddle-point-approximation}

In order to study the emergence of color superfluidity in the
presence of color-flip and color-orbit fields, 
we focus here on superfluid phases of paired states
with with zero center of mass momentum, that is, ${\bf Q} = {\bf 0}$. 
Thus, the only relevant pairing operator is 
$
a_{cc^{\prime}}^{\dagger}({\bf 0}) 
= 
\sum_{\bf k} 
f_c^{\dagger}({\bf k})
f_{c^{\prime}}^{\dagger}(-{\bf k}).
$ 
This implies that the emerging superfluid states are uniform throughout 
the sample volume, as it is discussed next, when 
we introduce the saddle-point approximation. This appoximation is
known to give excellent results at low temperatures
as it captures both the emergence of large Cooper pairs in the BCS region and
the emergence of tightly bound pairs (two-body bound states) 
in the BEC regime~\cite{sademelo-1997}. 

\subsection{Order Parameter and reduced Hamiltonian}
\label{sec:order-parameter-and-reduced-Hamiltonian}

Considering only pairing with zero center-of-mass momentum 
$({\bf Q} = {\bf 0})$, the order parameter for color superfluidity 
is defined by the tensor 
$
\Delta_{cc^\prime} 
= 
-g_{cc^\prime} \langle a_{cc^\prime} ({\bf 0}) \rangle/V,
$
with color indices $c \ne c^{\prime}$ describing paired states $RG$, $RB$ and $GB$. 
Using a mean-field (saddle-point) approximation for the interaction term 
in Eq.~(\ref{eqn:atom-atom-interaction-hamiltonian}) leads 
to the reduced Hamiltonian
\begin{equation}
{\hat H}_0 =
\frac{1}{2}
\sum_{\bf k}
{\bf f}_N^\dagger ({\bf k}) {\bf H}_0 ({\bf k}) {\bf f}_N ({\bf k})
+ 
V
\sum_{c \ne c^\prime}
\frac{\vert \Delta_{cc^\prime}\vert^2}{g_{cc^\prime}}
+
{\cal C}(\mu),
\end{equation}
where the six-dimensional field operator
$
{\bf f}_N^\dagger ({\bf k})
=
\left[
f_{\rm R}^\dagger ({\bf k}), 
f_{\rm G}^\dagger ({\bf k}),
f_{\rm B}^\dagger ({\bf k}),
f_{\rm R}({\bf -k}), 
f_{\rm G}({\bf -k}),
f_{\rm B}({\bf -k})
\right],
$
represents a colored-Nambu spinor, while the term 
$
{\cal C}(\mu) 
= 
\frac{1}{2} 
\sum_{{\bf k} c}
\xi_c ({\bf -k})
$
contains the kinetic energy of colored fermions 
$
\xi_c ({\bf k}) 
=
\varepsilon_c ({\bf k}) - \mu,
$
which contributes to the ground state energy. 

The saddle-point Hamiltonian matrix is
\begin{eqnarray}
\label{eqn:saddle-point-hamiltonian-matrix}
{\bf H}_{\rm 0}({\bf k})
=
\left(
\begin{array}{cc}
\overline {\bf H}_{\rm IP}({\bf k})	&	{\boldsymbol \Delta}		\\
{\boldsymbol \Delta}^{\dagger}	&    -\overline {\bf H}_{\rm IP}^* (-{\bf k})
\end{array}
\right),
\end{eqnarray}
where the $3 \times 3$ diagonal block matrix
$
\overline {\bf H}_{\rm IP}({\bf k})
=
{\bf H}_{\rm IP}({\bf k})
-\mu {\bf 1}
$
represents the independent particle Hamiltonian with 
respect to the chemical potential $\mu$
and the $3\times3$ off-diagonal block matrix
\begin{eqnarray}
\label{eqn:order-parameter-tensor}
{\boldsymbol \Delta}
=
\left(
\begin{array}{ccc}
0		&	\Delta_{RG}	&	\Delta_{RB}	\\
-\Delta_{RG}	&	0		&	\Delta_{GB}	\\
-\Delta_{RB}	&	-\Delta_{GB}	& 	0
\end{array}
\right)
\end{eqnarray}
represents the order parameter tensor $\Delta_{c c^\prime}$, 
which is clearly anti-symmetric since its transpose is equal to its
negative ${\bf \Delta}^{T} = - {\bf \Delta}$,
and thus traceless: ${\rm Tr} \left[{\bf \Delta} \right] = 0$.

The quasiparticle and quasihole excitation spectrum can be 
found by diagonalizing the matrix shown in 
Eq.~(\ref{eqn:saddle-point-hamiltonian-matrix}) 
or via the determinant 
$
P(\omega) 
= 
{\rm det}
\left[ \omega {\bf 1} - {\bf H}_0 ({\bf k}) \right].
$
The characteristic polynomial
$
P(\omega) 
= 
\prod_{j} 
\left[\omega - E_j ({\bf k})\right]
$ 
is of sixth degree, however, in the limit of zero detuning, where
the color-shift field $\delta = 0$, 
we can use both quasiparticle-quasihole and parity symmetries 
to reduce $P (\omega)$ to the bicubic polynomial
$
P(\omega) = 
\left[\omega^2 - E_1^2 ({\bf k})\right]
\left[\omega^2 - E_2^2 ({\bf k}) \right]
\left[\omega^2 - E_3^2 ({\bf k})\right],
$
that can be solved analytically using Cardano's method~\cite{cardano-2007}.
We show explicit solutions for 
$E_j ({\bf k})$ in section~\ref{sec:spectroscopic-properties}, 
but we warn the reader that the analytic solutions are not
particularly illuminating. 

In general, the six energy eigenvalues can be ordered as  
$
E_1({\bf k})
>
E_2({\bf k})
>
E_3({\bf k})
>
E_4({\bf k})
>
E_5({\bf k})
>
E_6({\bf k}),
$
and exhibit quasiparticle/quasihole symmetry
in momentum space for any chosen value of
the color-flip field (Rabi frequency) $\Omega$ or 
color-shift field (detuning) $\delta$. 
In this case, we can choose quasiparticle/quasihole partners as follows: 
$
E_6({\bf k}) = -E_1(-{\bf k}),
$
$
E_5({\bf k}) = -E_2(-{\bf k})
$
and
$
E_4({\bf k}) = -E_3(-{\bf k}).
$
However, each eigenergy $E_j ({\bf k})$ has well defined parity only
when the color-shift field (detuning) is zero, that is $\delta = 0$, 
in which case  $E_j ({\bf k}) = E_j ({\bf - k})$ is an even function 
of momentum ${\bf k}$.

Since the excitation spectrum $E_j ({\bf k})$ depends explicitly on 
the order parameter tensor $\Delta_{c c^\prime}$ and the chemical
potential $\mu$, we establish next self-consistency relations for 
both quantities at fixed total density of colored fermions.

\subsection{Self-consistency equations}
\label{sec:self-consistency-equations}

The excitation spectrum $E_j ({\bf k})$ is determined by solving
for the values of the order parameter amplitudes 
$\Delta_{RG}, \Delta_{RB}, \Delta_{GB}$ and the chemical potential $\mu$
self-consistently. Starting from the thermodynamic potential
$
\mathcal{Q}_{0}
= 
-T
\ln
\mathcal{Z},
$
where 
$
\mathcal{Z} 
= 
\int
\Pi_{c}D
\left[
f^{\dagger}_c ({\bf k})
,
f_c ({\bf k})
\right]
\exp
\left[
-S
\right]
$
is the grand-canonical partition function and 
$S$ is the action, we obtain the saddle point action to be
$$
T S_{0} 
= 
-
\frac{1}{2}
\sum_{n,{\bf k} }
{\bf f}_{N}^{\dagger} ({\bf k})
{\bf G}^{-1} 
{\bf f}_{N} ({\bf k})
+
V \sum_{c \ne c^\prime} \frac{\vert \Delta_{cc^\prime} \vert^2}{g_{cc^\prime}}
+
{\mathcal C} (\mu), 
$$
where 
$
{\bf G}^{-1} (i\omega_n, {\bf k})
=
\left[
i\omega_n {\bf 1}
-
{\bf H}_{0}({\bf k})
\right]
$
is the inverse of the resolvent (Green) matrix 
${\bf G} (i\omega_n, {\bf k})$. Here, 
$
\omega_n = (2n+1)\pi T
$
is the fermionic Matsubara frequency, and 
$T$ is the temperature. 
Integration over the fermionic fields gives 
the saddle-point thermodynamic potential
$
{\mathcal Q}_{\rm 0} 
= 
{\mathcal A}_{\rm 0} + {\mathcal C} (\mu),
$ 
with 
\begin{equation}
{\mathcal A}_{\rm 0}
=
-\frac{T}{2}
\sum_{\bf k}\sum_{j = 1}^3 
\ln 
\left\{
2 
+
2 \rm{cosh}\left[ \frac{E_j ({\bf k})}{T} \right] 
\right\}
+
V \sum_{c \ne c^\prime} \frac{\vert \Delta_{cc^\prime} \vert^2}{g_{cc^\prime}}
\end{equation}
where the sum over the index $j$ is over quasiparticles only, that
is $(j = \{1, 2, 3 \})$, given that we used the
quasiparticle/quasihole symmetry described above.

Minimizing ${\mathcal Q}_{0}$ with respect to $\Delta_{cc^\prime}^*$ 
via the condition $\delta {\mathcal Q}_{0}/\delta \Delta_{cc^\prime}^* = 0$
leads to three order parameter equations
\begin{equation}
\label{eqn:order-parameter-equation}
\frac{V}{g_{cc^\prime}} \Delta_{cc^\prime}
=
\frac{1}{2} 
\sum \limits_{\bf k}
\sum \limits_{j=1}^3
\tanh\left(
\frac{\beta E_j({\bf k})}{2}
\right) 
\frac{\partial E_j({\bf k})}{\partial \Delta_{cc^\prime}^*},
\end{equation}
for the available choices of 
$
\Delta_{cc^\prime} 
= \{ \Delta_{RG}, \Delta_{RB},\Delta_{GB} \},
$ 
since $c \ne c^\prime$.
The total number of particles is fixed
via the thermodynamic relation 
$N = - \partial{\mathcal Q}_{0}/\partial \mu \vert_{T,V}$
leading to the number equation
\begin{equation}
\label{eqn:number-equation}
N 
=  
\frac{1}{2}
\sum \limits_{\bf k}
\left[
\sum \limits_{j=1}^3
\tanh\left(
\frac{\beta E_j({\bf k})}{2}
\right) 
\frac{\partial E_j({\bf k})}{\partial \mu}
+
3
\right].
\end{equation}
In the current problem, we can only fix the total number of colored fermions,
because for arbitrarily small color-flip field $\Omega$ the number operator
$
{\hat N}_c 
= 
\sum_{\bf k} 
f^\dagger_c ({\bf k}) f_c ({\bf k})
$
for a given color $c$ does not commute with the full Hamiltonian ${\hat H}$
described in Eq.~(\ref{eqn:full-hamiltonian}). For instance, the commutator
of the independent particle Hamiltonian ${\hat H}_{\rm IP}$ and the color 
number operator ${\hat N}_m$ for color $m$ is 
$
\left[
{\hat H}_{\rm IP}, {\hat N}_m 
\right]
=
\sum_{{\bf k}, c} 
\{
\left[{\bf H}_{\rm IP}\right]_{m c}
f^\dagger_c ({\bf k}) f_m ({\bf k})
-
\left[{\bf H}_{\rm IP}\right]_{c m}
f^\dagger_m ({\bf k}) f_c ({\bf k})
\},
$
which only vanishes if the matrix 
${\bf H}_{\rm IP}$ is diagonal, that is, when the color-flip field
that causes transitions between different color states is zero: 
$h_x ({\bf k}) = 0$ (or $\Omega = 0$). 

Before solving the self-consistency equations derived above, 
we use the generalized Lippman-Schwinger relation
$
V/g_{cc^\prime} 
= 
-mV/(4 \pi a_{s, cc^\prime}) 
+ 
\sum_{\bf k} 
(
\varepsilon_c ({\bf k}) 
+ \varepsilon_{c^\prime} ({\bf k}) 
)^{-1},
$
to express the bare coupling constant $g_{cc^\prime}$ in terms 
of the scattering length $a_{s,cc^\prime}$ in the absence 
of the color-orbit and color-flip fields, 
where we assume that the masses and the energy dispersions 
of all colored fermions are the same, 
that is, in this expression we take explicitly
$
\varepsilon_c ({\bf k}) 
= 
\varepsilon_{c^\prime} ({\bf k}) 
=
{\bf k}^2/(2m).
$

A more general case in the context of ultra-cold atoms corresponds to the 
situation where the interactions between different colors 
are not the same, that is, $g_{RG} \ne g_{RB} \ne g_{GB}$, which leads to 
different scattering lengths $a_{RG} \ne a_{RB} \ne a_{GB}$.
This is indeed a more general situation, however, in fermionic isotopes 
of Ytterbium~\cite{takahashi-2010a,fallani-2016}, it is possible 
to select three internal atomic states such that the interactions 
are the same, that is, $g_{RG} = g_{RB} = g_{GB} = g$, which leads to 
$a_{RG} = a_{RB} = a_{GB} = a_s$, and to tune the scattering lengths via
optical Feshbach resonances~\cite{takahashi-2008, takahashi-2010b}. 
Such techniques may allow explorations of deep connections to 
SU(3) symmetric interactions in the context of color-superconductivity 
of quark-matter.  Furthermore, when color-flip
and color-orbit fields are considered in systems consisting of 
three internal states of fermionic isotopes of Lithium, Potassium or 
Ytterbium~\cite{kurkcuoglu-2016a,kurkcuoglu-2016b},
even the simple limits of: 
a) single-channel interactions $g_{RG} = g_{GB} = 0$ and $g_{RB} \ne 0$ 
$(a_{RG} = a_{GB} = 0$ and $a_{RB} \ne 0)$ 
can lead to color-superfluidity~\cite{kurkcuoglu-2016a} or 
b) no interactions at all $g_{RG} = g_{GB} = g_{RB} = 0$ 
$(a_{RG} = a_{GB} = a_{RB} = 0)$ can lead to 
non-trivial spinor physics~\cite{kurkcuoglu-2016b}. 
 
In this work, we consider the case where s-wave interactions 
between different colors are exactly the same, that is, 
$g_{RG} = g_{GB} = g_{RB} = g$, but with $g_{RR} = g_{GG} = g_{BB} = 0$.
This implies that the pairing amplitudes between fermions of different 
colors are also identical, that is, 
$\Delta_{RG} = \Delta_{GB} = \Delta_{RB} = \Delta$,
while $\Delta_{RR} = \Delta_{GG} = \Delta_{BB} = 0$.
In this case, the order parameter tensor $\Delta_{cc^\prime}$ is fully 
anti-symmetric, but characterized by a single complex scalar $\Delta$,
which is independent of momentum ${\bf k}$. In such a situation, 
self-consistency is achieved via a single order parameter equation. 
However, in general, 
interactions can also be different 
$g_{RG} \ne  g_{GB} \ne g_{RB} = g$ ($g_{RR} = g_{GG} = g_{BB} = 0$) 
leading automatically to non-identical order parameter components
$\Delta_{RG} \ne \Delta_{GB} \ne \Delta_{RB} $ 
(with $\Delta_{RR} = \Delta_{GG} = \Delta_{BB} = 0$) and 
three distinct order parameter equations described in 
Eq.~(\ref{eqn:order-parameter-equation}).

As mentioned above, in the case of identical s-wave interactions, 
the order parameter is characterized by a single complex component 
$\Delta$ in all pairing channel of colors $\{c, c^{\prime}\} = \{R, G, B\}$. 
However, we will see in Sec.~\ref{sec:hamitonian-mixed-color-basis} 
that the order parameter becomes a momentum dependent tensor 
$\Delta_{\alpha \beta} ({\bf k})$ 
with unequal components in all mixed color channels $\{\alpha, \beta\}$, 
reflecting the different interactions and momentum dependences that 
are induced by color-flip and color-orbit fields into the mixed color 
representation of the Hamiltonian. 

Before we present the phase diagram of colored fermions in the presence
of color-orbit and color-flip fields, it is important to look first 
at the limit where both terms are zero, 
that is, $h_z ({\bf k}) = 0$ ($k_T = 0$ and $\delta = 0$)
as well as $h_x ({\bf k}) = 0$ $(\Omega = 0)$. This limit 
serves as a reference and is discussed next.

\subsection{Zero color-flip and color-orbit coupling}
\label{sec:zero-color-flip-and-color-orbit-coupling}

In the limit where color-orbit and color-flip fields vanish the evolution 
of a color superfluid from the BCS to the BEC regime is relatively simple. 
The independent particle Hamiltonian matrix ${\bf H}_{\rm IP} ({\bf k})$ 
defined in Eq.~(\ref{eqn:independent-hamiltonian-matrix})
simplifies dramatically, since 
$h_z ({\bf k}) = 0$ ($k_T = 0$ and $\delta = 0$) 
and $h_x ({\bf k}) = 0$ $(\Omega = 0)$, making 
${\bf H}_{\rm IP} ({\bf k})$ diagonal and proportional to the 
unit matrix ${\bf 1}$, that is, 
${\bf H}_{\rm IP} ({\bf k}) = \varepsilon ({\bf k}) {\bf 1}$, 
with $\varepsilon ({\bf k}) = {\bf k}^2/(2m)$. This simplification 
makes the full Hamiltonian ${\hat H}$ defined 
in Eq.~(\ref{eqn:full-hamiltonian}) invariant under global $U(3)$ rotations
of the color states. This means that ${\hat H}$ commutes with the nine
generators of $U(3)$. Furthermore, it is possible to perform simultaneous
global $U(3)$ rotations in the independent particle 
$
\overline {\bf H}_{\rm IP} ({\bf k}) 
= {\bf H}_{\rm IP} ({\bf k}) - \mu {\bf 1}
$
and pairing ${\boldsymbol \Delta}$ sectors 
of the saddle point Hamiltonian 
matrix~(\ref{eqn:saddle-point-hamiltonian-matrix}),
such that the $U(3)$-rotated order parameter matrix
${\boldsymbol \Delta}_{U(3)}$ can be represented by a single complex
scalar $\Delta_{U(3)} = \Delta \sqrt{3}$, with only two non-vanishing
matrix entries, namely
$\left[{\boldsymbol \Delta}_{U(3)}\right]_{13} = \Delta \sqrt{3}$,
and 
$\left[{\boldsymbol \Delta}_{U(3)}\right]_{31} = -\Delta \sqrt{3}$.

In this limiting case, the quasiparticle spectrum is 
$
E_1 ({\bf k}) 
= 
E_2 ({\bf k}) 
= 
\sqrt{\xi^2 ({\bf k}) + 3 \vert \Delta \vert^2}
$
and $E_3 ({\bf k}) = \xi ({\bf k})$, where
$\xi ({\bf k}) = {\bf k}^2/(2m) - \mu$, corresponding to a fully gapped 
superfluid with two degenerate quasiparticle states and a third 
quasiparticle state which is passive, that is, it represents free 
non-interacting fermions. 
This occurs, because of the underlying global U(3) symmetry, which allows
rotations into a mixed color state, where only two mixed-colors 
are active in pairing, while the third one is passive. 
In this case, a standard BCS-to-BEC crossover occurs~\cite{baym-2010} 
similar to the standard case of two internal states~\cite{sademelo-1993}, 
but with the added feature that the third passive band provides 
a Fermi surface when the chemical potential lies above its minimum, 
and no Fermi surface when the chemical potential 
is below its minimum. This reference case is illustrated in the
phase diagram shown in Fig.~\ref{fig:three}b when $\Omega/E_F = 0$.
 
However, when color-orbit and color-flip fields are present the explicit 
global U(3) symmetry of the full Hamiltonian
${\hat H}$ in Eq.~(\ref{eqn:full-hamiltonian}) is broken 
and all colors are involved in pairing, 
producing a complex excitation spectrum that allows also 
for exotic gapless quantum phases and phase transitions between them, 
instead of a smooth crossover. These aspects are discussed next.

\subsection{Low temperature phase diagrams}
%
%
%

When color-orbit and color-flip fields are present, the global $U(3)$ symmetry 
is explicitly broken, and there is no longer an inert mixed-color band. 
This means that all mixed-color bands participate in pairing, and that the order
parameter tensor in the mixed-color representation has no longer a single entry above
the diagonal and a single entry below the diagonal. A detail analysis of the 
order parameter tensor in the mixed-color representation is performed in 
section~\ref{sec:hamitonian-mixed-color-basis}. 

In order to obtain the phase diagram and classify the emergent superfluid and 
normal phases it is sufficient to analyse the quasiparticle/quasihole excitation 
spectrum $E_j ({\bf k})$, since all phases seen from the $\{R, G, B\}$ color 
basis have the same order parameter tensor $\Delta_{c,c^\prime}$ and share 
the same color-symmetry controlled by a single complex component $\Delta$. 
However, as the amplitude of $\Delta$ and chemical potential $\mu$ vary as 
a function of the color-flip field $\Omega/E_F$ and interactions $1/(k_F a_s)$ 
for fixed color-orbit fields, the nodal structure of the spectrum 
$E_j ({\bf k})$ suffers dramatic changes in momentum space and 
Lifshitz-like transitions occur in the superfluid states.

\begin{figure} [hbt]
\centering
\epsfig{file=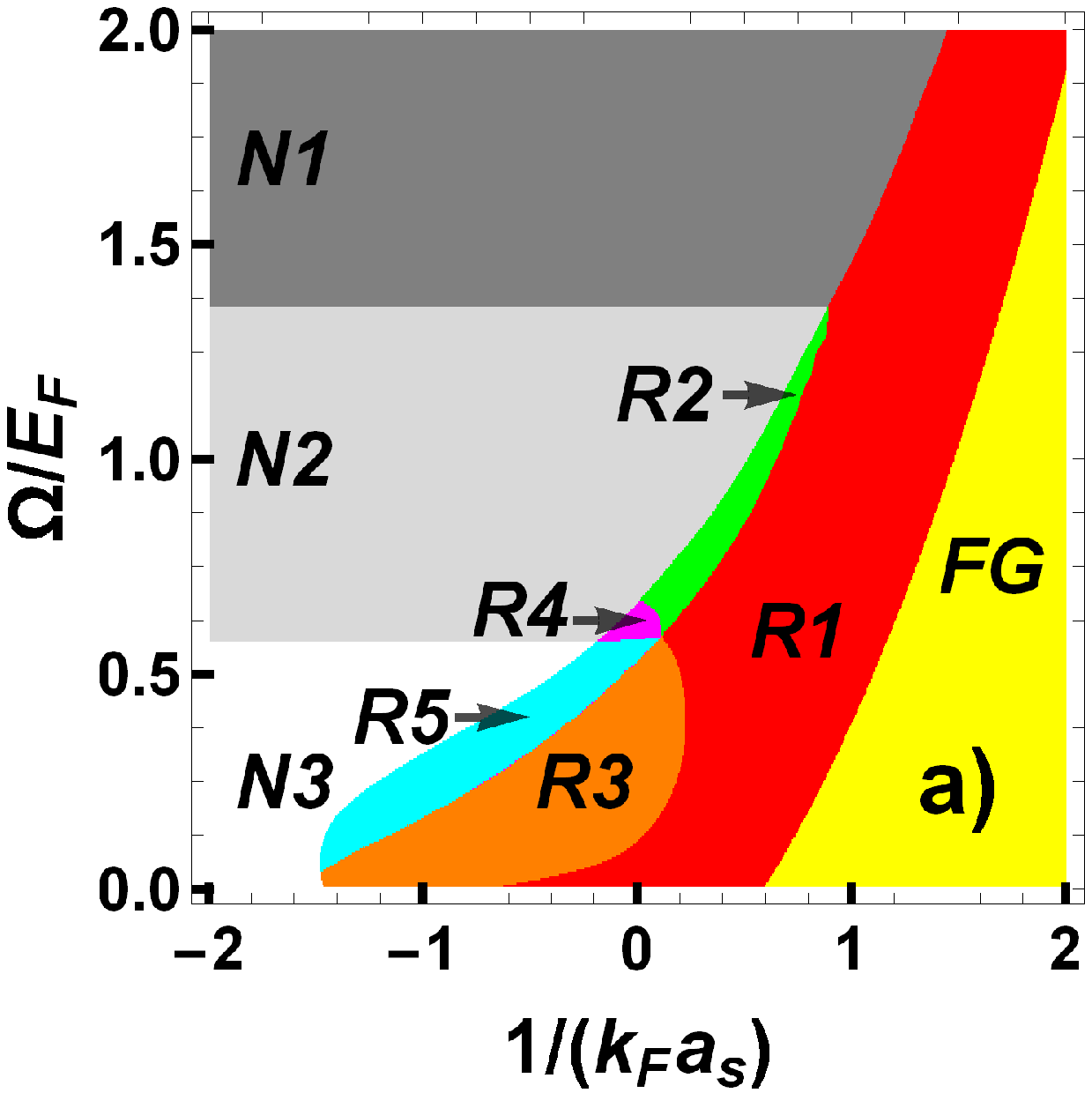,width=0.48\linewidth}
\epsfig{file=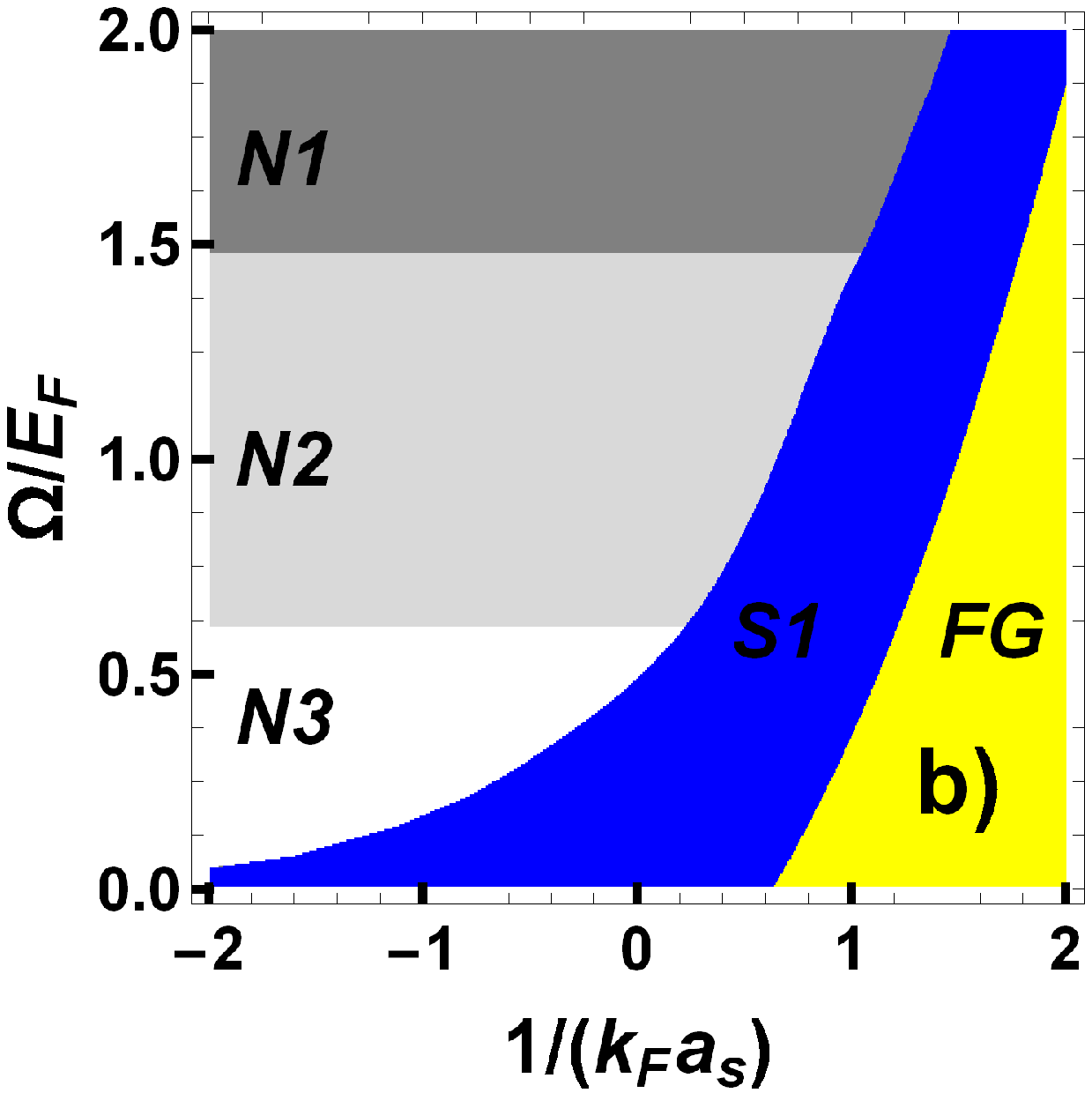,width=0.48 \linewidth}
\caption{ 
\label{fig:three}
(Color online) Phase diagrams of color-flip field $\Omega/E_F$ versus 
scattering parameter $1/(k_F a_s)$ for non-zero quadratic 
color-shift $b_z = k_T^2/2m$ $(\eta = 0)$ and two values color-orbit
coupling controlled by momentum transfer $k_T$. The temperature is $T/E_F = 0.01$.
In a) $k_T = 0.35 k_F$ and the superfluid phases are labelled according 
to the nodal structure of the quasiparticle excitation spectrum that 
the color-orbit coupling induces when the Rabi field $\Omega/E_F$ is non-zero. 
The normal phases $N1$, $N2$, and $N3$ are labelled according to the number of 
Fermi surfaces they possess. The superfluid phases are labeled according to the
number of nodal rings they possess ($R1$, $R2$, $R3$, $R4$, and $R5$). 
Phase transitions between various superfluid phases and between superfluid 
phases and normal states are continuous.
In b) $k_T = 0$ and the superfluid phases have either a fully nodal 
surface (S1) or a fully gapped (FG) phase. The fully nodal phase is 
reminiscent of the passive band when $\Omega = 0$, where only a crossover 
exist. The phase transition from 
the S1 phase to the normal phases is discontinous.
}
\end{figure}

In Fig.~\ref{fig:three}, we show the phase diagrams of color superfluids
in the color-flip field $\Omega/E_F$ versus interaction 
parameter $1/(k_F a_s)$ plane.
We consider only s-wave interactions between different colors and set  
the color-shift field to zero, that is, the detuning is set to zero $\delta = 0$, 
such that parity is preserved in the excitation spectrum. 
For Fig.~\ref{fig:three}a, the parameters are
$k_T = 0.35 k_F$, $b_z = k_T^2/(2m)$ $(\eta = 0)$ and 
for Fig.~\ref{fig:three}b, the parameters are $k_T = 0$, 
$b_z = k_T^2/(2m)$ $(\eta = 0)$. 
The contrast between the two figures is remarkable, indicating that the 
presence of color-orbit and color-flip couplings induce novel superfluids states 
as interactions are changed.
The phases $N1$, $N2$ and $N3$ correspond 
to the normal phases with one, two and three Fermi surfaces associated 
with the eigenvalues of ${\bf H}_{\rm IP} ({\bf k})$ and 
characterize the regime where the colored Fermi gas is degenerate. The lines 
separating these normal phases correspond to simple 
Lifshitz transitions~\cite{lifshitz-1960}. 

%

The superfluid phases in Fig.~\ref{fig:three}a 
are labeled according to their nodal structure, 
which in the present case, correspond to rings of zero-energy 
quasiparticles representing the residual Fermi surface of the 
starting degenerate colored Fermi gas. In the presence of color-orbit and 
color-flip fields, the quasiparticle excitation energies 
$ 
E_1({\bf k}), 
$
$
E_2({\bf k})
$
and 
$
E_3({\bf k})
$
have more complex momentum dependence, but only $E_3({\bf k})$ can 
have zeros. The zeros of $E_3 ({\bf k})$ define the 
{\it loci} (points, lines or surfaces) in momentum space,
where there is no energy cost to create quasiparticle excitations. 
The connectivity of these ${\it loci}$ of zero energy can be used to 
classify the topologically distinct superfluid phases of colored fermions. 

\begin{figure} [hbt]
\centering
\epsfig{file=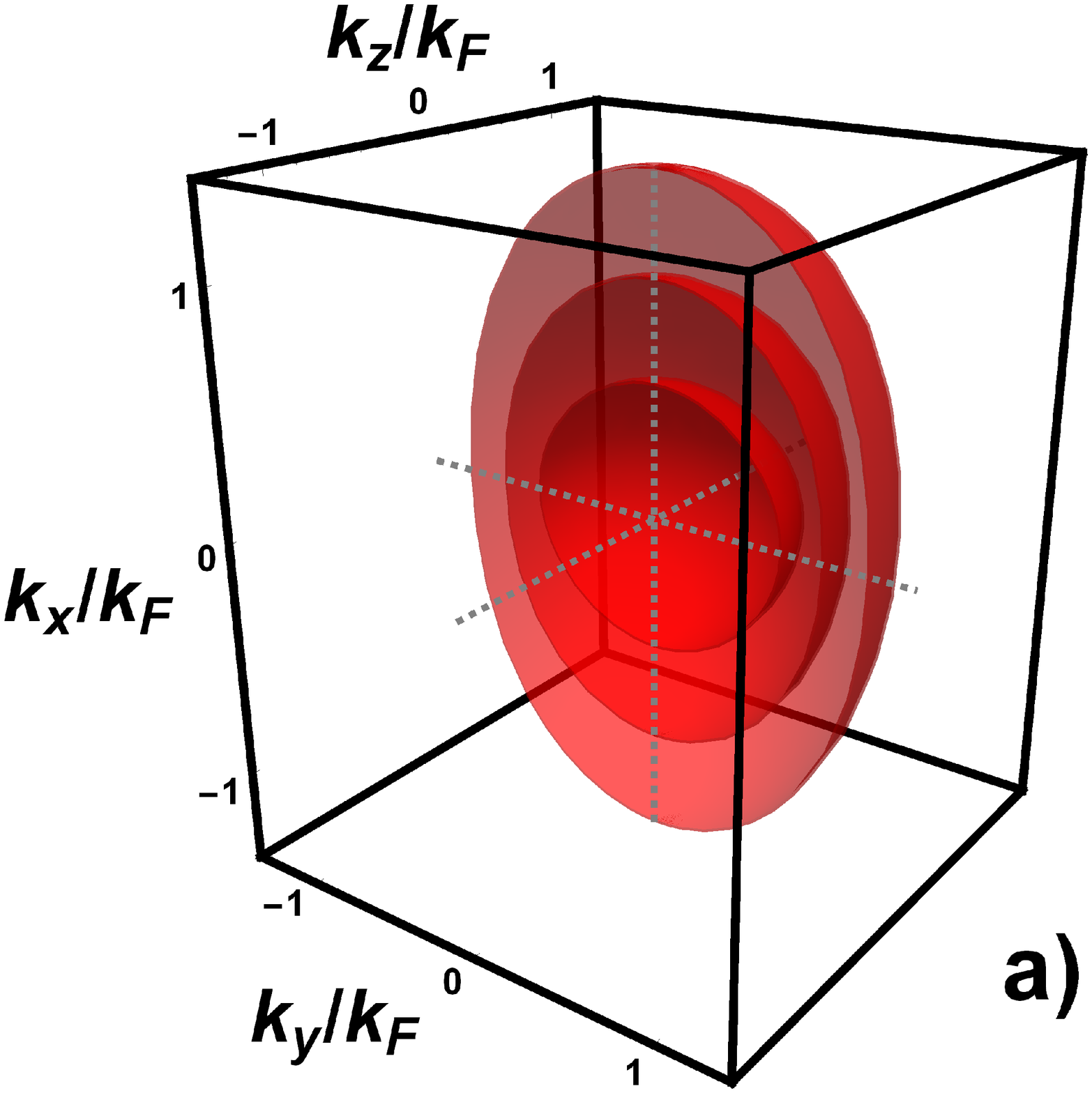,width=0.49 \linewidth}
\epsfig{file=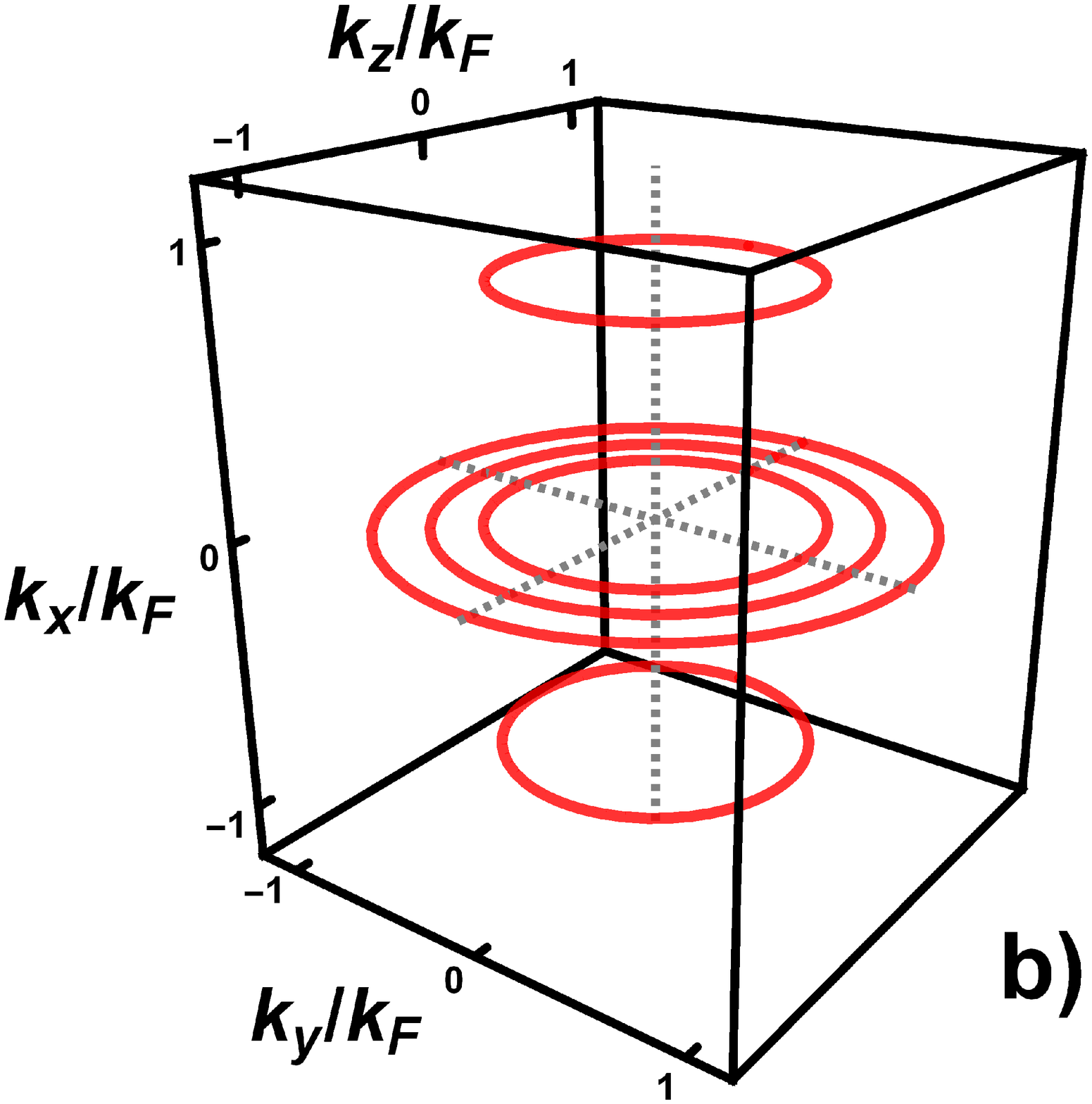,width=0.49 \linewidth}
\\
\epsfig{file=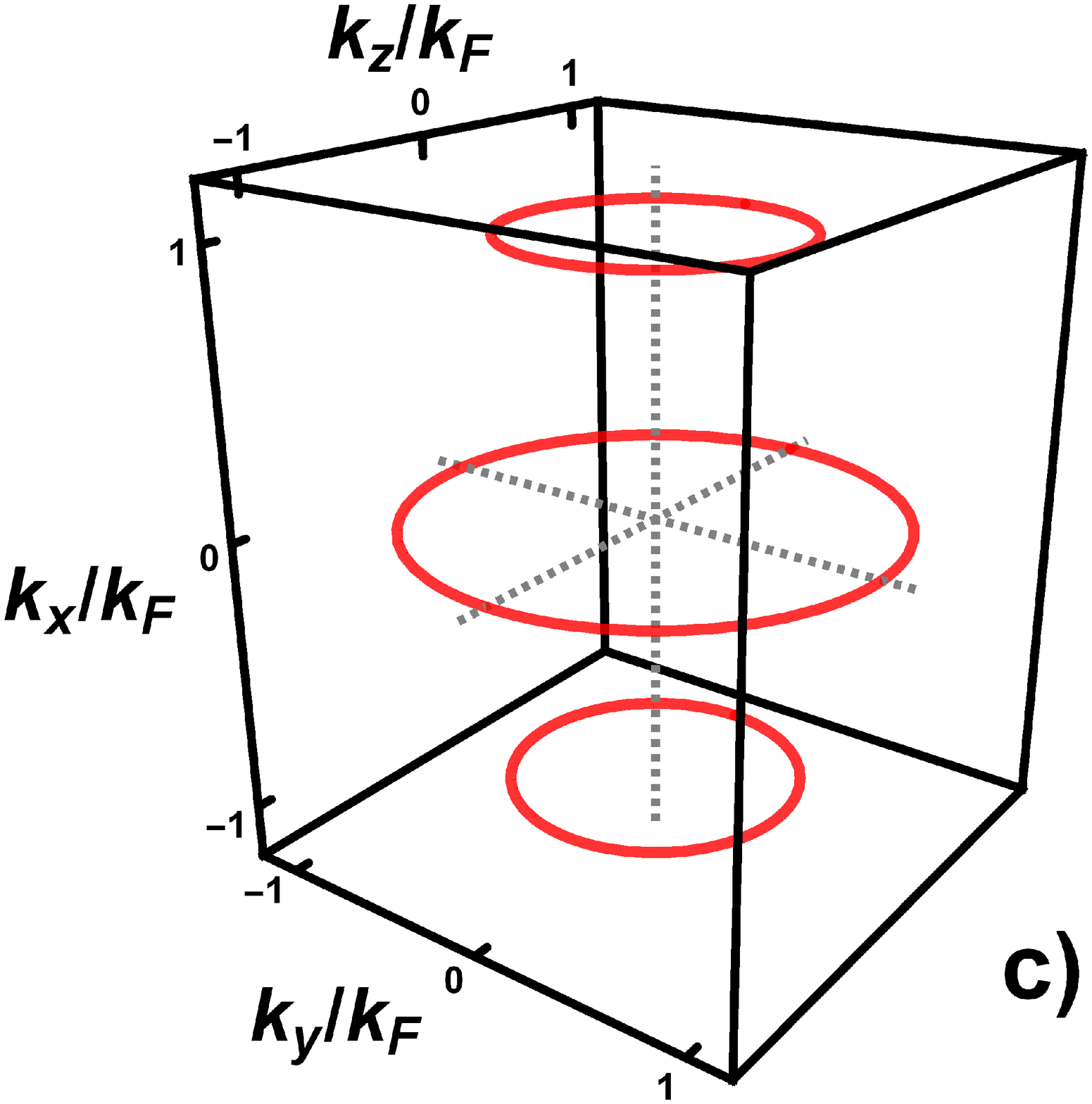,width=0.49 \linewidth}
\epsfig{file=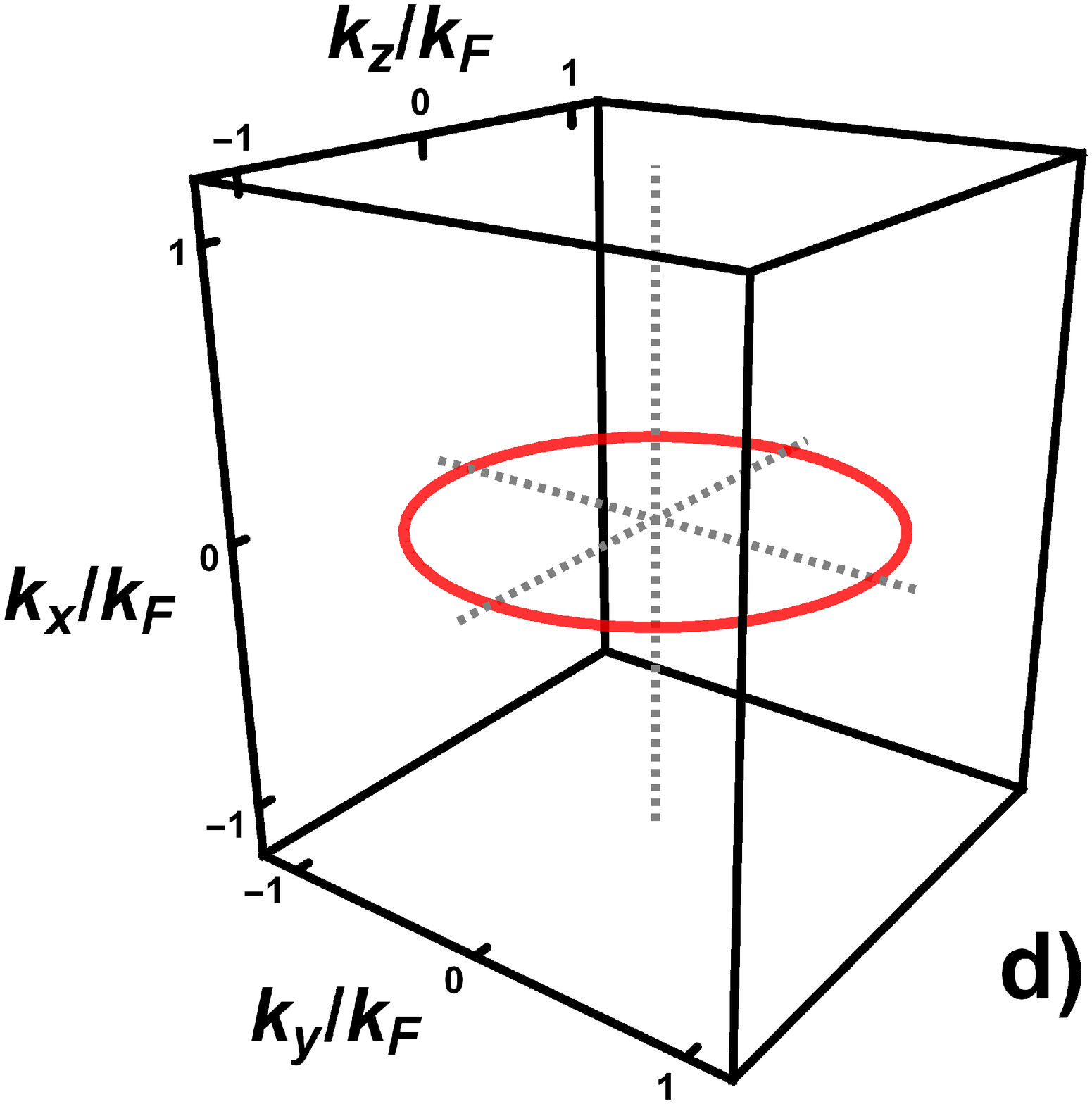,width=0.49 \linewidth}
\caption{ 
\label{fig:four}
(Color online) 
Nodal structure of the quasiparticle excitation spectrum
for phases $N3$, $R5$, $R3$ and $R1$ with parameters 
$\Omega = 0.29 E_F$, $k_T = 0.35 k_F$, and
$b_z = k_T^2/2m$ $(\eta = 0)$. 
a) Normal phase $N3$ with three Fermi surfaces, where
$1/(k_F a_s) = -1.31$, $\mu/E_F  = 0.97$ and $\vert \Delta \vert  = 0$;
b) Five-ringed superfluid phase $R5$, where 
$1/(k_F a_s) =  -1.03$, $\mu/E_F = 0.97$, $\vert \Delta \vert /E_F = 0.0056$; 
c) Three-ringed superfluid phase $R3$, where
$1/(k_F a_s)= -0.069$; $\mu/E_F = 0.81$, $\vert \Delta \vert /E_F = 0.31$;
d) One-ringed superfluid phase $R1$, where
$1/(k_F a_s) = 0.62$, $\mu/E_F = 0.19$, $\vert \Delta \vert/E_F = 0.73$.
}
\end{figure}

To analyse the phase diagram, we make use of the mixed color bands 
${\cal E}_\alpha ({\bf k})$ with $\alpha = \{ \Uparrow, 0, \Downarrow \}$,
as discussed in section~\ref{sec:mixed-color-representation}. 
For definiteness, we fix first the color-flip coupling
to $\Omega/E_F = 0.29$, in which case there are three mixed colored bands 
participating in pairing. 
As the scattering parameter $1/(k_F a_s)$ increases a nodal 
phase $R5$ with five rings, three in the outer $\Uparrow$, one the middle $0$, 
and one in the inner $\Downarrow$ band, gives way to
a nodal phase $R3$ with three rings in the outer $\Uparrow$ band, 
where the two internal rings {\it annihilate} at finite momenta 
in the $(0,k_y,k_z)$ plane. 
This leads to the opening of a gap at the $R5/R3$ boundary, but residual 
node lines persist. An additional increase of the scattering parameter 
leads to the one ring $R1$ phase in the outer $\Uparrow$ band, where the 
two other rings shrink to points at finite momenta along the $(k_x, 0, 0)$ 
direction when the phase boundary $R3/R1$ is reached. Finally, further 
increase in the scattering parameter transforms the $R1$ phase into 
a fully gapped $FG$ phase with no nodal regions. A similar analysis can be
done for different fixed values of $\Omega/E_F$ and varying \
scattering parameter. An important observation
is the existence of a quintuple point, where the five superfluid phases 
$R1, R2, R3, R4$ and $R5$ merge. The transitions between these topologically 
distinct superfluid phases are all continuous, therefore this quintuple 
point is also pentacritical. In Fig.~\ref{fig:four}, we plot the nodal structure 
of the $N3, R5, R3$ and $R1$ phases illustrated in the phase diagram 
of Fig.~\ref{fig:three}(a), for fixed value of $\Omega/E_F = 0.29$ and 
varying scattering parameter $1/(k_F a_s)$.

%
%

The richness of the phase diagram in 
Fig.~\ref{fig:three}a should be contrasted with simplicity of that in  
Fig.~\ref{fig:three}b, where the color-orbit coupling parameter is set
to zero, that is, $k_T = 0$. Indeed, in the case of Fig.~\ref{fig:three}b 
the phase diagram is much simpler. When the color-orbit 
and color-flip couplings are zero, that is, $k_T = 0$ and 
$\Omega = 0$, pairing occurs only between two mixed colors 
which produce a fully gapped superfluid, but the third mixed color 
is completely inert~\cite{baym-2010} and thus possesses the original
Fermi surface for non-interacting fermions when the chemical potential
lies above the minimum of the band. This situation corresponds to the line 
of $\Omega/E_F = 0$ in Fig.~\ref{fig:three}b and describes standard 
BCS-BEC crossover physics, where the superfluid
is always gapped as a function the interaction parameter 
$1/(k_F a_s)$. 

However, when $k_T$ is still zero, but the color-flip field 
$\Omega$ is non-zero, extra mixing of colors occur and the inert band 
becomes active. Nevertheless a single nodal surface continues to exist, 
as illustrated in the blue ($S1$) region. For $\Omega/E_F \ne 0$, the
quasi-particle dispersion $E_3 ({\bf k})$ is no longer identical 
to the independent particle energy $\xi ({\bf k})$, and the nodal structure
of $E_3 ({\bf k})$ is completely isotropic in momentum space. Thus, for finite
$\Omega/E_F$, there is a phase transition between a gapless superfluid with 
surface nodes $(S1)$ to a fully gapped $(FG)$ superfluid phase. 
For fixed $\Omega/E_F$, the line separating the $S1$ and $FG$ phases appears
when the degeneracy between the lowest energy mixed color band 
$\xi_{\Uparrow} ({\bf k }) = {\mathcal E}_{\Uparrow}({\bf k}) - \mu$
and its counterpart $-\xi_{\Uparrow} ({\bf k })$ at 
$\xi_{\Uparrow} ({\bf k }) = 0$ is lifted by non-zero 
color-flip fields $(\Omega \ne 0)$. This occurs when the chemical 
potential $\mu$ falls below the minimum value 
${\rm min}_{\bf k} \{ \mathcal{E}_{\Uparrow} ({\bf k}) \}$, 
when a full gap in the excitation spectrum $E_3 ({\bf k})$ emerges leading
to the yellow ($FG$) region in Fig.~\ref{fig:three}b. The
quasiparticle bands $E_1 ({\bf k})$ and $E_2 ({\bf k})$ are always gapped
in the present case.

In addition, when color-orbit coupling is zero $(k_T = 0)$, 
the transition from superfluid to normal phases is discontinuous 
(first order) at the low temperatures indicated 
in Fig.~\ref{fig:three}b and the phase boundary 
corresponds to the balancing of the color-flip ({\it magnetic}) 
energy $h_x \chi_{xx} h_x/2$, where $\chi_{xx}$ is the color-flip 
({\it magnetic}) susceptibility, and the condensation energy of the 
superfluid $\gamma \vert \Delta \vert^2$. This leads to 
$\Omega = \vert \Delta \vert \gamma/\chi_{xx}$ at the phase
boundary, where the order parameter amplitude $\vert \Delta \vert$ 
jumps discontinuously to zero. Such relation for color superfluids 
is a generalization of Clogston's result for Fermi superfluids paired with 
zero center of mass momentum in the singlet s-wave state of two spin-1/2 
fermions~\cite{clogston-1962}.

Another important difference between phase diagrams illustrated in 
Figs.~\ref{fig:three}a and~\ref{fig:three}b concerns the limit 
when the color-flip field vanishes, that is, when  
$\Omega \to 0$. In Fig.~\ref{fig:three}a the color-orbit coupling is 
non-zero, that is, $k_T \ne 0$, meaning that  
the Red-color (R) band is shifted to the left, the Green-color (G) 
band remains in the same place, and the Blue-color (B) band is shifted to 
the right, as seen in Fig.~\ref{fig:one}. Given that s-wave interactions only  
lead to $RG$, $RB$ and $GB$ pairs, this implies that pairing with zero center-of-mass 
momentum for $RB$ pairs can occur without energy cost, but pairing with zero 
center-of-mass momentum for $RG $ and $GB$ pairs cost energy of an amount
$k_T^2/(2m) \pm  k_x k_T/m$. Therefore, even for zero color-flip 
fields $(\Omega = 0)$,
a uniform zero center-of-mass superfluid phase is not favored until a critical
value of the interaction parameter $1/(k_F a_s)$ is reached. This is in sharp 
contrast with the case of two internal states, where the spin-orbit coupling 
shifts one band to the right and the other to the left and does not affect 
zero center-of-mass momentum pairing. This occurs because of the existence 
of a spin-gauge symmetry, which can be used to gauge away the momentum transfer
$k_T$ from the problem when $\Omega = 0$. The corresponding color-gauge symmetry for
the color superfluid problem is broken, and thus a color-gauge symmetry does not
exist even when $\Omega = 0$. This means that the cases of $\Omega = 0$ with 
$k_T = 0$ and with $k_T \ne 0$ are not equivalent. Furthermore, while in 
the two-state 
case uniform superfluidity is always present at zero temperature for any values of
$\Omega \ne 0$ and $k_T \ne 0$~\cite{seo-2012}, in the color problem at hand 
this is not the case, because of the energy cost associated with pairing in 
the $RG$ and $GB$ channels, and thus normal states phases may be present at 
zero temperature.  

Having discussed some general aspects of the phase diagram 
of color superfluids in the presence of color-orbit coupling 
and color-flip fields, we discuss next a few thermodynamic consequences
involving the thermodynamic potential and the equation of state
for the chemical potential.

\subsection{Thermodynamic potential}
\label{sec:thermodynamic-potential}

In the vicinity of the phase transition between the normal and superfluid 
phases, the thermodynamic potential in the superfluid phase 
${\cal Q}_0 [\Delta, \Delta^*]$ can be expanded about 
the normal state value ${\cal Q}_{\rm N}$ as 
\begin{equation}
\label{eqn:ginzburg-landau-expansion}
{\cal Q}_0 
=
{\cal Q}_N
+
a \vert \Delta \vert^2
+
\frac{b}{2} \vert \Delta \vert^4
+
\frac{c}{3} \vert \Delta \vert^6.
\end{equation}
The coefficients $a, b, c$ and ${\cal Q}_N$ depend on the microscopic 
parameters of the theory such as the color-flip field $\Omega$, 
the color-orbit momentum transfer $k_T$, and the scattering length 
$a_s$, as well as on thermodynamic parameters such as chemical potential $\mu$ and 
temperature $T$. The coefficient $c$ is found to be always positive in the
range of parameters investigated, and that guarantees the stability of the 
Ginzburg-Landau expansion shown in Eq.~(\ref{eqn:ginzburg-landau-expansion}).

%
\begin{figure} [hbt]
\centering 
\epsfig{file=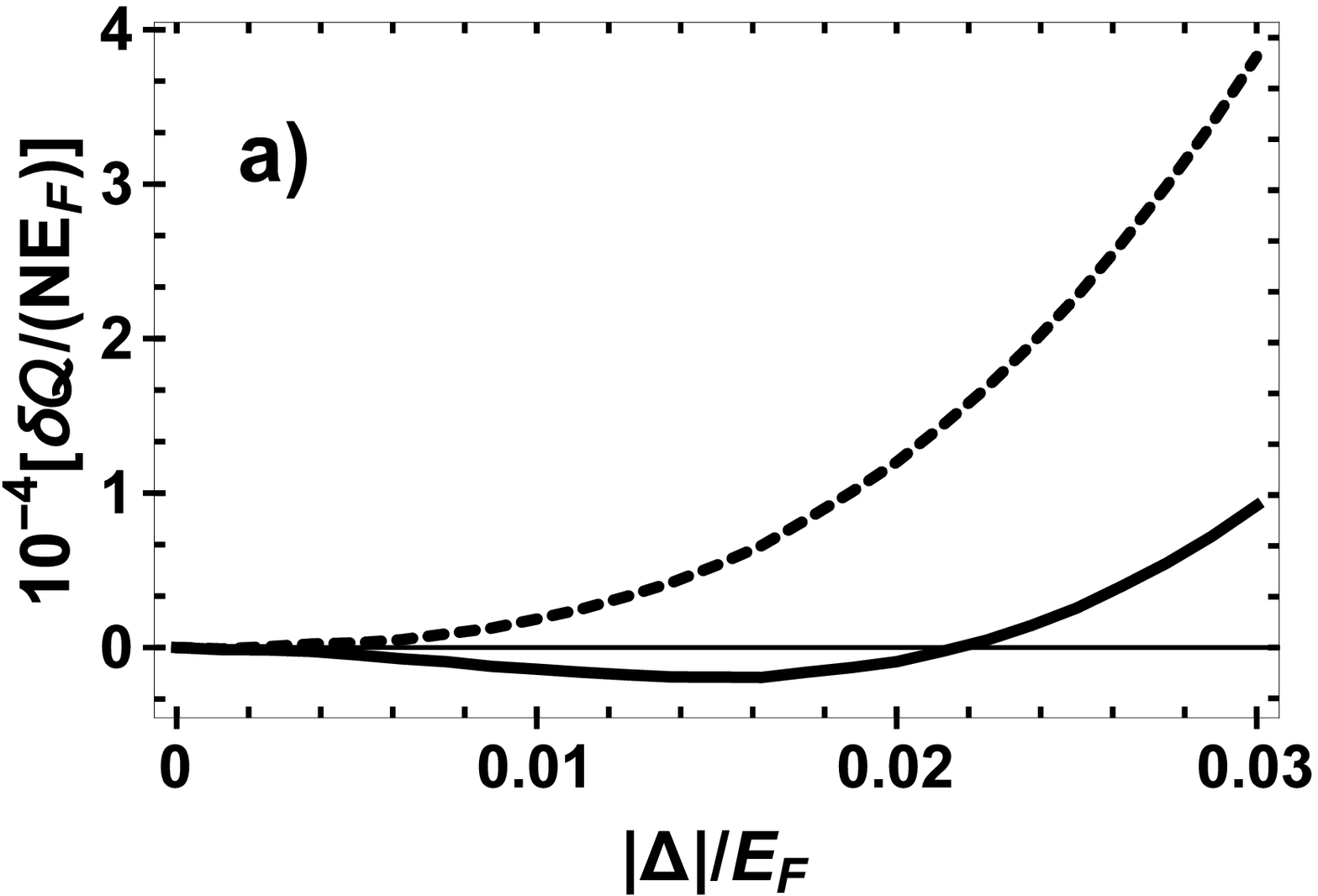,width=0.49 \linewidth}
\epsfig{file=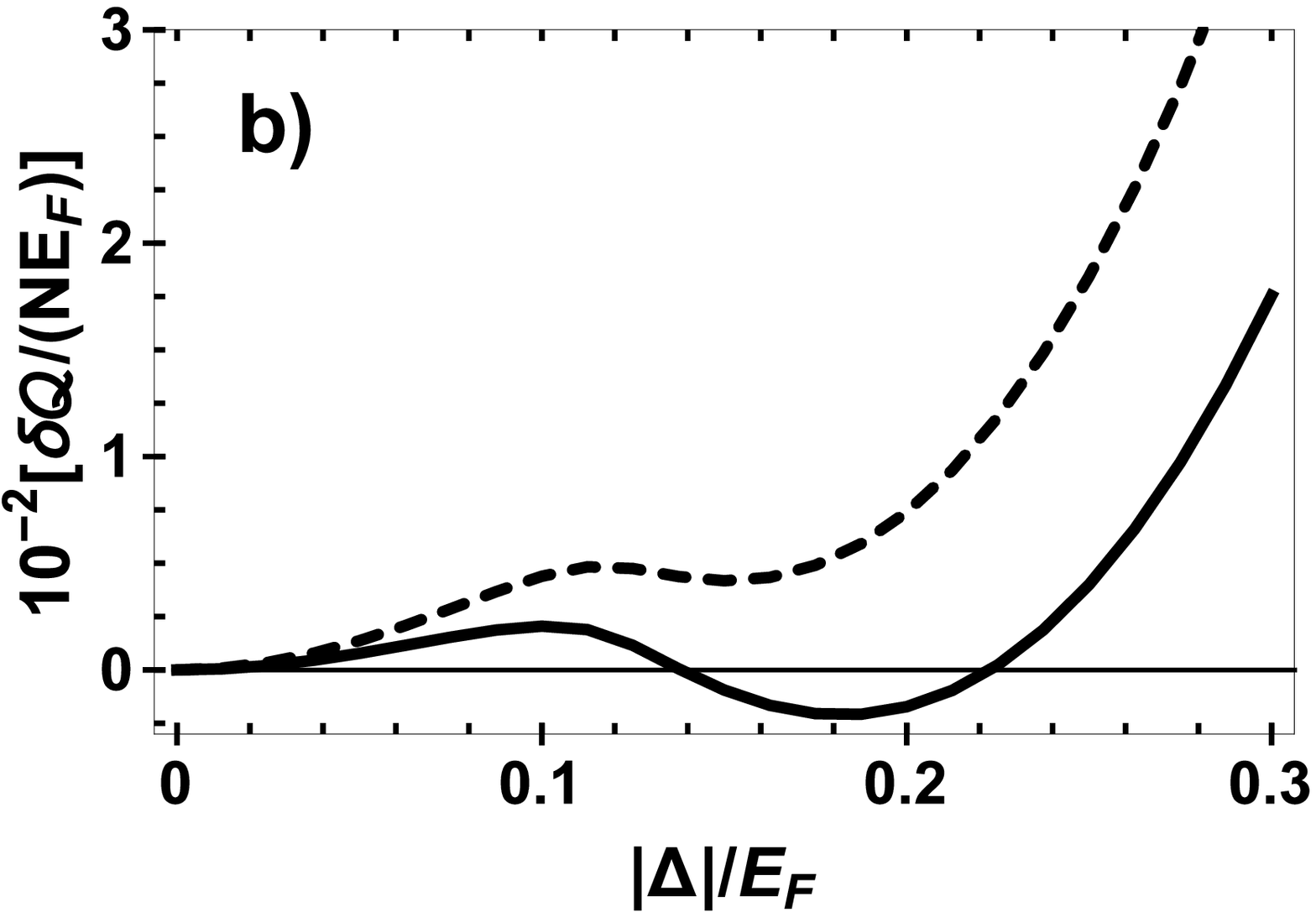,width=0.49 \linewidth}
\\
\epsfig{file=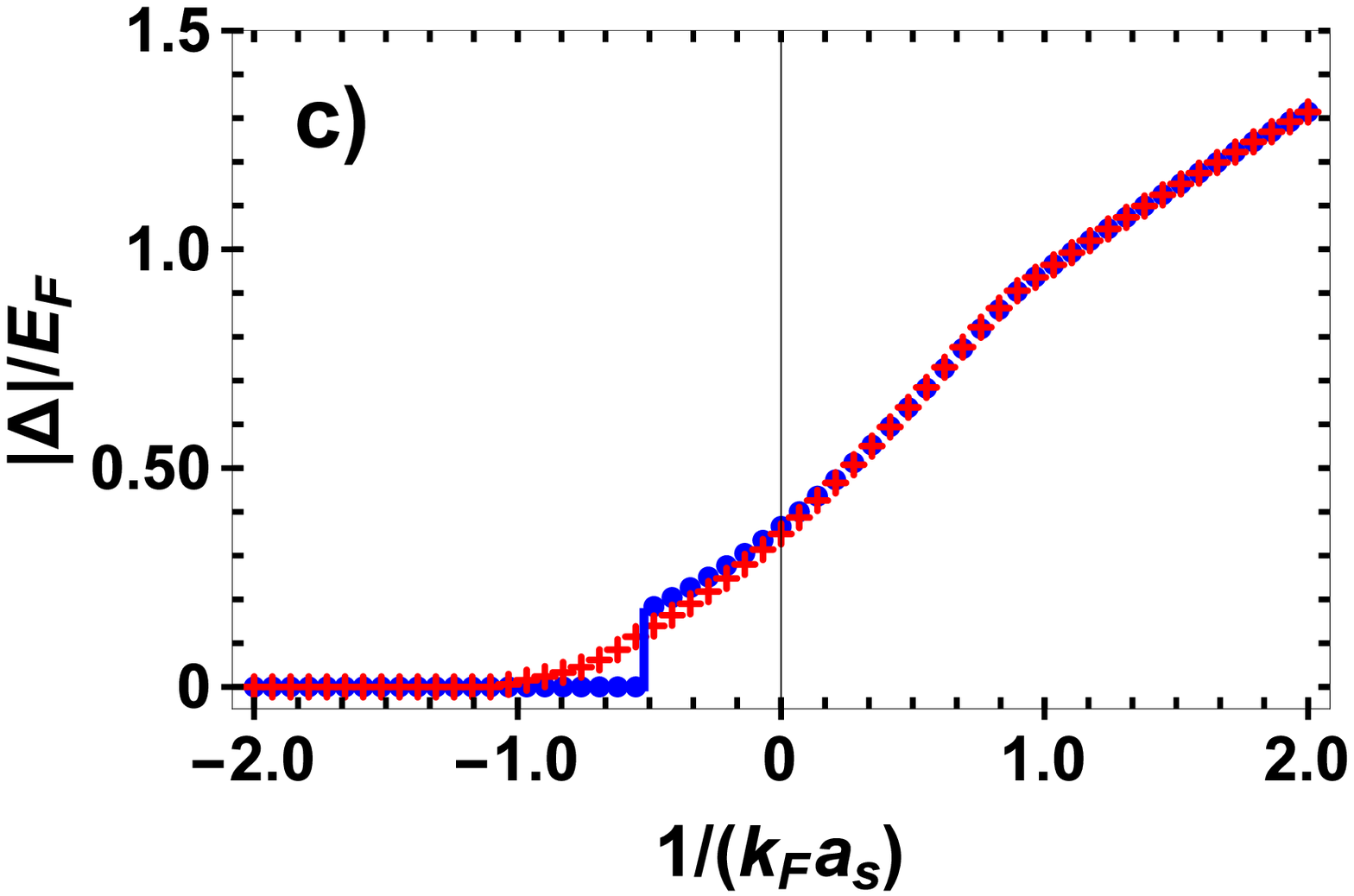,width=0.49 \linewidth}
\epsfig{file=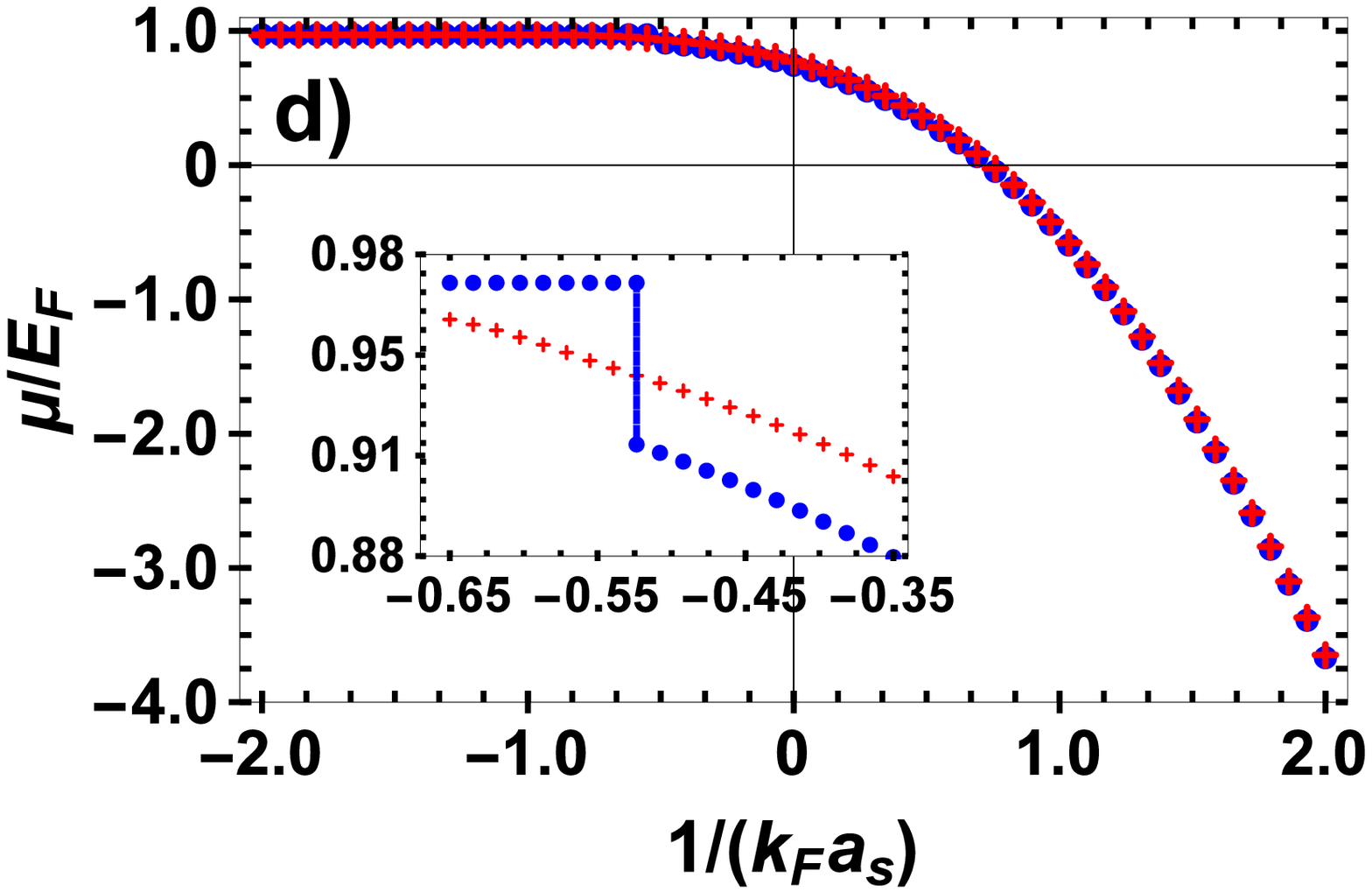,width=0.49 \linewidth}
\caption{ 
\label{fig:five}
(Color online) 
Difference $\delta {\cal Q}$ 
between thermodynamic potentials of superfluid ${\cal Q}_0$ and
normal state ${\cal Q}_N$ are shown in a) and b) at $T/E_F = 0.01$
and $\Omega/E_F = 0.29$.
In a) $\delta {\cal Q}$ is shown for $k_T/k_F  = 0.35$, 
the dashed line exhibits a global minimum at $\vert \Delta \vert = 0$, 
and describes the normal phase for parameter values
$1/(k_F a_s) = -1.17$ and $\mu/E_F = 0.97$.  
The solid line shows a global minimum at $\vert \Delta \vert / E_F = 0.015$, 
and describes a superfluid phase for parameter values 
$1/(k_F a_s) = -0.97$ and $\mu/E_F = 0.97$.
The transition from the superfluid to the normal state is continuous
(second-order).
In b) $\delta {\cal Q}$ is shown for $k_T/k_F  = 0$, 
the dashed line shows a global minimum at $\vert \Delta \vert = 0$, and 
describes the normal phase for parameter values
$1/(k_F a_s) = -0.62$ and $\mu/E_F = 0.97$.
The solid line shows a global minimum at $\vert \Delta \vert / E_F = 0.18$, 
and describes a superfluid phase for parameter values 
$1/(k_F a_s) = -0.48$ and $\mu/E_F = 0.91$.
In c) and d) the curves with blue circles describe the case of 
zero color-orbit coupling $(k_T = 0)$ and the curves with red crosses 
describe the case with $k_T \ne 0$.
In c) we show the order parameter amplitude $\vert \Delta \vert/E_F$ versus
scattering parameter $1/(k_F a_s)$ for $T/E_F = 0.01$ and $\Omega/E_F = 0.29$.
In d) we show the chemical potential $\mu/E_F$ versus scattering
parameter $1/(k_F a_s)$ for $T/E_F = 0.01$ and $\Omega/E_F = 0.29$.
Notice the discontinuous jumps in $\vert \Delta \vert/E_F$ 
and $\mu/E_F$ for the curves
with blue circles at the transition from the superfluid to normal state.
}
\end{figure} 

In Fig.~\ref{fig:five}, all plots correspond to fixed temperature
$T/E_F = 0.01$ and color-flip field $\Omega/E_F = 0.29$.
In Figs.~\ref{fig:five}a and~\ref{fig:five}b,
we show the thermodynamic potential difference 
$\delta {\cal Q} = {\cal Q}_0 - {\cal Q}_N$ 
for two values of scattering parameter $1/(k_F a_s)$ slightly before and 
after the transition from the normal to the superfluid state.
The difference $\delta {\cal Q}$ is shown in units of $N E_F$, where 
$N$ is the total particle number. 
In Fig.~\ref{fig:five}a (Fig.~\ref{fig:five}b), 
the color-orbit coupling momentum transfer is $k_T/k_F = 0.35$ $(k_T/k_F = 0)$ 
and the transition from the normal phase to the superfluid phase is 
continuous (discontinuous), according to Landau's classification, as can be 
seen from the plot of $\delta {\cal Q}$ versus $\vert \Delta \vert/E_F$. 
For the range of parameters investigated at temperature $T/E_F = 0.01$ 
the phase transition between the normal and superfluid phase is
always discontinous for the case of $k_T = 0$ and is always continuous 
for the case of $k_T/k_F = 0.35$. The order parameter amplitude 
$\vert \Delta \vert /E_F$ versus scattering parameter $1/(k_F a_s)$ are shown 
in Fig.~\ref{fig:five}c for $k_T = 0$ and  $k_T/k_F = 0.35$, 
and a clear discontinuity in 
$\vert \Delta \vert / E_F$ occurs at the normal/superfluid 
phase boundary when $k_T = 0$, while $ \vert \Delta \vert / E_F$
reaches zero continuously when $k_T/k_F = 0.35$. 
In Fig.~\ref{fig:five}d, we show the chemical potential $\mu/E_F$
for $k_T/k_F = 0$, and for $k_T/k_F = 0.35$. While for $k_T/k_F = 0.35$, 
the chemical potential $\mu/E_F$ evolves smoothly with scattering parameter 
$1/(k_F a_s)$, by contrast, there is a discontinuous jump
in $\mu/E_F$ in the case of $k_T/k_F = 0$, as the phase boundary between 
the normal and superfluid states is crossed.

\begin{figure} [hbt]
\centering 
\epsfig{file=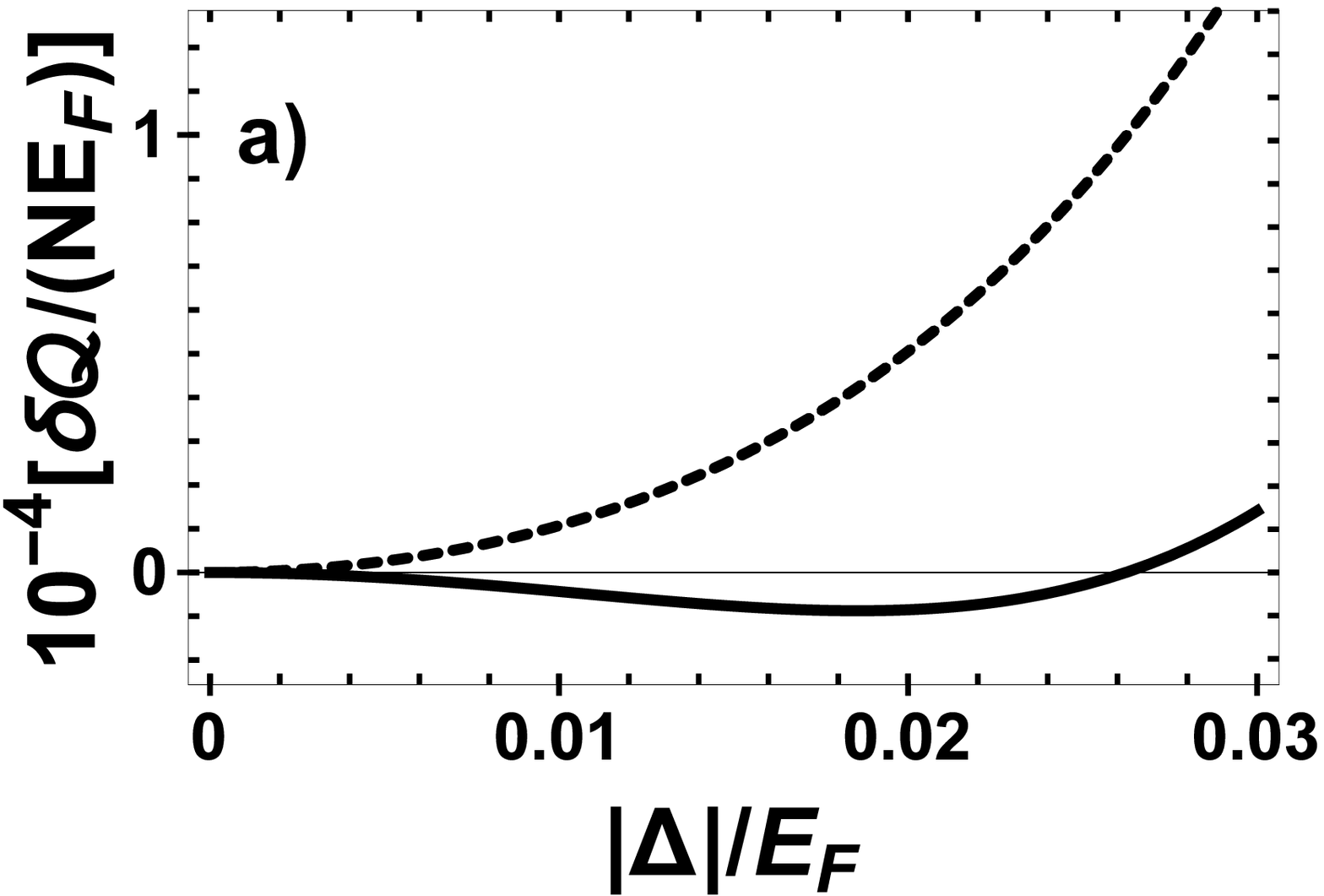,width=0.49 \linewidth}
\epsfig{file=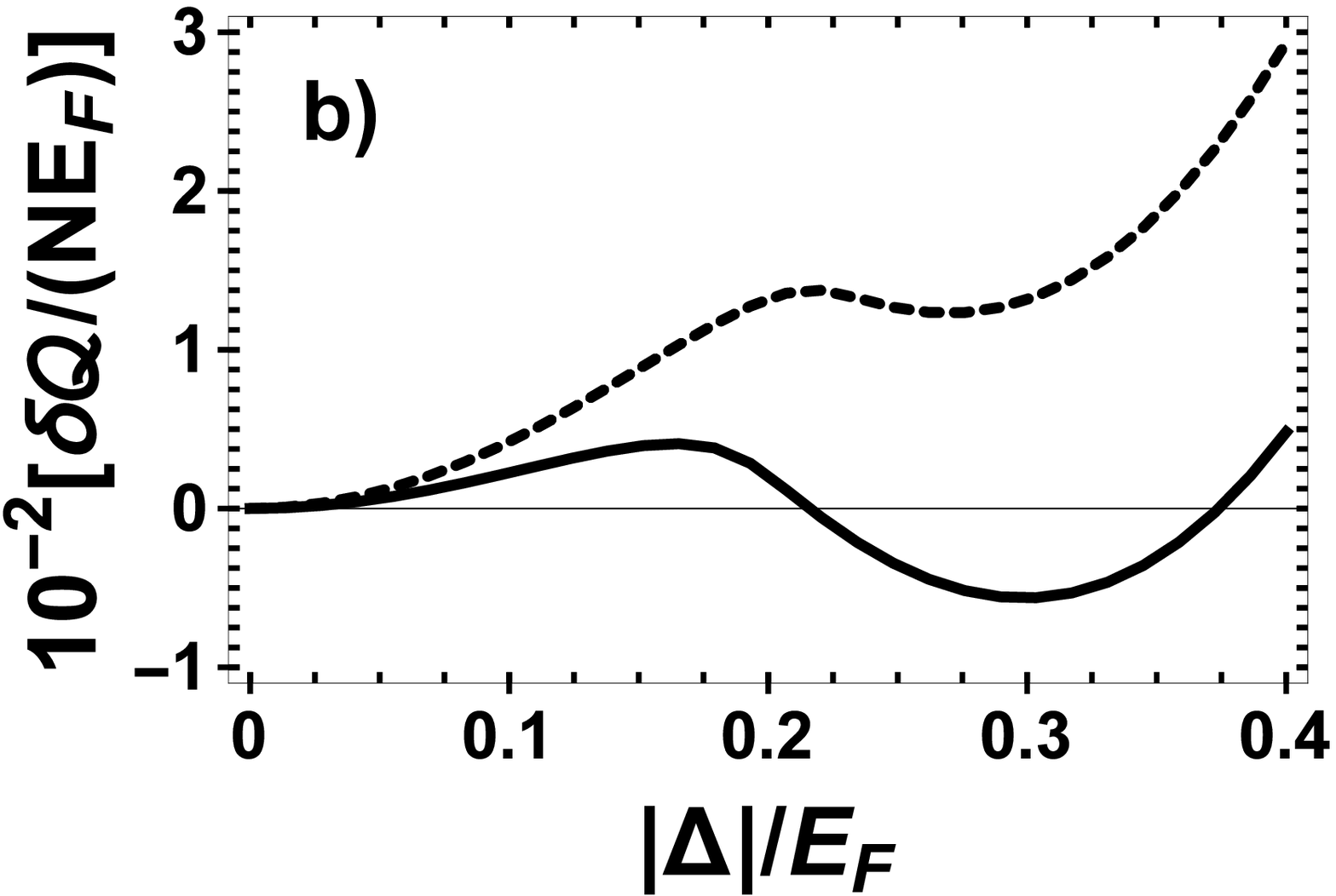,width=0.49 \linewidth}
\\
\epsfig{file=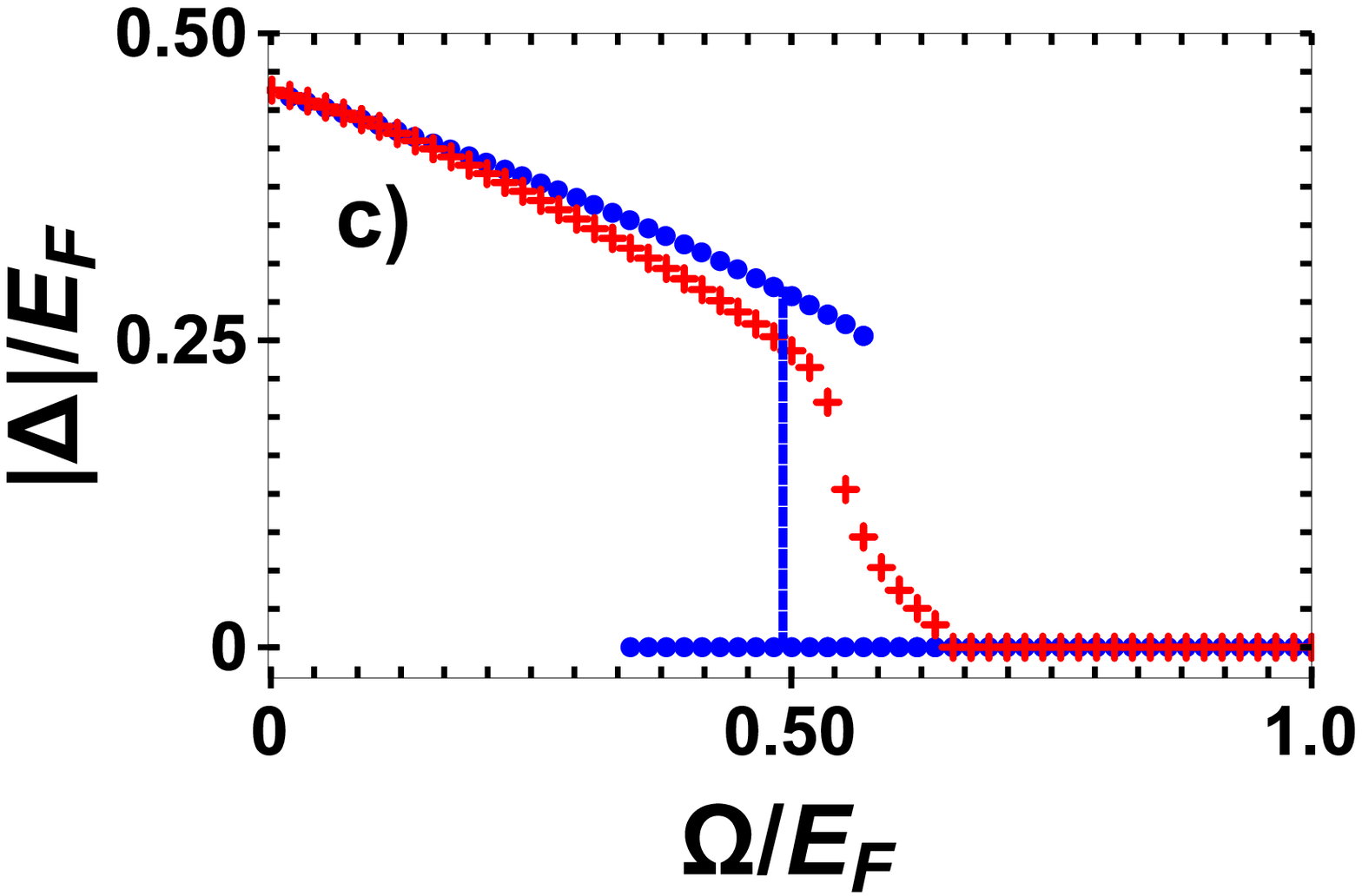,width=0.49 \linewidth}
\epsfig{file=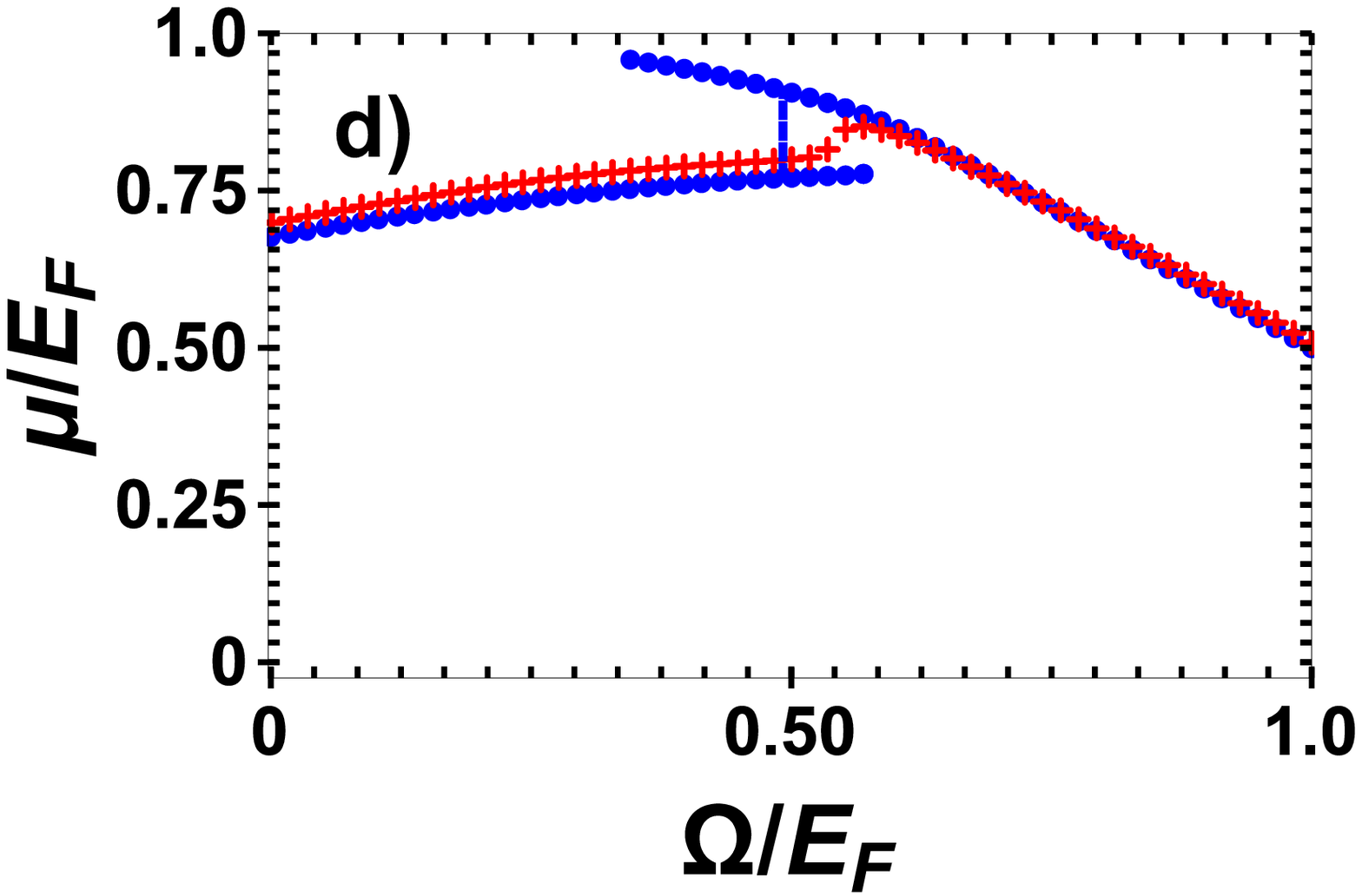,width=0.49 \linewidth}
\caption{ 
\label{fig:six}
(Color online) 
This figure refers to the unitarity limit
$1/(k_F a_s) = 0$ at temperatures $T/E_F = 0.01$.
The difference $\delta {\cal Q}$ 
between thermodynamic potentials of superfluid ${\cal Q}_0$ and
normal state ${\cal Q}_N$ is shown in a) and b).
In a) the color-orbit coupling is $k_T = 0.35 k_F$,
the solid line corresponds to parameters
$\Omega/E_F  = 0.64$ and $\mu/E_F = 0.81$,
while the dashed line corresponds to 
$\Omega/E_F = 0.67$ and $\mu/E_F = 0.79$.
In b) the color-orbit coupling is $k_T = 0$,
the solid line corresponds to parameters
$\Omega/E_F = 0.47$ and $\mu/E_F = 0.77$,
while the dashed line corresponds to 
$\Omega/E_F = 0.53$ and $\mu/E_F  = 0.77$.
In c) and d) the blue circles (red crosses) correspond
to $k_T/k_F = 0$ $(k_T/k_F = 0.35)$.
In c) the order parameter amplitude 
$\vert \Delta \vert/E_F$ versus $\Omega/E_F$ is
shown. In d) the chemical potential $\mu/E_F$ 
versus $\Omega/E_F$ is plotted. The vertical blue-solid lines 
indicate the location of the discontinuous transitions, and
the end points of the lines of blue-circles show the
hysteretic behavior in $\vert \Delta \vert/E_F$ and $\mu/E_F$ 
due to the existence of a metastable minimum in 
the thermodynamic potential shown in b).   
}
\end{figure}

In order to investigate the existence of the Clogston limit, 
we analyse our system at the unitarity limit where the scattering 
parameter $1/(k_F a_s) = 0$ and describe changes in the 
thermodynamic potential, order parameter amplitude
$\vert \Delta \vert/E_F$ and chemical potential $\mu/E_F$ versus 
the color-flip parameter $\Omega/E_F$. In Fig.~\ref{fig:six} we show
the difference $\delta {\cal Q}$ between thermodynamic potentials of 
superfluid ${\cal Q}_0$ and normal state ${\cal Q}_N$. 

%

The plots shown in Fig.~\ref{fig:six} refer to the unitarity limit
$1/(k_F a_s) = 0$ at temperatures $T/E_F = 0.01$.
In Fig.~\ref{fig:six}a, the thermodynamic potential difference 
$\delta {\cal Q}$ is plotted for 
color-orbit coupling $k_T = 0.35 k_F$:
the solid line corresponds to parameters
$\Omega/E_F  = 0.64$ and $\mu/E_F = 0.81$,
while the dashed line corresponds to 
$\Omega/E_F = 0.67$ and $\mu/E_F = 0.79$.
In Fig.~\ref{fig:six}b, $\delta {\cal Q}$ is plotted 
for color-orbit coupling $k_T = 0$:
the solid line corresponds to parameters
$\Omega/E_F = 0.47$ and $\mu/E_F = 0.77$,
while the dashed line corresponds to 
$\Omega/E_F = 0.53$ and $\mu/E_F  = 0.77$.
In Figs.~\ref{fig:six}c and d, the blue circles (red crosses) correspond
to $k_T/k_F = 0$ $(k_T/k_F = 0.35)$.
In Fig.~\ref{fig:six}c, the order parameter amplitude 
$\vert \Delta \vert/E_F$ versus $\Omega/E_F$ is
shown, and in Fig.~\ref{fig:six}d the chemical potential $\mu/E_F$ 
versus $\Omega/E_F$ is shown. Notice the hysteretic behaviour characteristic 
of discontinous (first order) phase transitions when  
the color-orbit coupling is $k_T/k_F = 0$.

Two important effects are illustrated in the panels of Fig.~\ref{fig:six}. 
First, there is a well defined Clogston limit when the color-orbit 
coupling $k_T/k_F = 0$ leading to a discontinuous (first order) 
transition from superfluid to normal phases.
Second, when the color-orbit parameter $k_T/k_F \ne 0$, the standard 
Clogston limit is exceeded and the transition
to the normal state becomes continuous (second-order).
The discontinuous phase transition from superfluid to normal phases for 
zero color-orbit coupling $(k_T/k_F = 0)$ and finite color-flip field
$\Omega \ne 0$ becomes continuous for arbitrarily small $k_T/k_F \ll 1$ 
and $\Omega \ne 0$, provided that a uniform superfluid is the
stable ground state. The discontinuity in the order parameter 
$\vert \Delta \vert/E_F$ at the phase boundary ceases to exist 
for arbitrarily small color-orbit coupling, suggesting that $k_T/k_F$ 
is a microscopic parameter that controls the non-uniform convergence 
of $\vert \Delta \vert /E_F$ versus the scattering parameter
$1/(k_F a_s)$ or versus the color-flip parameter $\Omega/E_F$. 

The non-uniform convergence of $ \vert \Delta \vert/E_F$ is similar 
mathematically to the non-uniform convergence of the Fermi function, where only
at zero temperature it develops a discontinuity as a function of momentum.
In the present case, the physical origin of the non-uniform convergence 
is quite different. In our current problem, attractive interactions are 
assumed to occur only between different colors, 
such that pairing can only exist in the 
s-wave channel. Thus, in the strict case where color-orbit coupling 
is zero $(k_T/k_F = 0)$ there is a cost in color-flip energy associated 
with pairing and thus there is a Clogston limit even in the color problem
at low temperatures. However, when a uniform superfluid solution is the ground
state for $k_T/k_F \ne 0$, then higher-order angular momentum pairing is 
induced by the color-orbit coupling as suggested by the nodal structures 
shown in Fig.~\ref{fig:four}. Thus, pairing in the mixed-color states may 
occur not only in the singlet channel, but also in the triplet or quintet 
channels. This allows the color superfluid to respond to a color-flip field
by simply rotating the triplet or quintet mixed-color state without breaking
pairs, and thus beating the standard Clogston limit. An analysis of the order 
parameter in the mixed color basis is therefore important for a deeper 
understanding of the phase diagram obtained in Fig.~\ref{fig:three}.

Now that we have analysed a few thermodynamic properties of color superfluids,
and characterized the transitions between the normal and superfluid states, 
we are ready to investigate in detail the structure of the order parameter 
in each one of the superfluid phases found.

\section{Hamiltonian in mixed color basis}
\label{sec:hamitonian-mixed-color-basis}

In order to understand in more detail the different superfluid phases 
that emerge, it is important to analyse the microscopic Hamiltonian 
in a mixed color basis that diagonalizes the independent particle 
Hamiltonian discussed in section~\ref{sec:independent-particle-hamiltonian}.

The excitation spectrum and the momentum space topology of 
colored quasiparticles and quasiholes 
can be understood by writing the saddle-point Hamiltonian 
${\bf H}_{0} ({\bf k})$ 
defined in Eq.~(\ref{eqn:saddle-point-hamiltonian-matrix}) as 
\begin{eqnarray}
\label{eqn:saddle-point-hamiltonian-matrix-helicity}
\widetilde{\bf H}_{0}({\bf k})
=
\left(
\begin{array}{cc}
{\bf H}_{M} ({\bf k})	&	{\boldsymbol \Delta}_M
\\
{\boldsymbol \Delta}_{M}^{\dagger} &	-{\bf H}_{M}^* (-{\bf k})
\end{array}
\right)
\end{eqnarray}
in the mixed color basis.
The matrix elements of ${\bf H}_{M}({\bf k})$ 
represent the mixed color energy bands, and are given by  
$
{\bf H}_{M,\alpha \beta} ({\bf k})
=
\xi_{\alpha} ({\bf k}) \delta_{\alpha \beta}
$ 
with energies
$
\xi_{\alpha} ({\bf k})
=
{\mathcal E}_{\alpha} ({\bf k})
-
\mu
$
measured with respect to the chemical potential $\mu$.
While the matrix elements of 
$
{\boldsymbol \Delta}_{M}
$
are 
$
\Delta_{M,\alpha \beta} ({\bf k})
=
\Delta_{\alpha \beta} ({\bf k}),
$	
representing the order parameter tensor in the mixed color basis
labeled by indices $\{\alpha, \beta\} = \{\Uparrow, 0, \Downarrow\}.$ 
The elements $\Delta_{\alpha \beta} ({\bf k})$
are strongly momentum {\it dependent} in sharp contrast
to the elements $\Delta_{c c^\prime} ({\bf k})$  of the original matrix
$
{\boldsymbol \Delta},
$
defined in Eq.~(\ref{eqn:order-parameter-tensor}),
which are {\it independent} of momentum. 
In order to ellucidate the symmetry properties of fermion pairs, we analyse
next the order parameter tensor in the mixed color basis.

\subsection{Order parameter in mixed color basis}

The order parameter tensor in the mixed color basis
can be written as 
\begin{equation}
\Delta_{\alpha \beta} ({\bf k})
=
R_{\alpha c} ({\bf k}) \Delta_{c c^\prime} R_{c^\prime \beta} ({\bf -k}),
\end{equation}
where Einstein's summation convention of repeated indices is understood
and $R_{\alpha c} ({\bf k})$ are matrix elements of the color mixing 
matrix ${\bf R}({\bf k})$ in Eq.~(\ref{eqn:unitary-matrix}), 
where the $\alpha$-row elements ${\bf R}_{\alpha} ({\bf k}) 
= 
\left[
R_{\alpha R} ({\bf k}), R_{\alpha G} ({\bf k}), R_{\alpha B} ({\bf k})
\right]
$
are the eigenvectors of the independent particle
Hamiltonian matrix ${\bf H}_{\rm IP} ({\bf k})$.
The order parameter matrix in the mixed color basis has the property 
$
\Delta_{\alpha \beta} ({\bf k}) 
= 
- \Delta_{\beta \alpha} ({\bf - k})
$
due to Fermi statistics. Such condition ensures that the diagonal elements
$\Delta_{\alpha \alpha} ({\bf k})$ have odd parity, as required 
by the Pauli exclusion principle. However, in general, this is not sufficient 
to force the off-diagonal elements to have well defined parity.

\begin{figure} [hbt]
\centering 
\epsfig{file=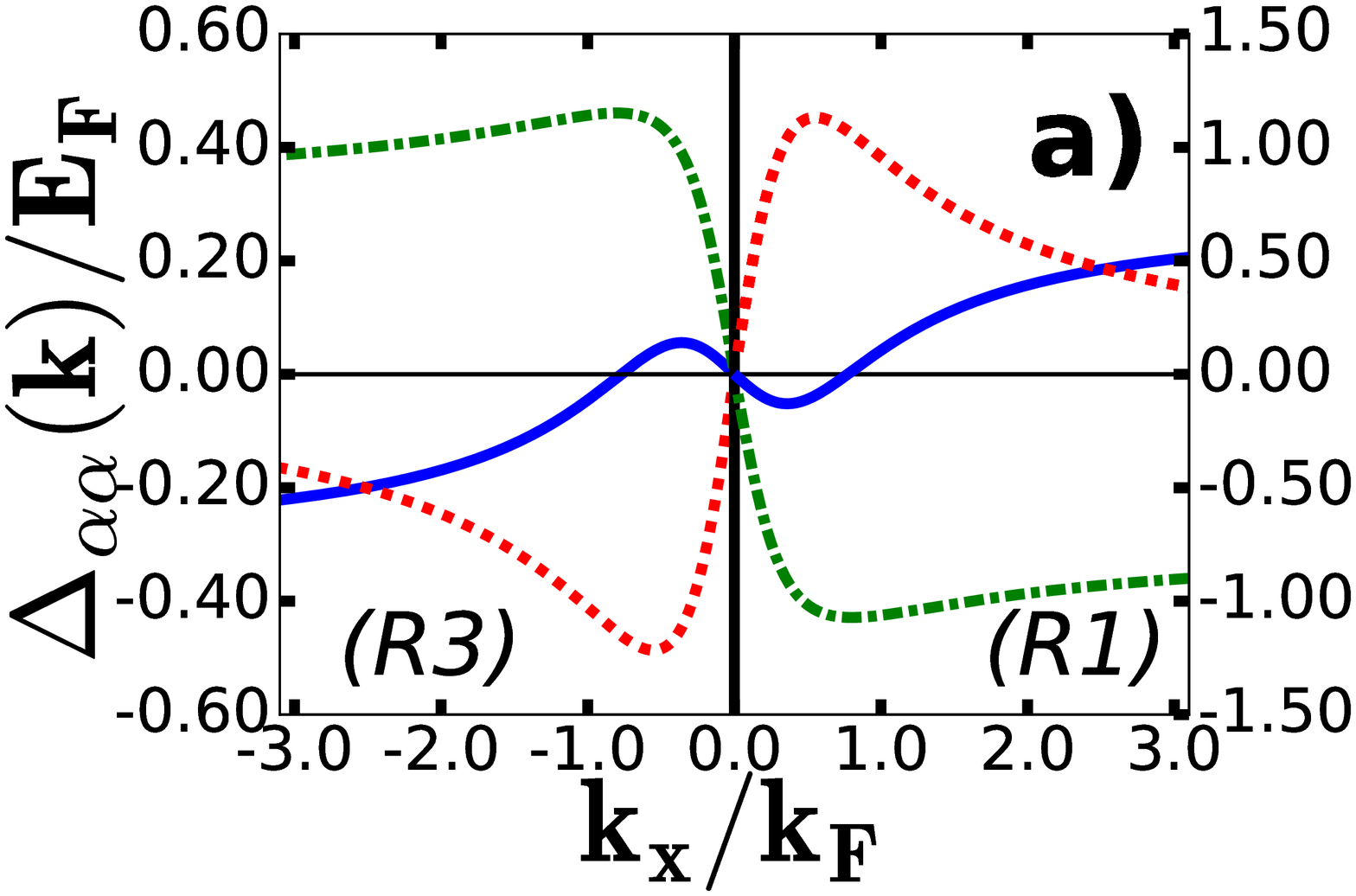,width=0.49 \linewidth}
\epsfig{file=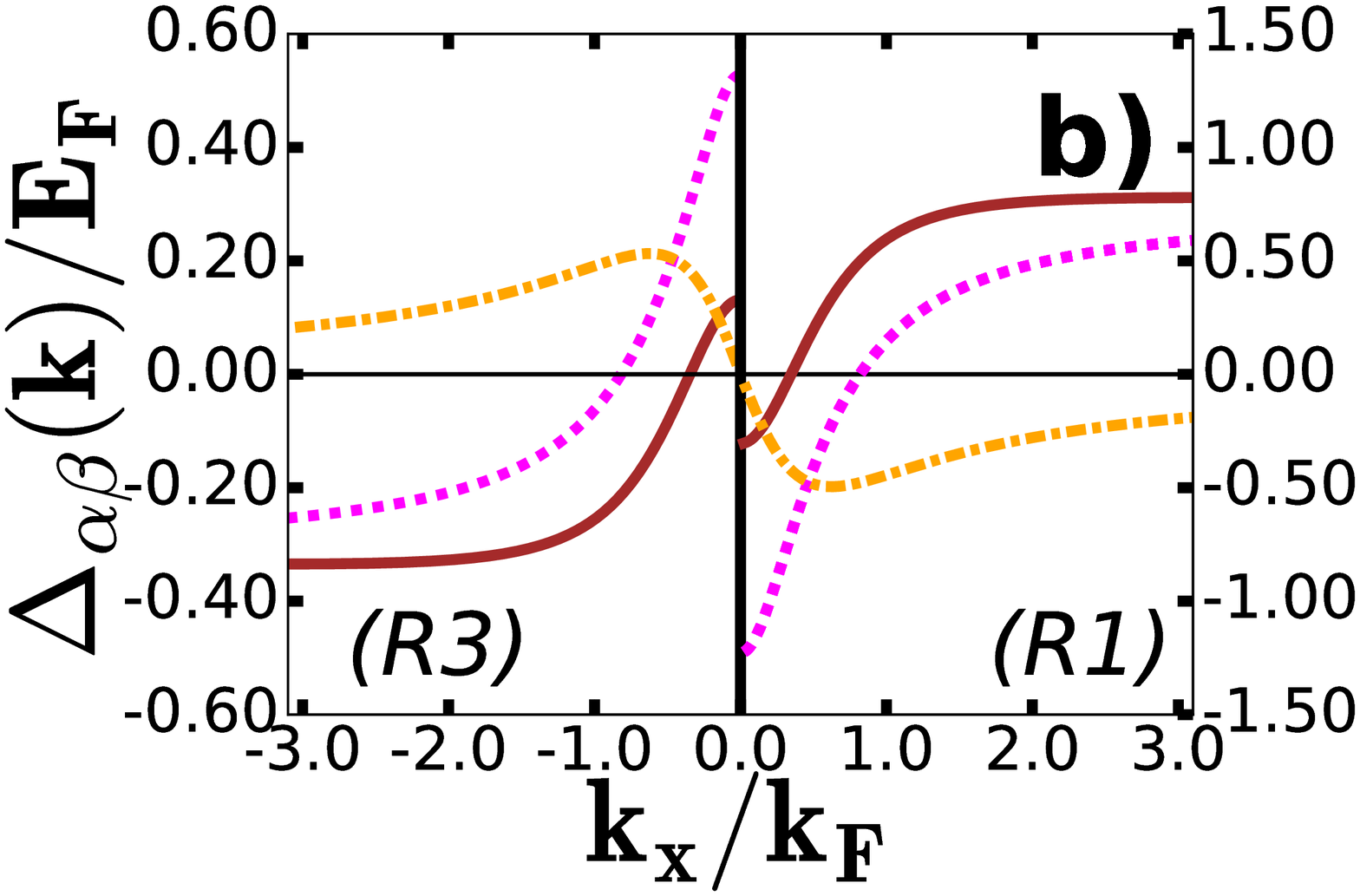,width=0.49 \linewidth}
\\
\epsfig{file=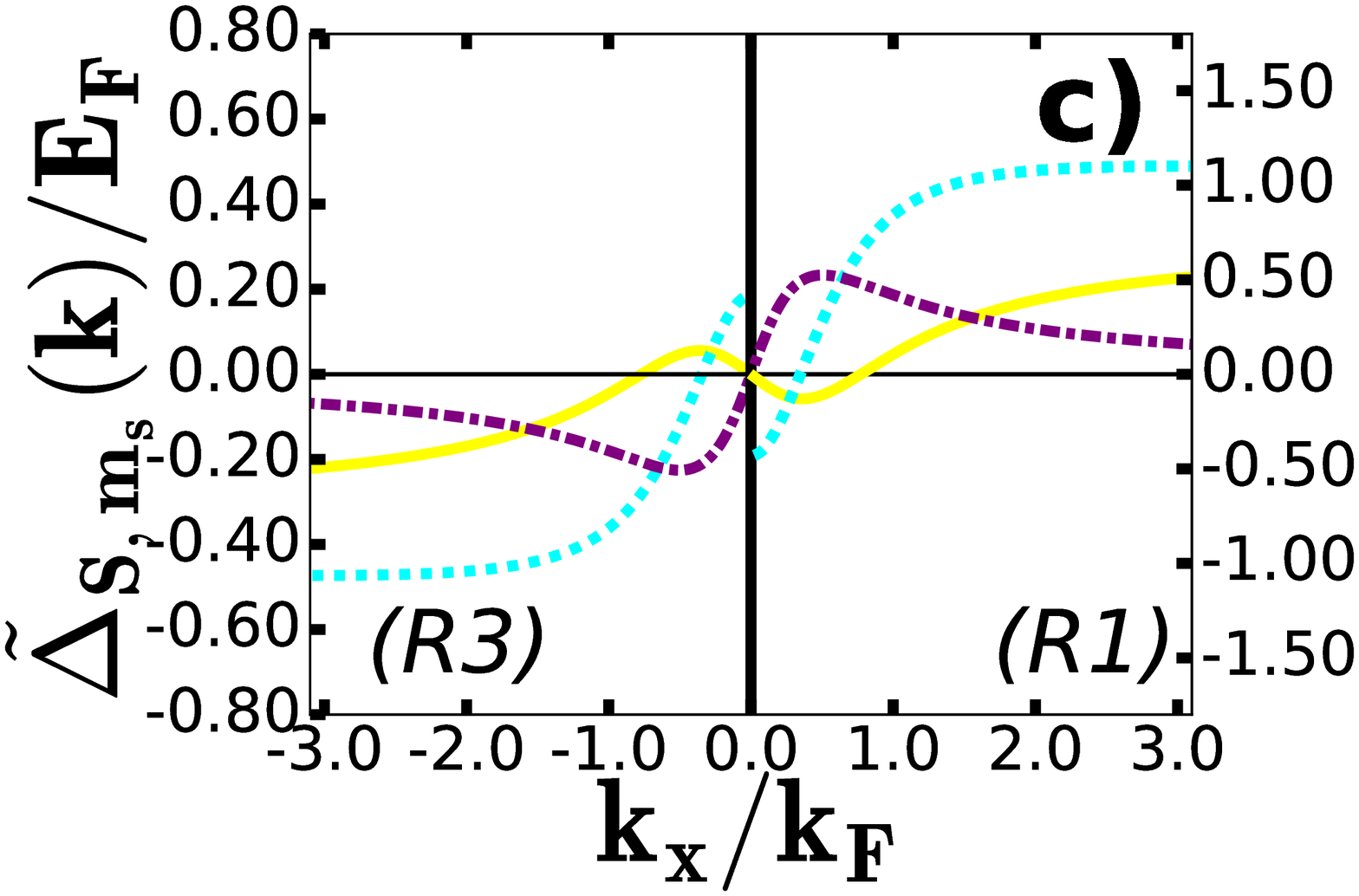,width=0.49 \linewidth}
\epsfig{file=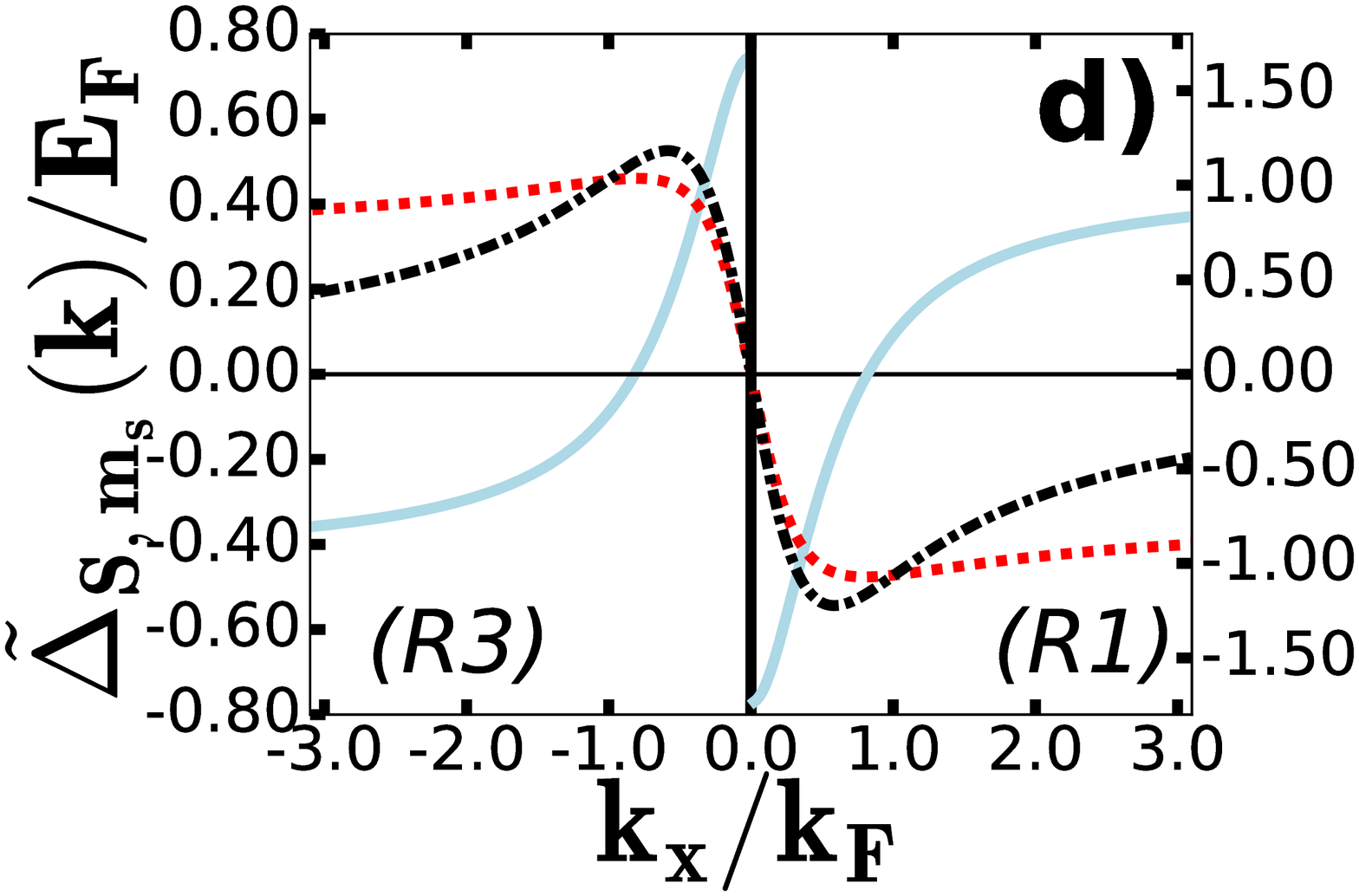,width=0.49 \linewidth}
\caption{ 
\label{fig:seven}
(Color online) 
Plots of the components of the order parameter tensor 
$\Delta_{\alpha \beta} ({\bf k})$ and 
${\widetilde \Delta}_{S m_s} ({\bf k})$ versus 
momentum $(k_x,0, 0)$ are shown in a,b and c,d respectively. 
We show plots for the $R3$ phase with scattering parameter
$1/(k_F a_s) = -0.07$ ($\mu/E_F = 0.81$ and 
$\vert \Delta \vert/E_F = 0.31$) corresponding to the vertical scale 
on the left. We also show plots for the $R1$ phase with 
scattering parameter $1/(k_F a_s) = 0.62$ ($\mu/E_F = 0.19$ and 
$\vert \Delta \vert/E_F = 0.73$) corresponding to the 
vertical scale on the right.
In a) the solid blue curve describes 
$\Delta_{\Uparrow\Uparrow} ({\bf k})$, 
the dashed red curve describes $\Delta_{00} ({\bf k})$, 
and the dot-dashed green curve describes 
$\Delta_{\Downarrow\Downarrow} ({\bf k})$. 
In b) the solid brown line represents $\Delta_{\Uparrow 0} ({\bf k})$, 
the dashed magenta line represents $\Delta_{\Downarrow 0} ({\bf k})$, 
the dot-dashed orange line represents $\Delta_{\Uparrow\Downarrow} ({\bf k})$.
In c) the solid yellow curve corresponds to $\widetilde \Delta_{22} ({\bf k})$, 
the dashed cyan curve corresponds to $\widetilde \Delta_{21} ({\bf k})$,
the dot-dashed purple curve corresponds to $\widetilde \Delta_{20} ({\bf k})$.
In d) the solid light-blue line indicates $\widetilde \Delta_{2{\bar 1}} ({\bf k})$,
the dashed red line indicates $\widetilde \Delta_{2{\bar 2}} ({\bf k})$,
the dot-dashed black line indicates $\widetilde \Delta_{00} ({\bf k})$.
}
\end{figure}

In Fig.~\ref{fig:seven}, we describe in detail the momentum dependence of the
order parameter tensor along $(k_x, 0, 0)$ 
in the mixed color basis (Figs.~\ref{fig:seven}a and b),
and in the total pseudo-spin basis (Figs.~\ref{fig:seven}c and d)
for parameters $\Omega/E_F = 0.29$, $k_T/k_F = 0.35$, 
$b_z = k_T^2/(2m)$ $(\eta = 0)$. The order parameter tensor on either 
basis depends on momentum only along the $k_x$ direction and 
is independent of momentum along $k_y$ and $k_z$, due to the 
one dimensional nature of the color-orbit coupling.
In Figs.~\ref{fig:seven}a, b, c and d, we show only two cases. 
The first one corresponds to the $R3$ phase with scattering parameter
$1/(k_F a_s) = -0.07$ ($\mu/E_F = 0.81$ and 
$\vert \Delta \vert/E_F = 0.31$) 
and the plots correspond to the vertical scale on the left. 
The second case corresponds to the $R1$ phase with scattering parameter 
$1/(k_F a_s) = 0.62$ ($\mu/E_F = 0.19$ and 
$\vert \Delta \vert/E_F = 0.73$) 
and the plots correspond to the vertical scale on the right. 

In Fig.~\ref{fig:seven}a, we show the momentum dependence of the 
diagonal components $\Delta_{\alpha \alpha} ({\bf k})$ 
of the order parameter tensor in two cases.  
From these plots it is evident that the nodal structure of the order 
parameter tensor components $\Delta_{\alpha \alpha} ({\bf k})$ 
is exactly the same for the $R3$ and $R1$ phases. The only difference 
between the two cases is the overall magnitude of the amplitude 
$\vert \Delta \vert$ reflected in the two different scales. 
This implies that the nodal structure of the lowest quasiparticle band 
$E_3 ({\bf k})$ does not coincide with the nodal structure of the 
order parameter matrix elements $\Delta_{\alpha \alpha} ({\bf k})$.
The solid blue curve describes $\Delta_{\Uparrow\Uparrow} ({\bf k})$,
which has an f-wave character (three nodes); the dashed
red curve describes $\Delta_{00} ({\bf k})$, which has a p-wave
character (one node); the dot-dashed green curve describes
$\Delta_{\Downarrow\Downarrow} ({\bf k})$, which also has a p-wave
character (one node).

In Fig.~\ref{fig:seven}b, we show the momentum dependence of the 
off-diagonal components $\Delta_{\alpha \beta} ({\bf k})$  
of the order parameter tensor, with 
$\alpha \ne \beta$, in two cases. 
From these plots it is also evident that the nodal structure of the 
order parameter tensor components $\Delta_{\alpha \beta} ({\bf k})$ 
is exactly the same for the $R3$ and $R1$ phases. Again, the only 
difference between the two cases is the overall 
magnitude of the amplitude $\vert \Delta \vert$. As in the case of 
diagonal components, this implies that the nodal structure of the lowest 
quasiparticle band $E_3 ({\bf k})$ does not coincide with the 
nodal structure of the off-diagonal matrix elements 
$\Delta_{\alpha \beta} ({\bf k})$.
The solid brown curve describes $\Delta_{\Uparrow 0} ({\bf k})$,
which has an f-wave character (two nodes and a discontinous sign change); 
the dashed magenta curve describes $\Delta_{\Downarrow 0} ({\bf k})$, 
which has an f-wave character (two nodes and a discontinous sign change); 
the dot-dashed orange curve describes
$\Delta_{\Uparrow\Downarrow} ({\bf k})$, which has a p-wave
character (one node).

A very important property that emerges from Figs.~\ref{fig:seven}a and b is that the 
order parameter tensor $\Delta_{\alpha \beta} ({\bf k})$ for color superfluids
(with three colors) is always an odd function of momentum ${\bf k}$ when 
the one-dimensional color-orbit coupling $h_z ({\bf k}) = 2 k_T k_x/(2m)$ 
is present with zero color shift (detuning $\delta = 0$), 
meaning that the condition 
$\Delta_{\alpha \beta} ({\bf k}) = - \Delta_{\alpha \beta} ({\bf - k})$ 
is satisfied. The odd parity condition combined with the Fermi statistics 
property $\Delta_{\alpha \beta} ({\bf k}) = - \Delta_{\beta \alpha} ({\bf - k})$ 
leads to a symmetric order parameter tensor 
$\Delta_{\alpha \beta} ({\bf k}) = \Delta_{\beta \alpha} ({\bf k})$ under 
mixed color exchange $\alpha \leftrightarrow \beta$, when $k_T \ne 0$. 
Furthermore, the order parameter tensor is odd under reflection along
the $k_x$ direction, but even under reflections along the $k_y$ and $k_z$ directions.

%

As seen in Fig.~\ref{fig:four}c, the nodal structure of quasiparticle
excitations in the $R3$ phase is similar to that of the excitation 
spectrum of a fully spin-polarized triplet f-wave superfluid of 
spin-$1/2$ fermions with energy band
$\xi_{\bf k} = {\bf k}^2/(2m) - \mu$ and 
order parameter amplitude 
$\vert \Delta_{\bf k} \vert = \vert a_f k_x^3 - a_p k_x \vert$, 
with $a_f$ and $a_p$ being positive. In this case,
the quasiparticle excitation spectrum is simply 
$
E ({\bf k}) 
= 
\sqrt{ \xi_{\bf k}^2 + \vert \Delta_{\bf k} \vert^2 },
$
and the nodes $E_{\bf k}$ occur at the intersection of the surfaces
$\xi_{\bf k} = 0$ and $\vert \Delta_{\bf k} \vert = 0$. Since the zeros 
of the order parameter are located at $k_x = 0$, and 
$k_x = \pm \sqrt{(a_p/a_f)}$, and the zeros of $\xi_{\bf k}$ 
are located at $(k_y^2 + k_z^2)/(2m) = \mu - k_x^2/(2m)$, the {\it loci}
of zero quasiparticle energy has a three ring structure.
Additionally, the nodal structure of quasiparticle excitations in 
the superfluid state $R1$, shown in Fig.~\ref{fig:four}d, is similar
to that of a triplet p-wave superfludid of fully spin-polarized 
spin-$1/2$ fermions with order parameter amplitude 
$\vert \Delta_{\bf k} \vert = \vert a_p k_x \vert$. In this case, 
the nodes in $E ({\bf k})$ occur at the intersection of the surfaces
$k_x = 0$ and $(k_y^2 + k_z^2)/(2m) = \mu - k_x^2/(2m)$, thus leading
to a single ring nodal strucure for the {\it loci} of zero quasiparticle
energy. It is important to emphasize, though, that in the color problem
the location of zeros of the quasiparticle band $E_3 ({\bf k})$ 
does not coincide with the simultaneous zeros of the order parameter 
tensor components $\Delta_{\alpha \beta} ({\bf k}) = 0 $ and 
band dispersions $\xi_{\alpha}({\bf k}) - \mu$.

It is also important to compare the momentum dependence of 
the order parameter tensor $\Delta_{\alpha \beta} ({\bf k})$ 
for the color problem with color-orbit and color-flip fields, 
and the corresponding spin-$1/2$ problem with spin-orbit and Zeeman
fields~\cite{seo-2012}. As shown in Figs.~\ref{fig:seven}a and b 
the momentum dependence of $\Delta_{\alpha \beta} ({\bf k})$ shows
higher angular momentum pairing in the color indices. It does so in 
a similar, but more complicated fashion in comparison to 
the spin-1/2 case~\cite{seo-2012}, where the $2\times 2$ order parameter 
tensor $\Delta_{\alpha \beta} ({\bf k})$ in the generalized helicity basis
$\{ \Uparrow, \Downarrow \}$ acquires also a triplet component 
with dominant p-wave character.
A particularly notable difference between the color case and 
the spin-$1/2$ case is that the order parameter 
tensor $\Delta_{\alpha \beta} ({\bf k})$ is a symmetric tensor 
when both color-orbit and color-flip fields are non-zero,
that is 
$
\Delta_{\alpha \beta} ({\bf k}) 
= 
\Delta_{\beta \alpha} ({\bf k}),
$ 
while in the spin-$1/2$ the corresponding order parameter tensor 
$\Delta_{\alpha \beta} ({\bf k})$ is neither anti-symmetric or 
symmetric, that is, the off-diagonal elements 
$\Delta_{\Uparrow\Downarrow} ({\bf k})$ and 
$\Delta_{\Downarrow\Uparrow} ({\bf k})$ are neither equal 
or opposite in sign.
Therefore the order parameter tensor has symmetric components 
$
\Delta_{\Uparrow\Uparrow} ({\bf k}),
$ 
$
\left[ 
\Delta_{\Uparrow\Downarrow} ({\bf k})
+ 
\Delta_{\Downarrow\Uparrow} ({\bf k})
\right]/2,
$
and 
$
\Delta_{\Downarrow\Downarrow} ({\bf k}),
$
corresponding to the triplet sector and 
an antisymmetric component 
$
\left[ 
\Delta_{\Uparrow\Downarrow} ({\bf k})
- 
\Delta_{\Downarrow\Uparrow} ({\bf k})
\right]/2,
$
corresponing to the singlet sector. 

In the color problem discussed here, the tensor 
$\Delta_{\alpha \beta} ({\bf k})$ is only antisymmetric when  
the color-orbit field $h_z ({\bf k}) = 2 k_T k_x/(2m)$ is zero,
that is, the color-dependent momentum transfer $k_T = 0$.  
This jump from an antisymmetric tensor for zero 
color-orbit fields to a symmetric tensor for non-zero color-orbit fields, 
arises due to the absence of color-gauge symmetry 
associated with the three-color states even when the color-flip field is zero.
This singular perturbation caused by the color-orbit field $h_z ({\bf k})$ 
introduces only parity odd momentum dependences in 
the mixed color representation of the order parameter tensor,
provided that the color shift field $\delta = 0$, as it is the case throughout
this manuscript.

To highlight further the structure of the order parameter tensor 
in the color problem with color-flip and color-orbit fields, 
we discuss next its structure in the total pseudo-spin basis,
where singlet, triplet and quintet sectors emerge in a similar fashion
to the singlet and triplet sectors that arise for the order parameter
tensor of spin-1/2 Fermi superfluids with spin-orbit coupling and Zeeman fields.

\subsection{Order parameter in the total pseudo-spin basis}

In order to understand further the order parameter structure in the color 
problem, we also analyse it in the total pseudo-spin basis 
$\vert S, m_s \rangle$  built from the colored mixed states 
$
\{ 
\vert \Uparrow \rangle, 
\vert 0 \rangle, 
\vert \Downarrow \rangle
\}
\otimes
\{ 
\vert \Uparrow \rangle, 
\vert 0 \rangle, 
\vert \Downarrow \rangle
\}.
$ 
The order parameter tensor in the total pseudo-spin basis 
${\widetilde \Delta}_{S m_s} ({\bf k})$ can be separated into singlet, 
triplet and quintet sectors, and can be written as 
$
{\widetilde \Delta}_{S m_s} ({\bf k})
= 
M_{\alpha \beta}^{S m_s} 
\Delta_{\alpha \beta ({\bf k})},
$ 
where $M_{\alpha \beta}^{S m_s}$ is a tensor whose elements represent
generalized Clebysh-Gordon coefficients. 
The singlet sector is described by fermion pairs in the state 
$\vert S m_s \rangle = \vert 00 \rangle$ 
with order parameter element
\begin{equation}
{\widetilde \Delta}_{00} ({\bf k})
= 
\frac{1}{\sqrt{3}} \Delta_{\Uparrow\Downarrow} ({\bf k}) 
-
\frac{1}{\sqrt{3}} \Delta_{00} ({\bf k}) 
+
\frac{1}{\sqrt{3}} \Delta_{\Downarrow\Uparrow} ({\bf k}), 
\nonumber
\end{equation}
while the triplet sector is characterized by fermions pairs in the
states $\{\vert S m_s \rangle \} = 
\{\vert 11 \rangle, \vert 10 \rangle, \vert 1{\bar 1} \rangle \},$
with order parameter elements
\begin{eqnarray}
{\widetilde\Delta}_{11} ({\bf k}) 
& = &
\frac{1}{\sqrt{2}} \Delta_{\Uparrow 0} ({\bf k}) 
-
\frac{1}{\sqrt{2}} \Delta_{0 \Uparrow} ({\bf k}),
\nonumber
\\
{\widetilde\Delta}_{10} ({\bf k}) 
& = &
\frac{1}{\sqrt{2}} \Delta_{\Uparrow \Downarrow} ({\bf k}) 
-
\frac{1}{\sqrt{2}} \Delta_{\Downarrow \Uparrow} ({\bf k}),
\nonumber
\\
{\widetilde\Delta}_{1{\bar 1}} ({\bf k}) 
& = &
\frac{1}{\sqrt{2}} \Delta_{0 \Downarrow} ({\bf k}) 
-
\frac{1}{\sqrt{2}} \Delta_{\Downarrow 0} ({\bf k}), 
\nonumber
\end{eqnarray}
where we used the notation ${\bar 1} = -1$. Finally, the quintet sector 
is described by fermion pairs in states
$\{\vert S m_s \rangle \} = 
\{\vert 22 \rangle, \vert 21 \rangle, \vert 20 \rangle, 
\vert 2{\bar 1} \rangle, \vert 2{\bar 2} \rangle \},$ 
with order parameter elements
\begin{eqnarray}
{\widetilde\Delta}_{22} ({\bf k}) 
& = &
\Delta_{\Uparrow \Uparrow} ({\bf k}), 
\nonumber
\\
{\widetilde\Delta}_{21} ({\bf k}) 
& = &
\frac{1}{\sqrt{2}} \Delta_{\Uparrow 0} ({\bf k}) 
+
\frac{1}{\sqrt{2}} \Delta_{0 \Uparrow} ({\bf k}), 
\nonumber
\\
{\widetilde\Delta}_{20} ({\bf k}) 
& = &
\frac{1}{\sqrt{6}} \Delta_{\Uparrow \Downarrow} ({\bf k}) 
+
\sqrt{\frac{2}{3}} \Delta_{0 0} ({\bf k}) 
+
\frac{1}{\sqrt{6}} \Delta_{\Downarrow \Uparrow} ({\bf k}), 
\nonumber
\\
{\widetilde\Delta}_{2 {\bar 1}} ({\bf k}) 
& = &
\frac{1}{\sqrt{2}} \Delta_{0\Downarrow} ({\bf k}) 
+
\frac{1}{\sqrt{2}} \Delta_{\Downarrow 0} ({\bf k}), 
\nonumber
\\
{\widetilde\Delta}_{2 {\bar 2} } ({\bf k}) 
& = &
\Delta_{\Downarrow \Downarrow} ({\bf k}),
\nonumber 
\end{eqnarray}
where we used the notation ${\bar m}_s = - m_s$.

From the linear combinations given above, we can see that the order 
parameter components in the singlet and quintet sectors are symmetric
with respect to mixed-color exchange, while those in the triplet 
sector are anti-symmetric with respect to mixed-color exchange.
However, for non-zero color-orbit coupling and color-flip field, but
zero color-shift ($\delta = 0$), the tensor $\Delta_{\alpha \beta} ({\bf k})$ 
is symmetric in mixed-color indices, and thus the only non-vanishing 
components of $\widetilde\Delta_{S m_s} ({\bf k})$ occur in the singlet or quintet 
sectors, while all the components in the triplet sector vanish 
identically. In Figs.~\ref{fig:seven}c and d, we show the non-vanishing 
components of $\widetilde\Delta_{S m_s} ({\bf k})$. 

In Figs.~\ref{fig:seven}c and d, the momentum dependences of the order
parameter tensor $\widetilde\Delta_{S m_s} ({\bf k})$ are the same for the R3 and R1 
superfluid phases for fixed $\Omega/E_F$ and varying $1/(k_F a_s)$, but 
the overall magnitude is different. Again, this shows that the nodes in the 
quasiparticle band $E_3 ({\bf k})$ are not trivially related to the nodes of 
the order parameter tensor $\widetilde\Delta_{S m_s} ({\bf k})$.
In Fig.~\ref{fig:seven}c the solid yellow curve corresponds to 
$\widetilde \Delta_{22} ({\bf k})$, 
the dashed cyan curve indicates $\widetilde \Delta_{21} ({\bf k})$,
the dot-dashed purple curve describes $\widetilde \Delta_{20} ({\bf k})$.
Notice that $\widetilde \Delta_{22} ({\bf k})$ has an f-wave character (three nodes),
$\widetilde \Delta_{21} ({\bf k})$ has also an f-wave character (two nodes and
one discontinous sign change), and $\widetilde \Delta_{20} ({\bf k})$ 
has a p-wave character (one node). 
In Fig.~\ref{fig:seven}d the solid light-blue line indicates 
$\widetilde \Delta_{2{\bar 1}} ({\bf k})$,
the dashed red line shows $\widetilde \Delta_{2{\bar 2}} ({\bf k})$,
the dot-dashed black line describes $\widetilde \Delta_{00} ({\bf k})$.
Notice that $\widetilde \Delta_{2{\bar 1}} ({\bf k})$ has 
an f-wave character (two nodes and a discontinuous sign change),
$\widetilde \Delta_{2{\bar 2}} ({\bf k})$ has a p-wave character (one node), 
and $\widetilde \Delta_{00} ({\bf k})$ has also a p-wave character (one node). 
Lastly, since the mixed-color order parameter tensor 
$\Delta_{\alpha \beta} ({\bf k})$ is symmetric in $\alpha \leftrightarrow \beta$, 
the triplet sector tensor components
$\widetilde\Delta_{1,1} ({\bf k})$, 
$\widetilde\Delta_{1,0} ({\bf k})$ and 
$\widetilde\Delta_{1,\bar 1} ({\bf k})$ all vanish identically, and thus these 
components are not shown in Fig.~\ref{fig:seven}.

In order to understand better color pairing phenomena and color superfluid phases,
we discuss next spectroscopic properties, such as 
quasiparticle excitation spectrum  obtained via pairing in 
the mixed-color basis, as well as momentum distributions and density
of states of fermions in their original colors $\{R, G, B\}$.

\section{Spectroscopic Properties}
\label{sec:spectroscopic-properties}

In this section, we discuss several spectroscopic properties of 
color superfluids in the presence of color-orbit coupling and color-flip fields. 
These spectroscopic properties can help characterize the different topological
phases that emerge for fixed color-orbit coupling, but changing 
color-flip fields $\Omega/E_F$ and interactions 
$1/(k_F a_s)$. We begin our discussion
by analysing the quasiparticle and quasihole excitation spectrum. 

\subsection{Quasiparticle energy spectrum}
\label{sec:quasiparticle-energy-spectrum}

To investigate in detail the quasiparticle and quasihole excitation spectrum, 
it is easier to start from the Hamiltonian written in the mixed color basis
$\{ \Uparrow, 0, \Downarrow \}$ as described in 
Eq.~(\ref{eqn:saddle-point-hamiltonian-matrix-helicity}). 
In this case, the matrix ${\bf H}_M ({\bf k})$ has only diagonal elements
$
\{ 
\xi_{\Uparrow} ({\bf k}),  \xi_{0} ({\bf k}),  \xi_{\Downarrow} ({\bf k})  
\}
$
corresponding to mixed-color particle energies. 
Furthermore, the matrix 
$
-{\bf H}_M^* (-{\bf k})
$ 
has also only diagonal elements
$
\{ 
-\xi_{\Uparrow} (-{\bf k}), -\xi_{0} (-{\bf k}), -\xi_{\Downarrow} (-{\bf k})  
\}
$
corresponding to mixed-color hole energies.
The matrices $\Delta_M$ and $\Delta_M^{\dagger}$ are characterized
by the elements 
$
\left[
\Delta_{M}
\right]_{\alpha \beta}
=
\Delta_{\alpha \beta} ({\bf k})
$ 
and
$
\left[
\Delta_{M}^{\dagger}
\right]_{\alpha \beta}
=
\Delta_{\beta \alpha}^* ({\bf k}),
$
which couple mixed-color bands with indices 
$
\{ \alpha, \beta \} 
= 
\{ \Uparrow, 0 , \Downarrow \},
$
and thus tend to lift degeneracies between 
particle $\xi_{\alpha} ({\bf k})$ and 
hole $-\xi_{\beta} (- {\bf k})$
mixed-color bands.

The quasiparticle and quasihole excitation spectrum can be found analytically 
from the secular equation 
$
{\rm det} 
\left[
\omega {\bf 1} 
- {\widetilde H}_0 ({\bf k})
\right]
=
0.
$
Notice that 
$P(\omega) = 
{\rm det} 
\left[
\omega {\bf 1} 
- {\widetilde H}_0 ({\bf k})
\right]
=
\prod_{j} 
\left[
\omega - E_j ({\bf k})
\right]
$
is in general a polynomial of order six, and admits six eigenvalues
$E_j ({\bf k})$, three of them with positive energy corresponding 
to quasiparticles, and three with negative energy corresponding to quasiholes. 
Since the eigenvalues of the Hamiltonian matrix are independent of the basis 
representation that is used, we recover the same eigenvalues as those 
from the direct diagonalization of the Hamiltonian matrix in the original
color basis $\{c, c^\prime\} = \{R, G, B\}$.

Recall that the Hamiltonian ${\widetilde H}_0 ({\bf k})$ is particle-hole
symmetric, implying that its eigenvalues
$E_j ({\bf k})$ satisfy the quasiparticle-quasihole symmetry
$E_j ({\bf k}) = -E_{7-j} (-{\bf k})$, 
and in the case of zero color-shift field $\delta = 0$, where
parity is a good quantum number, the eigenvalues are 
parity even satisfying the relation $E_j ({\bf k}) = E_j (-{\bf k})$.
In this case, the characteristic polynomial becomes
$
P(\omega) 
= 
a_0 ({\bf k}) 
+ 
a_2 ({\bf k}) \omega^2 
+ 
a_4 ({\bf k}) \omega^4
+ 
\omega^6.
$
The nodal structure can be found by setting $\omega = 0$, and coincides
with the results for the original Hamiltonian matrix, that is 
$a_0 ({\bf k}) = 0$ leads to the same nodal structure previously obtained.
At this point it is illustrative to compare the present situation in 
color Fermi superfluids with that of spin-1/2 Fermi superfluids, where
the nodal structure in the quasiparticle energies is directly 
related to the nodal structure of the $2 \times 2$ order parameter 
tensor $\Delta_{\alpha \beta} ({\bf k})$ or more precisely related 
to the nodal structure of $\Delta_{S m_s} ({\bf k})$ in the singlet 
and triplet sectors~\cite{seo-2012}. The situation for color Fermi 
superfluids is very different, given that the coefficient $a_0 ({\bf k})$
depends in a non-trivial way not only on the components of the 
$\Delta_{\alpha \beta} ({\bf k})$, but also on the eigenergies 
$\xi_{\alpha} ({\bf k})$ of the mixed color states 
$\{ \Uparrow, 0, \Downarrow \}$. 

Given that the Hamiltonian matrix ${\widetilde H}_0 ({\bf k})$
is Hermitian, its eigenvalues are guaranteed to be real, 
so the discriminant ${\mathcal D}$ of the cubic equation obtained 
with the substitution $z = \omega^2$ is always non-positive, that is, 
${\mathcal D} \le 0$. The cubic equation obtained can be written as
$P_3 (z) = c + b z + a z^2 + z^3$, where 
$c = a_0 ({\bf k})$, $b = a_2 ({\bf k})$ and $a = a_4 ({\bf k})$,
and can be solved exactly using the Cardano method~\cite{cardano-2007}.
The discriminant can be obtained from the
auxiliary functions $Q = (3b - a^2)/9$ and 
$R = (9ab - 27c - 2a^3)/54$ as ${\mathcal D} =  Q^3 + R^2$. 
If ${\mathcal D} \le  0$, it is clear that
$Q^3 = {\mathcal D} - R^2$ is also negative, and thus
both $-Q^3$ and $-Q$ are positive. If we let 
$\cos (\theta) = R/\sqrt{-Q^3}$, then the three real 
roots of $P_3 (z)$ are 
$z_1 = 2 \sqrt{-Q} \cos (\theta/3) - a/3$,
$z_2 = 2 \sqrt{-Q} \cos \left[ (\theta + 2\pi)/3 \right] - a/3$,
and 
$z_3 = 2 \sqrt{-Q} \cos \left[ (\theta + 4\pi)/3)\right] - a/3$. The three 
roots of the cubic polynomial $P_3 (z)$ correspond to the squares of
the excitations energies $E_j^2 ({\bf k})$, and thus lead to the 
six solutions $E_j ({\bf k})$ that we are seeking. The positive energy solutions
$E_1 ({\bf k}), E_2 ({\bf k}), E_3 ({\bf k})$ correspond to quasiparticle 
excitations and the negative energy solutions 
$E_4 ({\bf k}), E_5 ({\bf k}), E_6 ({\bf k})$ correspond to quasihole
excitations. The analytic solutions for $E_j ({\bf k})$ agree with the 
direct numerical diagonalization of either ${\bf H}_0 ({\bf k})$ in the
color basis $\{ R, G, B \}$ 
defined in Eq.~(\ref{eqn:saddle-point-hamiltonian-matrix})
or $\widetilde{\bf H}_0 ({\bf k})$ 
in the mixed color basis $\{\Uparrow, 0, \Downarrow \}$
defined in Eq.~(\ref{eqn:saddle-point-hamiltonian-matrix-helicity}). 
However, the reader must agree that these analytic solutions are not 
particularly illuminating.

\begin{figure} [hbt]
\centering 
\epsfig{file=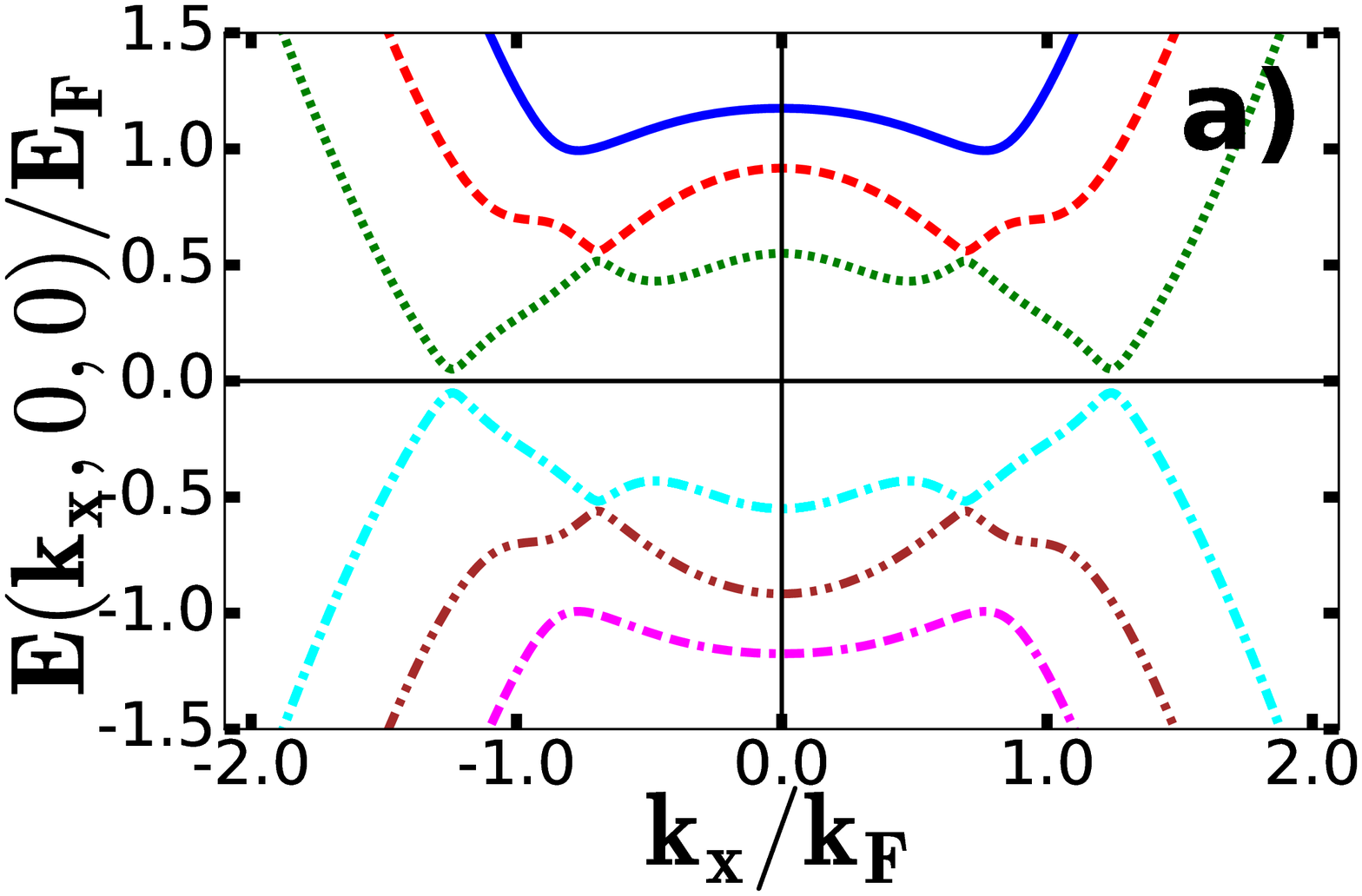,width=0.49 \linewidth}
\epsfig{file=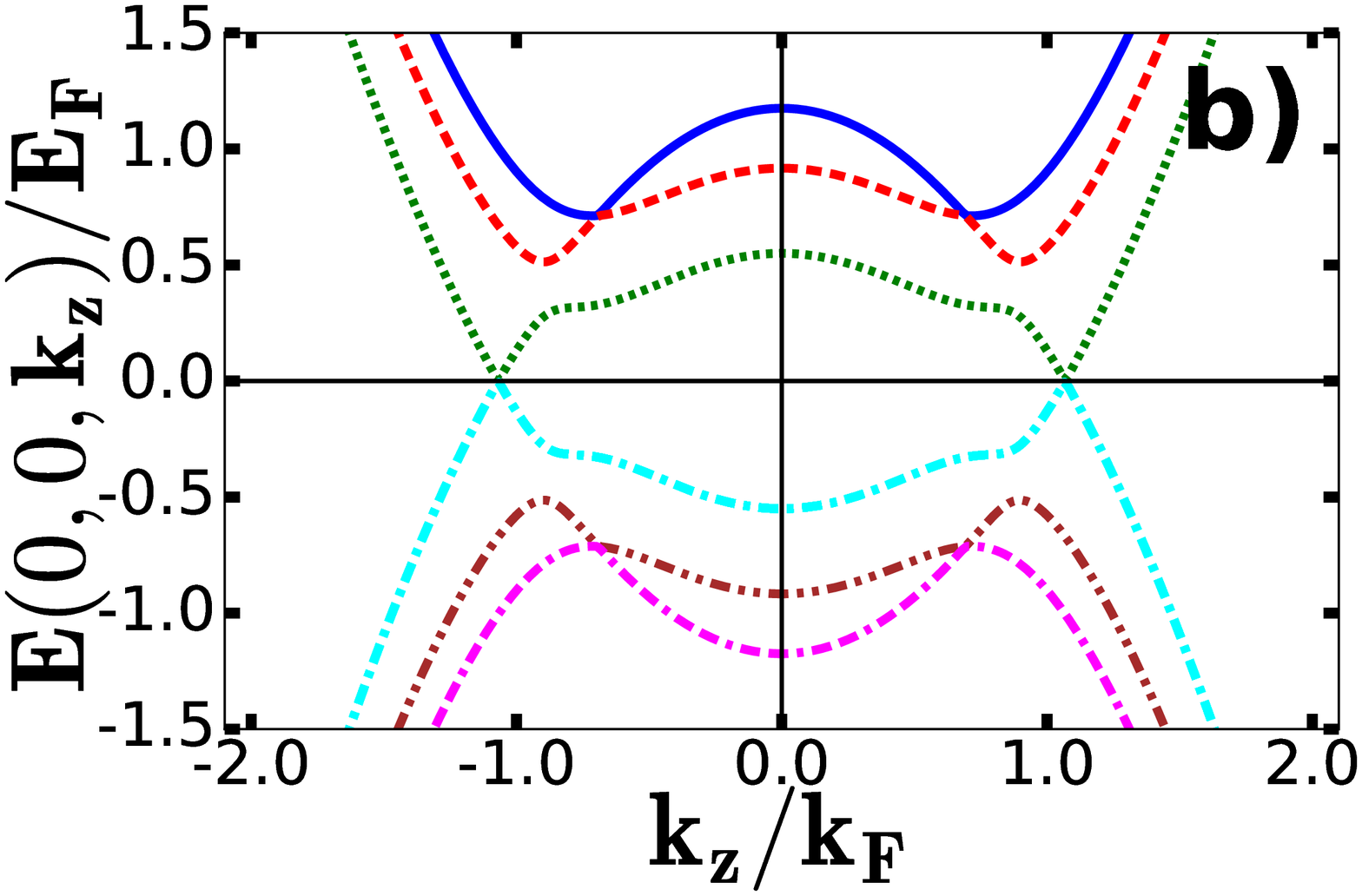,width=0.49 \linewidth}
\\
\epsfig{file=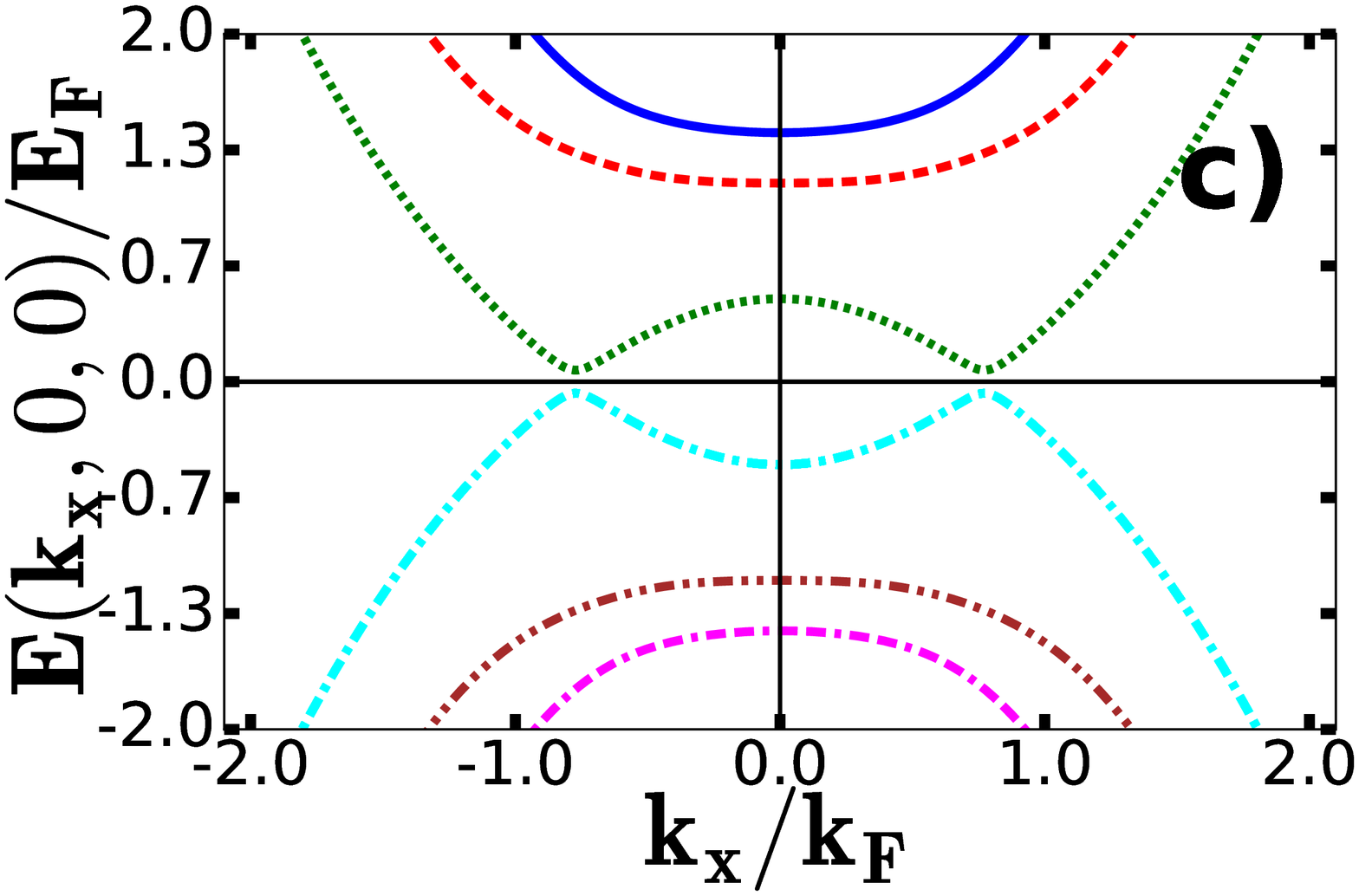,width=0.49 \linewidth}
\epsfig{file=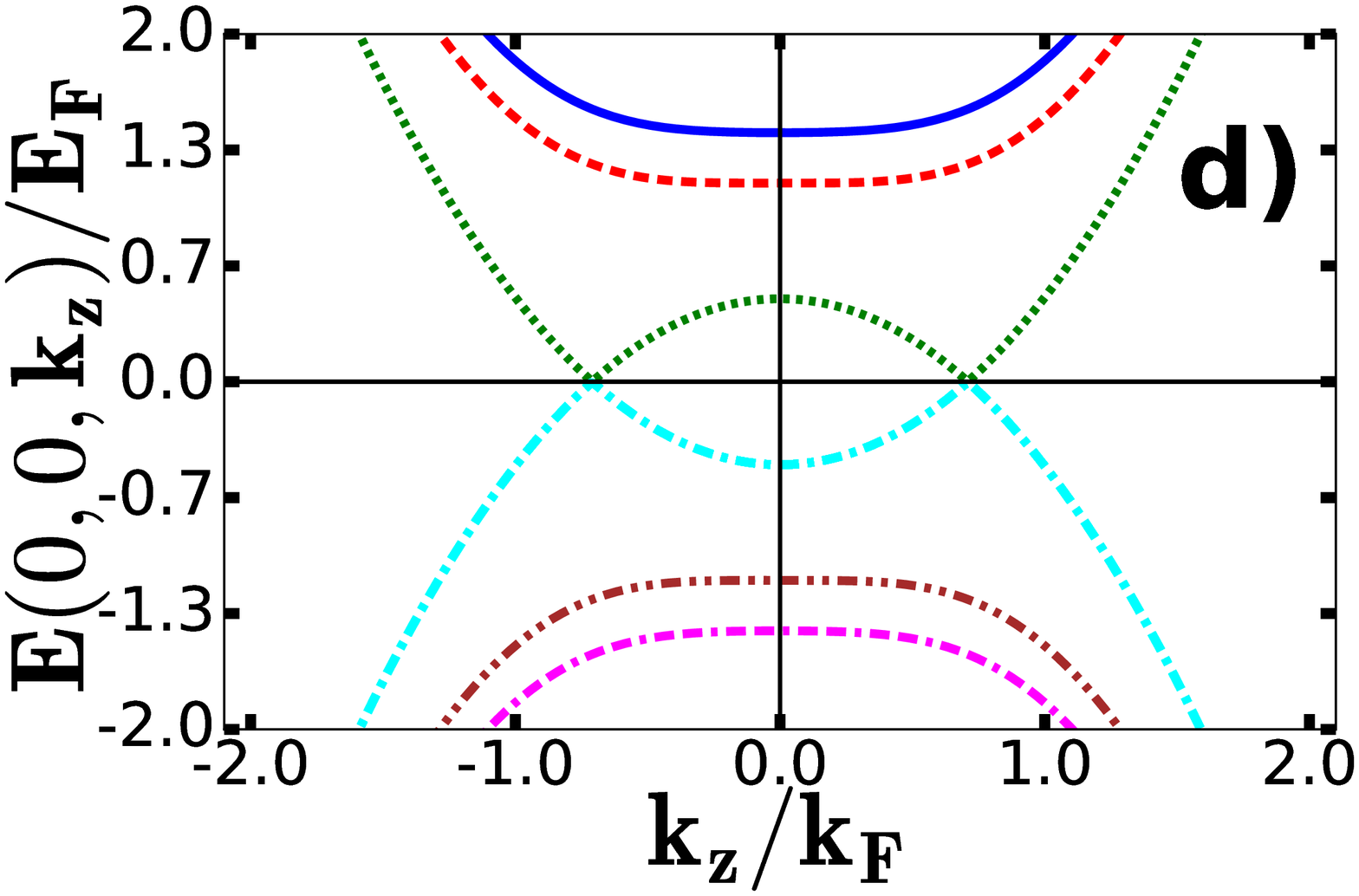,width=0.49 \linewidth}
\caption{ 
\label{fig:eight}
(Color online)
Quasiparticle and quasihole excitation spectra $E_j ({\bf k})$
in superfluid phases $R3$ and $R1$ for fixed parameters $\Omega/E_F = 0.29$,
$k_T = 0.35 k_F$, $T/E_F = 0.01$, 
and $b_z = k_T^2/(2m)$.
The style-color code for energies $E_j ({\bf k})$ is: 
$E_1 ({\bf k})$ is solid blue, 
$E_2 ({\bf k})$ is dashed red, 
$E_3 ({\bf k})$ is dotted green,
$E_4 ({\bf k})$ is dash-dotted cyan,
$E_5 ({\bf k})$ is dashed-double-dotted brown,
and $E_6 ({\bf k})$ is double-dashed-dotted magenta. 
In a) and b), 
we show $E_j ({\bf k})$ versus $(k_x, 0, 0)$ 
and $(0,0, k_z)$, respectively for the $R3$ phase with parameters 
$1/(k_F a_s) = -0.069$ $(\mu/E_F = 0.81, \vert \Delta \vert/E_F = 0.31)$.
In c) and d), 
we show $E_j ({\bf k})$ versus $(k_x, 0, 0)$ 
and $(k_x = 0, 0, k_z)$, respectively, 
for the $R1$ phase with parameters
$1/(k_F a_s) = 0.62$ $(\mu/E_F = 0.19, \vert \Delta \vert/E_F = 0.73)$.
}
\end{figure}

In order to understand the excitation spectrum obtained on physical grounds,
it is more convenient to work in the mixed-color basis 
$\{\Uparrow, 0, \Downarrow \}$. The excitation
spectra shown in Fig.~\ref{fig:eight} can be generically understood
as resulting from the coupling of mixed-color particle states with
energies ${\cal E}_\alpha ({\bf k})$ and mixed-color hole states with
energies $-{\cal E}_\beta ({\bf - k})$ via the order parameter 
$\Delta_{\alpha \beta} ({\bf k})$. 
Thus, wherever the energies ${\cal E}_\alpha ({\bf k})$ and  
$-{\cal E}_\beta ({\bf - k})$ cross in momentum space, 
the order parameter matrix elements $\Delta_{\alpha \beta} ({\bf k})$ 
can lift the degeneracies between the bands labeled by $\alpha$ and $\beta$
at the crossing {\it loci} (points, lines, surfaces) and impound additional 
momentum dependence. 

In Fig.~\ref{fig:eight}, we show representative quasiparticle and 
quasihole energies $E_j ({\bf k})$ versus momentum ${\bf k}$ 
in superfluid phases R3 and R1 for fixed parameters 
$\Omega/E_F = 0.29$, $k_T = 0.35 k_F$, $T/E_F = 0.01$, and $b_z = k_T^2/(2m)$.
The spectrum is sorted out such that $E_1 ({\bf k})$ is the highest energy 
and $E_6 ({\bf k})$ is the lowest energy for fixed ${\bf k}$. 
The solid blue curves correspond to $E_1 ({\bf k})$, 
the dashed red plots describe $E_2 ({\bf k})$,
the dotted green lines show $E_3 ({\bf k})$, 
the dash-dotted cyan curves correspond to $E_4 ({\bf k})$, 
the dash-double-dotted brown plots describe $E_5 ({\bf k})$,
and the double-dashed-dotted magenta lines show $E_6 ({\bf k})$. 
The excitation spectrum $E_j ({\bf k})$ has cylindrical symmetry around the
$k_x$ axis, and thus its momentum dependence $(k_x, k_y, k_z)$ is
characterized only by the coordinates $(k_x, k_\perp)$, 
where $k_{\perp} = \sqrt{k_y^2 + k_z^2}$ is the magnitude
of momentum in the $k_y k_z$ plane.

In Figs.~\ref{fig:eight}a and b, 
we show $E_j ({\bf k})$ versus $(k_x, 0, 0)$ 
and $(0, 0, k_z)$, respectively, for the R3 phase with parameters 
$1/(k_F a_s) = -0.069$ $(\mu/E_F = 0.81, \vert \Delta \vert /E_F = 0.31)$.
Notice, that only one ring of nodes is illustrated in 
Fig.~\ref{fig:eight}b where $k_x = 0$, since the nodal points in $k_z$ correspond
to a ring of nodes in the $(k_y, k_z)$ plane due to cylindrical symmetry. 
The other two rings of nodes for the $R3$ phase occur at characteristic values 
$k_x = \pm k_x^* \ne 0$ and $k_{\perp} = k_{\perp}^* \ne 0$ 
as found in Fig.~\ref{fig:four}c. The additional 
rings of the R3 phase are not seen in the spectrum shown in Fig.~\ref{fig:eight},
because in Fig.~\ref{fig:eight}a the magnitude of the momentum 
in the $(k_y, k_z)$ plane is $k_{\perp} = 0$ and in 
Fig.~\ref{fig:eight}b the momentum along the $k_x$ direction is $k_x = 0$.

In Figs.~\ref{fig:eight}c and d, 
we show $E_j ({\bf k})$ versus $(k_x, 0, 0)$ 
and $(0, 0, k_z)$, respectively, for the $R1$ phase with parameters
$1/(k_F a_s) = 0.62$ $(\mu/E_F = 0.19, \vert \Delta \vert/E_F = 0.73)$.
The $R1$ phase has only one ring of nodes (see Fig.~\ref{fig:four}d), and 
this ring is illustrated in the spectrum shown in Fig.~\ref{fig:eight}d,
where $k_x = 0$ and the nodal points in $k_z$ correspond to a ring
of nodes in the $(k_y, k_z)$ plane. 

%

We would like to point out that photoemission spectroscopy has been already used 
to probe directly the elementary excitations and energy dispersion in a 
strongly interacting Fermi gas of $^{40}$K atoms with two internal 
states throughout the evolution from the BCS to the BEC limits~\cite{jin-2008}, 
but without spin-orbit coupling or Zeeman fields. 
The use of the same technique for the color problem with $^6$Li, $^{40}$K or 
$^{173}$Yb should reveal the rich nodal structure of the excitation spectrum 
when color-orbit and color-flip fields are present, and thus
provide direct evidence of the various topological superfluid phases shown
in the phase diagram of Fig.~\ref{fig:three}a.

In addition to measuring the quasiparticle dispersions, there are other
auxiliary experiments than can help characterize the quantum phases found in the
color problem. Therefore, we analyse next the momentum distribution 
$n_c ({\bf k})$ for color states $c = \{R, G, B \}$ as color-flip fields 
and interactions are changed for fixed color-orbit coupling. 

\subsection{Momentum Distribution}
\label{sec:momentum-distribution}

The measurement of momentum distributions is an easy experiment 
to do in systems of cold-atoms, and it is routinely realized
in atomic fermions and bosons. More recently these types of measurements
have also been performed in ultracold fermions 
such as $^{40}$K with two internal states 
and spin-orbit coupling~\cite{spielman-2013},
as well as $^{173}$Yb with three or more
internal states and spin-orbit orbit coupling~\cite{fallani-2016}.

Therefore, in this section, we describe how the momentum distribution 
$n_c ({\bf k})$ for different color states $c = \{ R, G, B \}$ 
can be obtained directly from the resolvent operator
\begin{equation}
\hat{\bf G}(z) 
= 
\left[ z \hat{\bf I} - \hat{\bf H} \right]^{-1},
\end{equation}
where z is a complex energy, $\hat {\bf I}$ is the 
identity operator and $\hat{\bf H}$ is the full Hamiltonian of the system. 
In the present case, the resolvent operator can be written in energy and 
momentum space as the $6 \times 6$ matrix
\begin{equation}
{\bf G}_{\nu \nu^\prime} (z, {\bf k}) 
=
\left[ z{\bf I} - {\bf H}_0 ({\bf k}) \right]^{-1}_{\nu \nu^\prime},
\end{equation}
where ${\bf H}_0 ({\bf k})$ is the Hamiltonian matrix defined
in Eq.~(\ref{eqn:saddle-point-hamiltonian-matrix}) describing the color superfluid 
phases at low temperatures, and $\{ \nu, \nu^\prime \}$ are Nambu color indices 
representing the states created by the six-dimensional colored-Nambu spinor
$
{\bf f}_N^\dagger ({\bf k})
=
\left[
f_{\rm R}^\dagger ({\bf k}), 
f_{\rm G}^\dagger ({\bf k}),
f_{\rm B}^\dagger ({\bf k}),
f_{\rm R}({\bf -k}), 
f_{\rm G}({\bf -k}),
f_{\rm B}({\bf -k})
\right],
$
which was defined in section~\ref{sec:order-parameter-and-reduced-Hamiltonian}.

Writing the eigenstates of the Hamiltonian matrix 
${\bf H}_0 ({\bf k})$ 
as
$
\vert j, {\bf k}\rangle 
= 
{\bf M}_{j\nu} ({\bf k}) \vert \nu, {\bf k} \rangle
$
in terms of the Nambu color states 
$\vert \nu, {\bf k} \rangle$ 
and using the corresponding eigenvalues $E_j ({\bf k})$,
leads to the Green's function matrix
\begin{equation}
\label{eqn:greens-function}
{\bf G}_{\nu \nu^\prime} (z, {\bf k}) 
=
\sum_j 
\frac{M_{j\nu} ({\bf k}) M_{\nu^\prime j}^* ({\bf k})}{z - E_j ({\bf k})}.
\end{equation}
The momentum distribution for color state $c$ can 
be written as
$
n_c ({\bf k})
= 
- T
\sum_{i\omega_n} 
{\bf G}_{cc} (i\omega_n, {\bf k}),
$
where ${\bf G}_{cc} (i\omega_n, {\bf k})$ are the
first three diagonal elements of 
${\bf G}_{\nu \nu^\prime} (z = i \omega_n, {\bf k})$,
where $T$ is the temperature and $\omega_n = (2n+1)\pi/T$ being the 
fermionic Matsubara frequencies. Performing the Matsubara sums leads to the 
momentum distribution 
\begin{equation}
\label{eqn:color-momentum-distributions}
n_c ({\bf k}) 
=
\sum_{j} 
\vert 
{\bf M}_{jc} ({\bf k}) 
\vert^2
F\left[ E_j({\bf k}) \right],
\end{equation}
where the matrix element 
$M_{jc} ({\bf k})$ represents the probability amplitude of finding
the color state $\vert c, {\bf k} \rangle$ as part of the eigenstate
$\vert j, {\bf k} \rangle$ and 
$
\vert 
{\bf M}_{jc} ({\bf k}) 
\vert^2
=
{\bf M}_{jc} ({\bf k}) {\bf M}_{cj}^* ({\bf k}).
$
Here, $F\left[E_j ({\bf k})\right]$ is the
Fermi function associated with eigenergy $E_j ({\bf k})$, and the summation 
over $j$ includes both quasiparticle and quasihole states, that is, 
$j$ runs from 1 to 6.

\begin{figure} [tbh]
\centering 
\epsfig{file=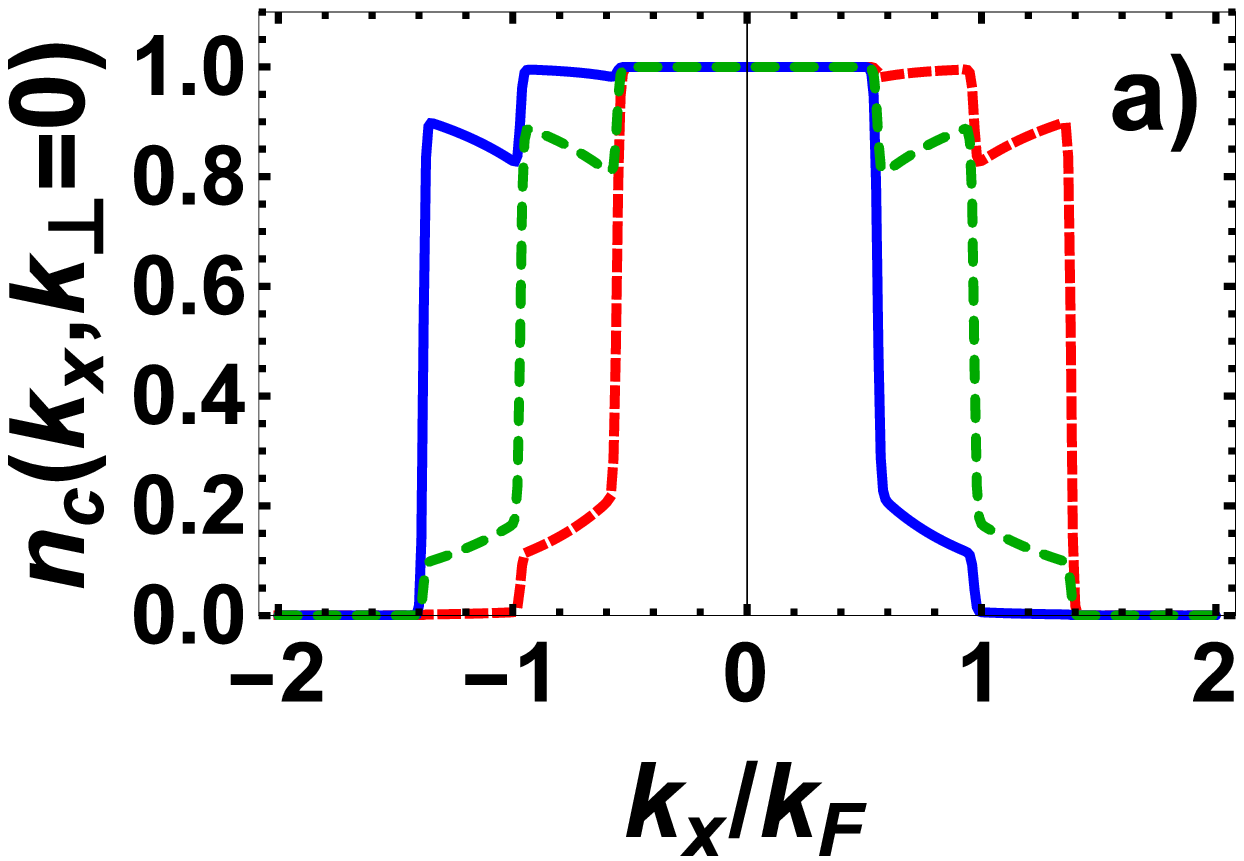,width=0.49 \linewidth}
\epsfig{file=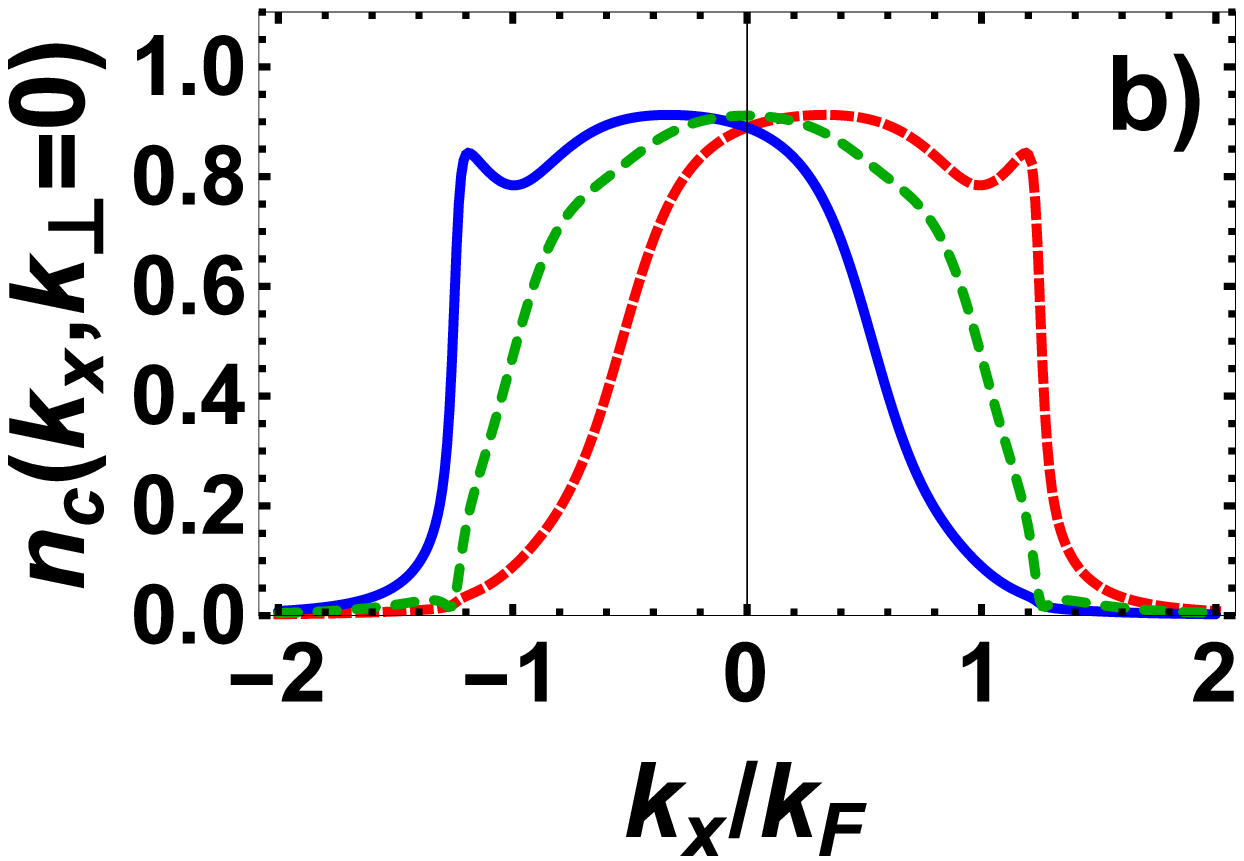,width=0.49 \linewidth}
\\
\epsfig{file=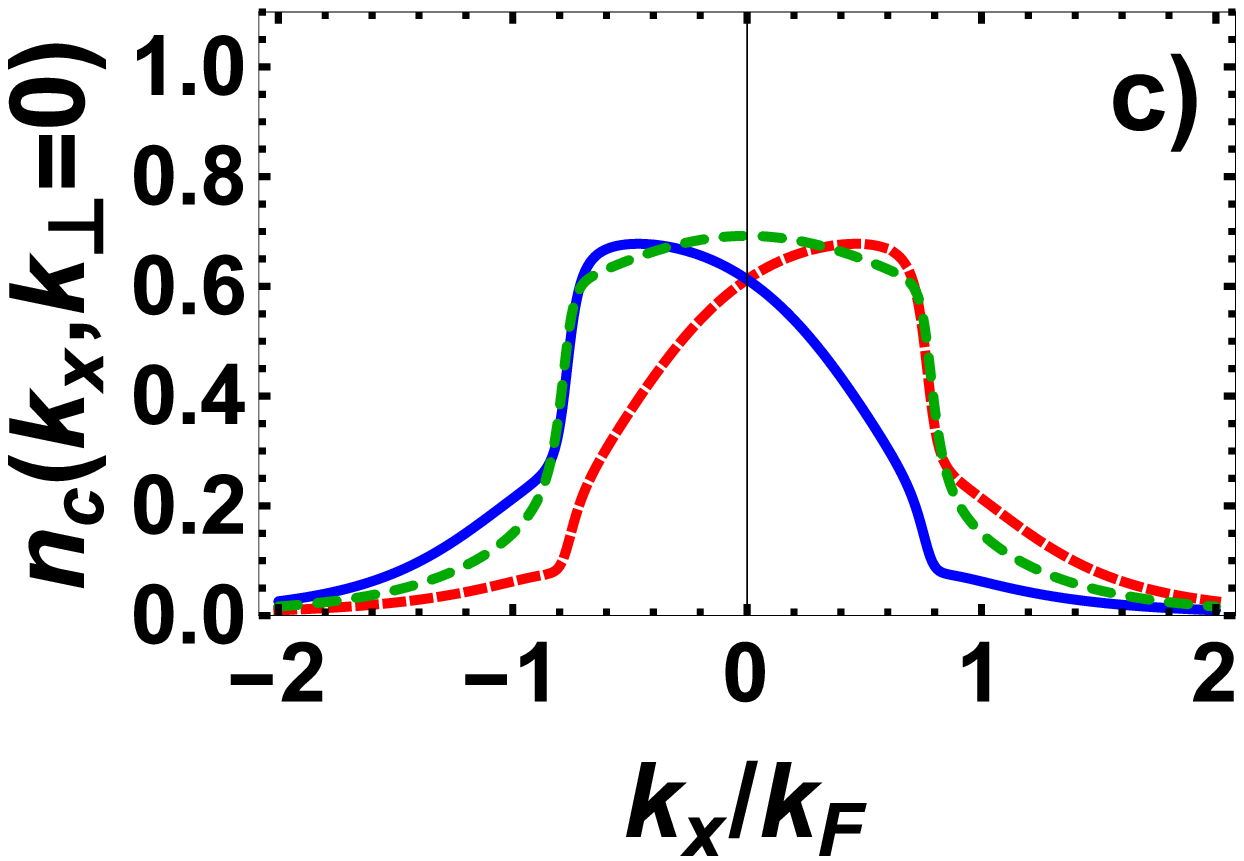,width=0.49 \linewidth}
\epsfig{file=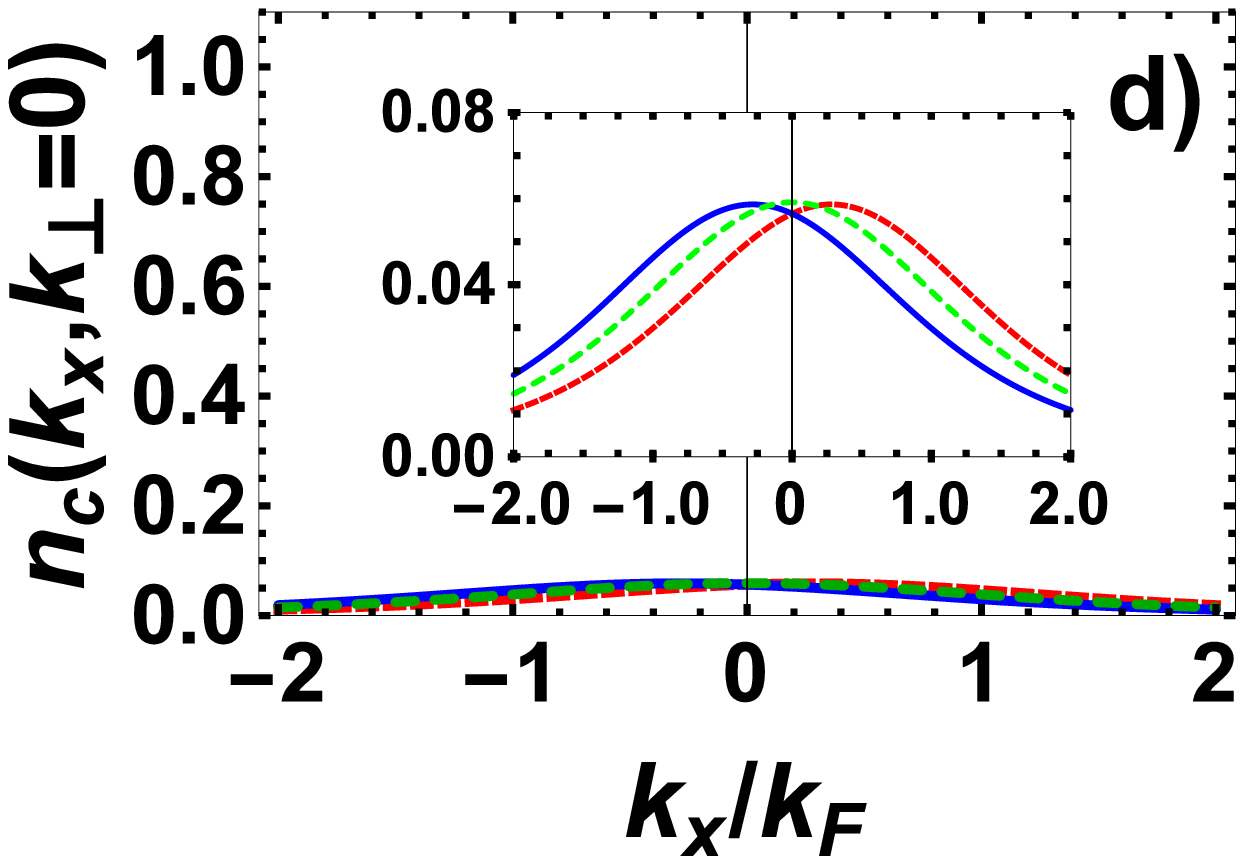,width=0.49 \linewidth}
\caption{ 
\label{fig:nine}
(Color online)
Momentum distributions $n_c ({\bf k})$ for
for $\Omega = 0.29E_F$, $k_T = 0.35 k_F$, 
$b_z = k_T^2/(2m)$ and $T = 0.01 E_F$
along direction $(k_x, 0, 0)$ for different color indices 
$c = \{R,G,B \}$. 
The solid blue lines correspond to Blue states, 
the long-dashed green lines to Green states,
and the short-dashed red lines to Red states. 
In a) we show plots for the the normal phase $N3$ 
with scattering parameter $1/(k_F a_s) = -1.8$ 
($\mu/E_F = 0.97$ and $\vert \Delta \vert/E_F = 0.0$).
In b) we show plots for the three-rings superfluid phase $R3$ 
with scattering parameter $1/(k_F a_s) = -0.069$
($\mu/E_F = 0.81$ and $\vert \Delta \vert/E_F = 0.31$).
In c) we show plots for the one-ring superfluid phase $R1$ 
with scattering parameter $1/(k_F a_s) = 0.62$
($\mu/E_F = 0.19$ and $\vert \Delta \vert/E_F = 0.73$).
In d) we show plots for the fully gapped superfluid phase $FG$
with scattering parameter $1/(k_F a_s) = 1.8$
($ \mu/E_F = -2.88$ and $\vert \Delta \vert/E_F = 1.25$).
}
\end{figure}

In Figs.~\ref{fig:nine}a-d, we show momentum distributions $n_c ({\bf k})$
for $\Omega = 0.29E_F$, $k_T = 0.35 k_F$, $b_z = k_T^2/(2m)$ and $T = 0.01 E_F$
that describe the normal phase $N3$ in 
a) with scattering parameter $1/(k_F a_s) = -1.8$ 
($\mu/E_F = 0.97$ and $\vert \Delta \vert/E_F = 0.0$);
the three-rings superfluid phase $R3$ in 
b) with scattering parameter $1/(k_F a_s) = -0.069$
($\mu/E_F = 0.81$ and $\Delta/E_F = 0.31$);
the one-ring superfluid phase $R1$ in
c) with scattering parameter $1/(k_F a_s) = 0.62$
($\mu/E_F = 0.19$ and $\vert \Delta \vert/E_F = 0.73$);
the fully gapped superfluid phase $FG$ in
d) with scattering parameter $1/(k_F a_s) = 1.8$
($ \mu/E_F = -2.88$ and $\vert \Delta \vert/E_F = 1.25$).

A general key feature of the momentum distributions shown
in Fig.~\ref{fig:nine} is that the distributions of
Red fermions are shifted to the right (towards positive $k_x$), 
and that of the Blue fermions are shifted to the left
(towards negative $k_x$), while the distribution 
of Green fermions remain centered at zero momentum. 
A second general feature revealed by the panels 
of Fig.~\ref{fig:nine} is that the momentum distributions 
get smeared and broadened along $k_x$ by the emergence 
of order parameter of the superfluid state and by increasing
scattering parameter. This leads to an overall reduction of 
the maximum values of the momentum distributions, making 
the fermionic system less degenerate, similarly to the case 
of two internal states, that is, the spin-1/2 case.
Another important observation about Fig.~\ref{fig:nine} is
that the momentum distributions $n_{R} ({\bf k})$ of the
Red ($R$) states and $n_{B} ({\bf k})$ of the Blue ($B$) states satisfy
the relation $n_{R} ({\bf k}) = n_{B} (-{\bf k})$, because 
the quasiparticle and quasihole energies are even functions of momentum
$E_j (-{\bf k}) = E_j ({\bf k})$ and the matrix elements 
$M_{j R } ({\bf k}) = M_{j B} (-{\bf k})$. The latter symmetry 
relation follows from the fact that the R and B states 
experience momentum shifts $k_T$ in opposite directions.

\begin{figure} [tbh]
\centering 
\epsfig{file=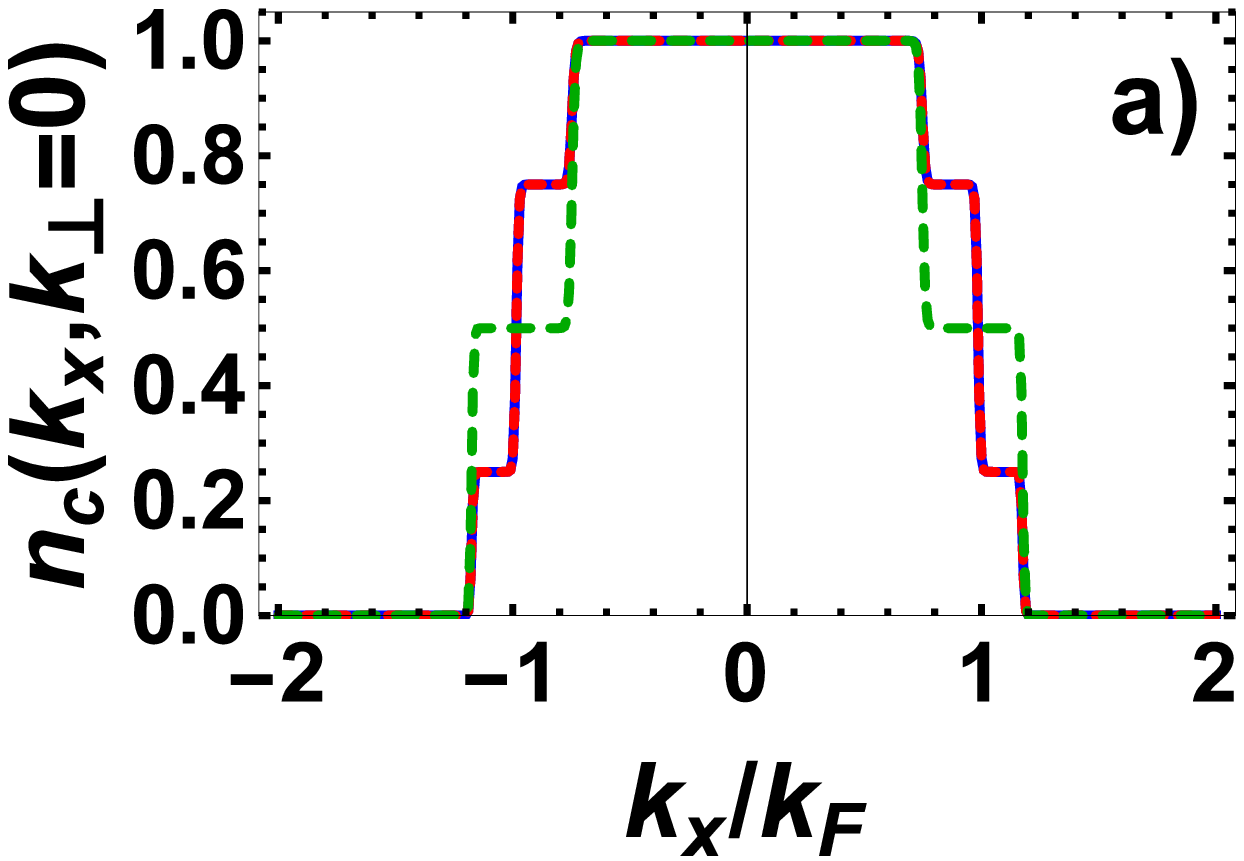,width=0.49 \linewidth}
\epsfig{file=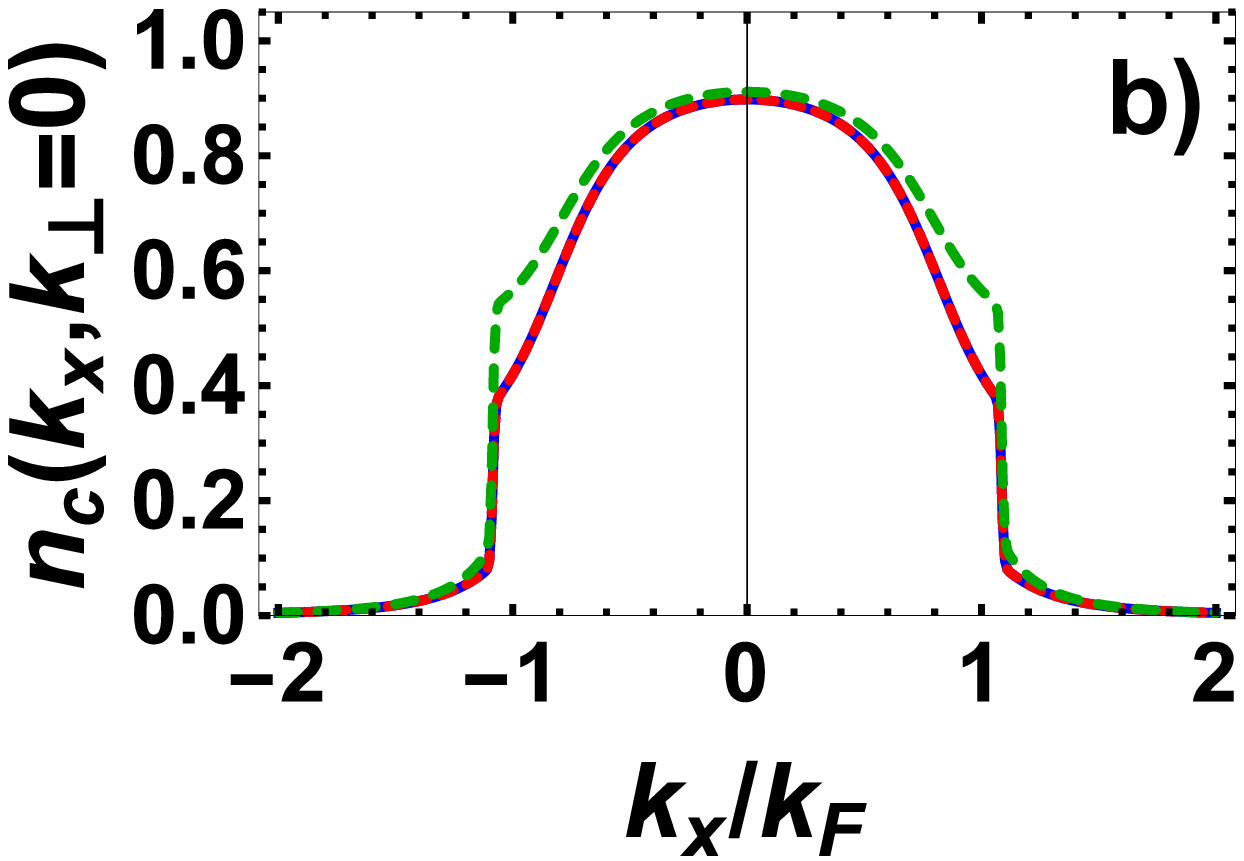,width=0.49 \linewidth}
\\
\epsfig{file=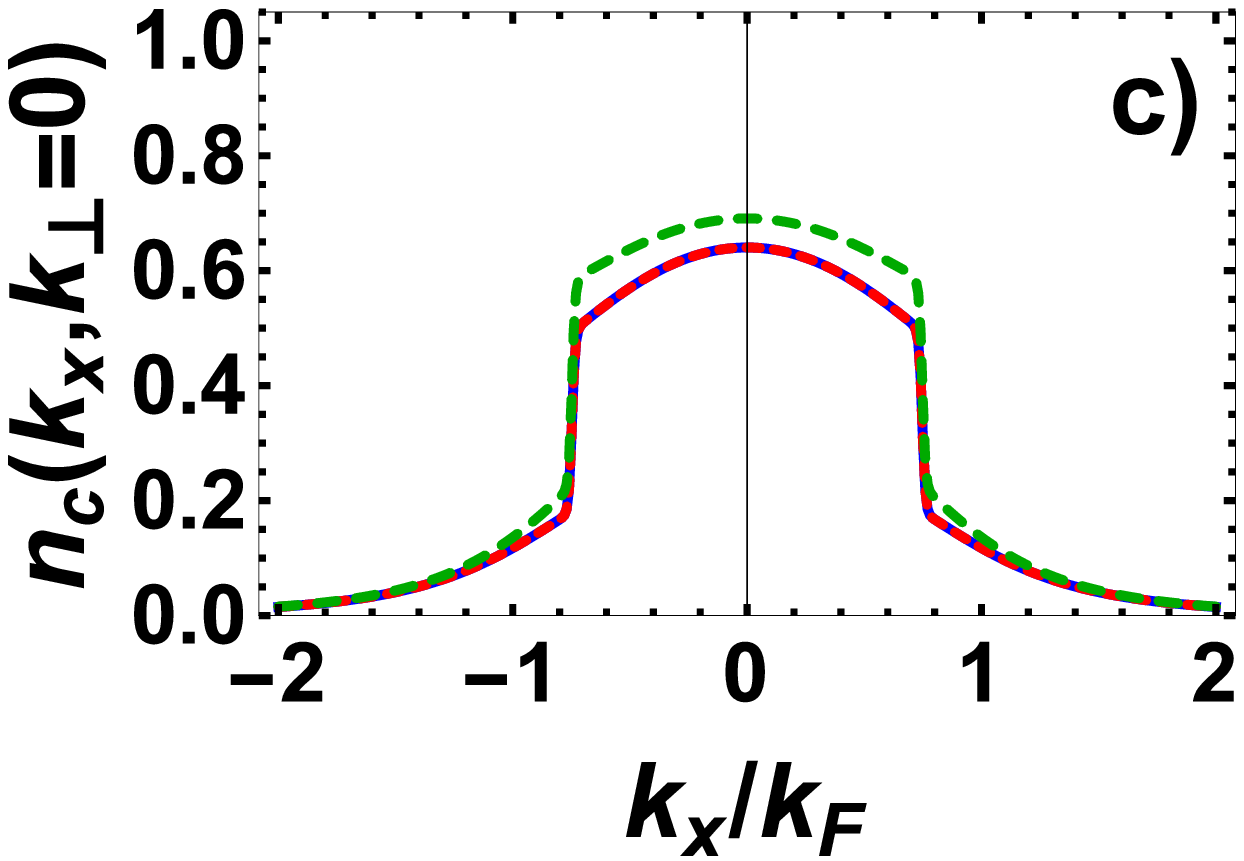,width=0.49 \linewidth}
\epsfig{file=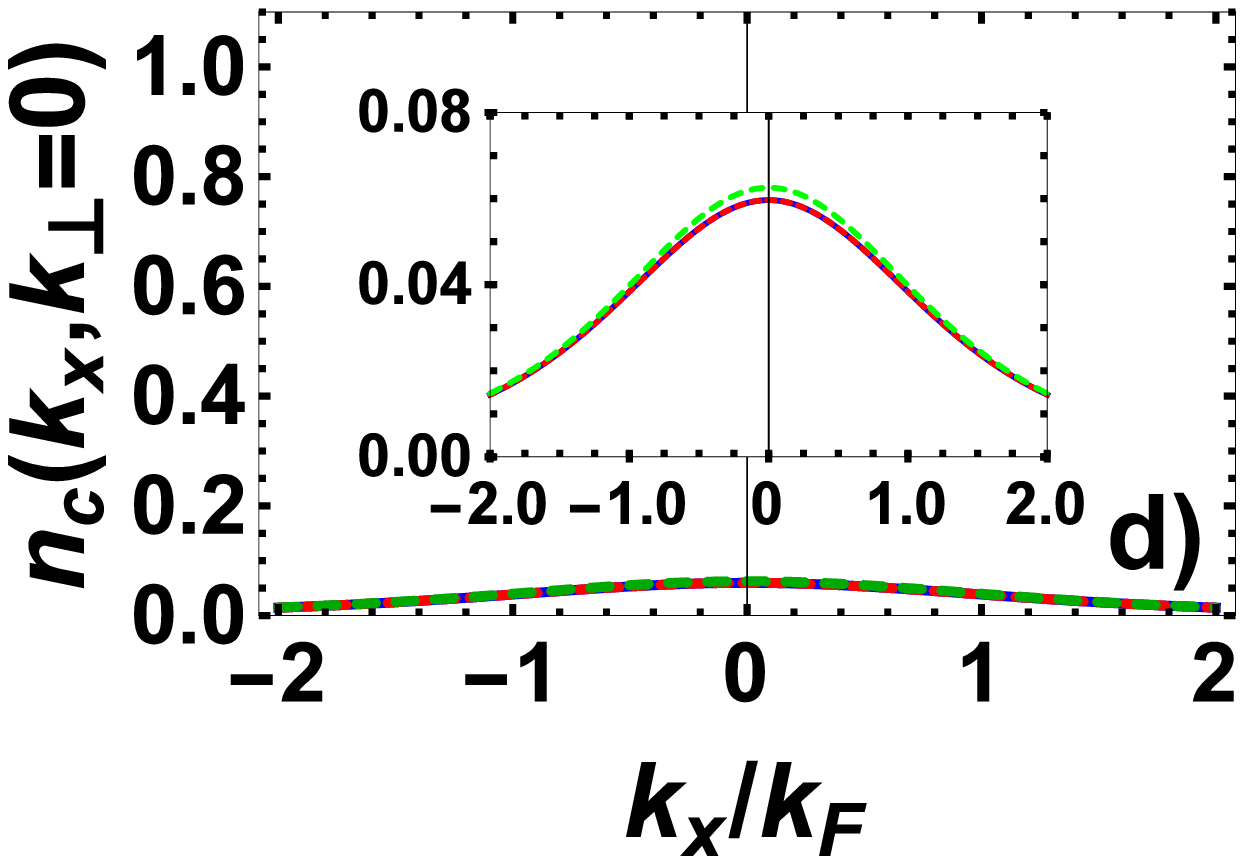,width=0.49 \linewidth}
\caption{ 
\label{fig:ten}
(Color online)
Momentum distributions $n_c ({\bf k})$ for
$k_T = 0$ and $b_z = k_T^2/(2m) = 0$ at $T = 0.01E_F$ 
along direction $(k_x, 0, 0)$ for different color indices 
$c = \{R,G,B \}$. 
The solid blue lines corresponds to the Blue states, 
the long-dashed green lines to the Green states,
and the short-dashed red lines to the Red states. 
In a) we show plots for the normal phase $N3$ 
with scattering parameter $1/(k_F a_s) = -1.8$ 
($\mu/E_F = 0.97$ and $\vert \Delta \vert/E_F = 0.0$).
In b) we show plots for the superfluid phase $S1$ 
with scattering parameter $1/(k_F a_s) = -0.069$
($\mu/E_F = 0.78$ and $\vert \Delta \vert/E_F = 0.33$).
In c) we show plots for the superfluid phase $S1$ 
with scattering parameter $1/(k_F a_s) = 0.62$
($\mu/E_F = 0.17$ and $\vert \Delta \vert/E_F = 0.73$).
In d) we show plots for the fully gapped superfluid phase $FG$
with scattering parameter $1/(k_F a_s) = 1.8$
($ \mu/E_F = -2.88$ and $\vert \Delta \vert/E_F = 1.25$).
}
\end{figure}

In Fig.~\ref{fig:ten}, we show the momentum distributions 
$n_c ({\bf k})$ for zero color-orbit coupling $k_T = 0$, 
zero quadratic color-shift $b_z = k_T^2/(2m) = 0$, 
but for finite color-flip field $\Omega/E_F = 0.29$ 
at $T = 0.01E_F$. The distributions are 
shown along direction $(k_x, 0, 0)$ for different color 
indices $c = \{R,G,B \}$ and can 
be constrasted with those of Fig.~\ref{fig:nine}, 
where the color-orbit coupling is $k_T = 0.35 k_F$. 
To facilitate comparison between Figs.~\ref{fig:nine}
and~\ref{fig:ten}, we use the same scattering parameters
for corresponding panels. 
In a) we show plots for the normal phase $N3$ 
with scattering parameter $1/(k_F a_s) = -1.8$ 
($\mu/E_F = 0.97$ and $\vert \Delta \vert/E_F = 0.0$).
In b) we show plots for the superfluid phase $S1$ 
with scattering parameter $1/(k_F a_s) = -0.069$
($\mu/E_F = 0.78$ and $\vert \Delta \vert/E_F = 0.33$).
In c) we show plots for the superfluid phase $S1$ 
with scattering parameter $1/(k_F a_s) = 0.62$
($\mu/E_F = 0.17$ and $\vert \Delta \vert/E_F = 0.73$).
In d) we show plots for the fully gapped superfluid phase $FG$
with scattering parameter $1/(k_F a_s) = 1.8$
($ \mu/E_F = -2.88$ and $\vert \Delta \vert/E_F = 1.25$).
Notice that the interaction parameters and color-flip fields 
are exactly the same as those of the 
corresponding panels in Fig.~\ref{fig:nine} and were chosen 
as such in order to illustrate the effect of
the color-orbit coupling $k_T$. 

Momentum distributions are relatively easy to measure experimentally 
in the case of two internal states, and there should be no 
additional difficulties in measuring them for the case of color 
states. However, it is important to emphasize that only measurements 
of the quasiparticle and quasihole excitation spectrum 
can identify fully each of the topological superfluid 
phases and their nodal structure, as discussed in 
section~\ref{sec:quasiparticle-energy-spectrum}.

%

A general feature of the momentum distributions shown
in Fig.~\ref{fig:ten} is that the distributions of
Red and Blue fermions are not shifted with respect
to that of the Green fermions, because there is no color-orbit
coupling $(k_T = 0)$, and thus no color-dependent momentum 
transfer. A second general feature revealed by the panels 
of Fig.~\ref{fig:ten} is that the momentum distributions 
get smeared and broadened along $k_x$ by the emergence 
of the order parameter of the superfluid and by the increasing
scattering parameter. This leads to an overall reduction of 
the maximum values of the momentum distributions, making 
the fermionic system less degenerate, similarly to the case 
of Fig.~\ref{fig:nine}, where the color-orbit coupling is
non-zero $(k_T \ne 0)$. Another important observation about 
Fig.~\ref{fig:ten} is that the momentum distributions 
$n_{R} ({\bf k})$ of the Red ($R$) states and $n_{B} ({\bf k})$ 
of the Blue ($B$) states are identical, that is, 
$n_{R} ({\bf k}) = n_{B} ({\bf k})$, because 
the matrix elements $M_{jR} ({\bf k})$ and 
$M_{jB} ({\bf k})$ in Eq.~(\ref{eqn:color-momentum-distributions})
satisfy the relation $\vert M_{j R } ({\bf k})\vert^2 
= \vert M_{j B} ({\bf k}) \vert^2$. This implies 
that the R and B states can no longer be distinguished 
by measurements of their momentum distribution since they 
do not experience momentum shifts in opposite directions, 
that is, $k_T = 0$.

In order to sharpen our understanding of spectroscopic properties of 
color superfluids in the presence of color-orbit coupling and
color-flip fields, we will discuss next the density of states for 
each color.

\subsection{Color Density of States}
\label{sec:color-density-of-states}

The density of states $\rho_c (\omega)$ 
for each color $c = \{R, G, B \}$ can be obtained from the 
Green's function defined in Eq.~(\ref{eqn:greens-function}) as
\begin{equation}
\rho_c (\omega) 
=
- 
\frac{1}{\pi} \sum_{{\bf k}}
\lim_{\delta \to 0} {\cal I} 
\left[ 
G_{cc} (z = \omega + i\delta, {\bf k})
\right],
\end{equation}
where ${\cal I}$ denotes the imaginary part. 
Taking the limit of $\delta \to 0$ leads to the simplified
expression
\begin{equation}
\rho_c (\omega) 
=
\sum_{{\bf k},j} 
\vert {\bf M}_{jc} ({\bf k}) \vert^2
\delta \left[ \omega - E_j ({\bf k}) \right], 
\end{equation}
where the sum over $j = \{1, ..., 6\}$ spans over all 
quasiparticle and quasihole states.

\begin{figure} [tbh]
\centering 
\epsfig{file=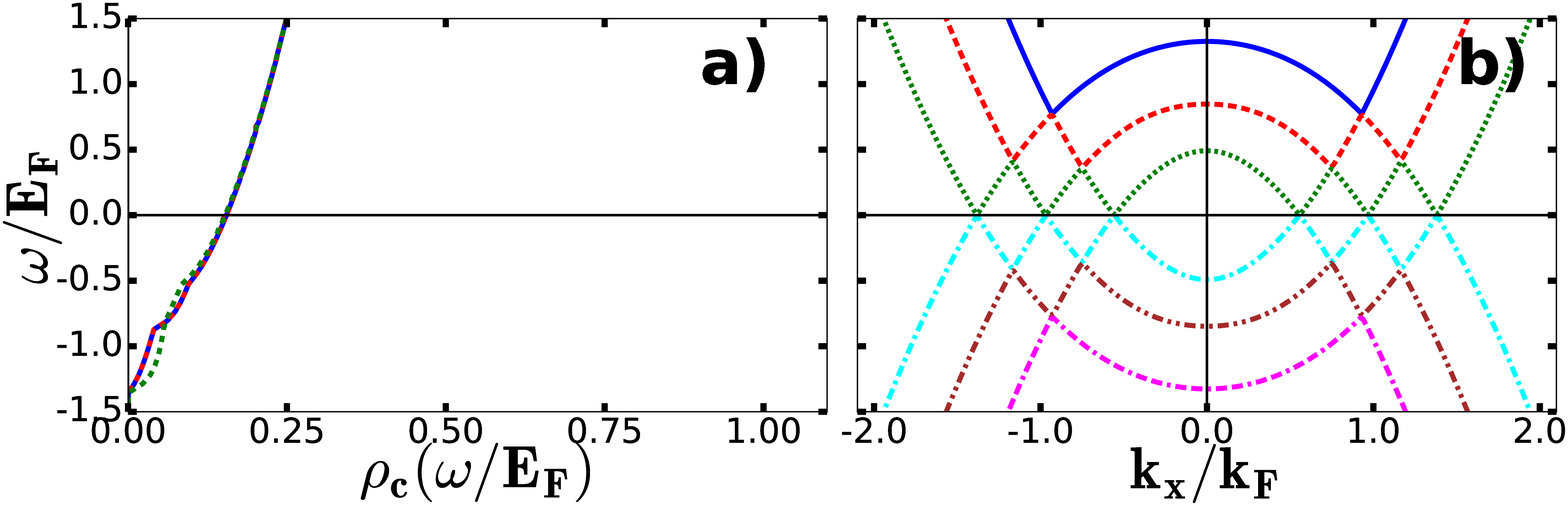,width=0.98 \linewidth} \\
\epsfig{file=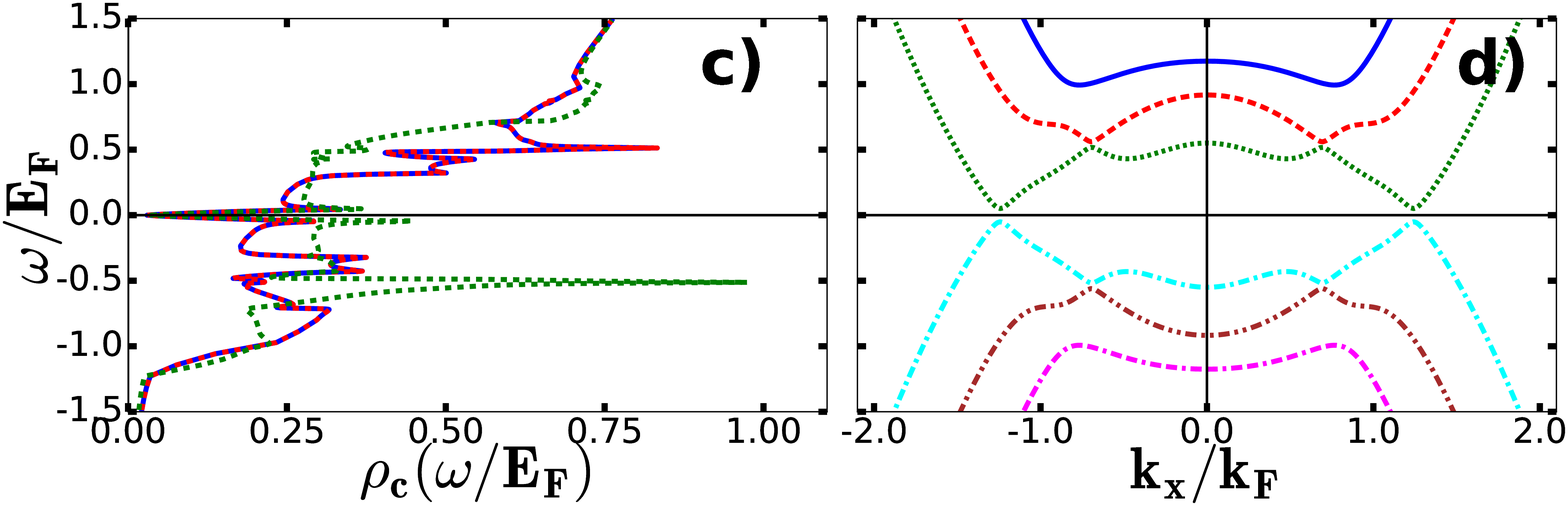,width=0.98 \linewidth} \\
\epsfig{file=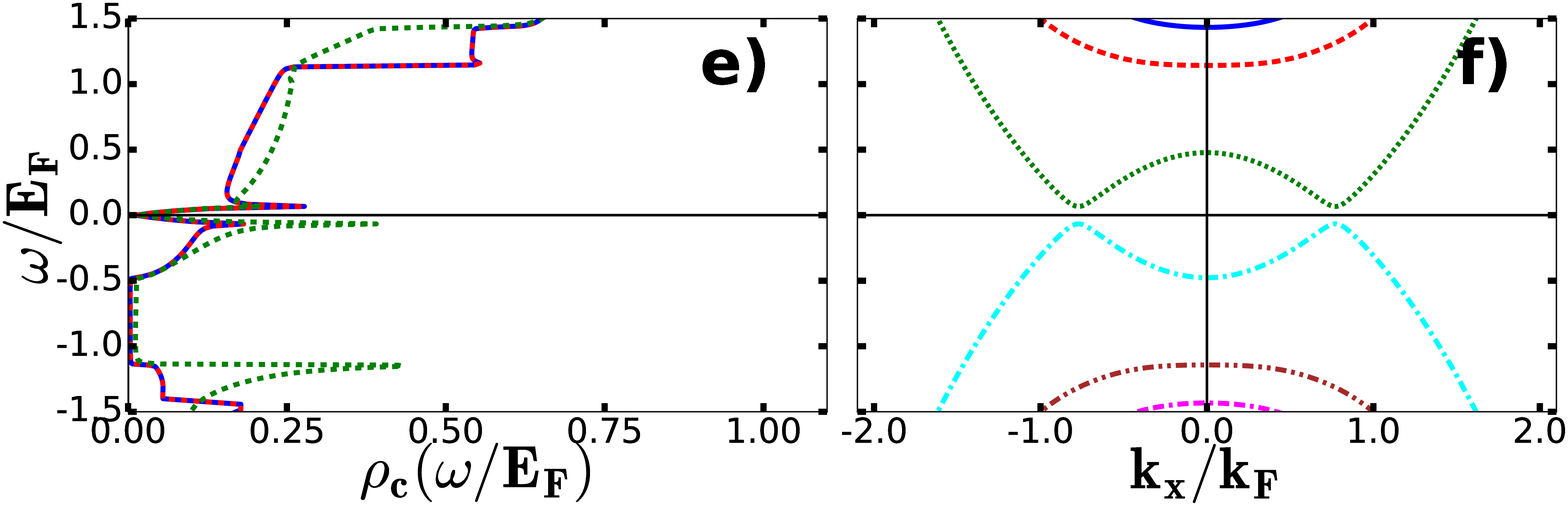,width=0.98 \linewidth} \\
\epsfig{file=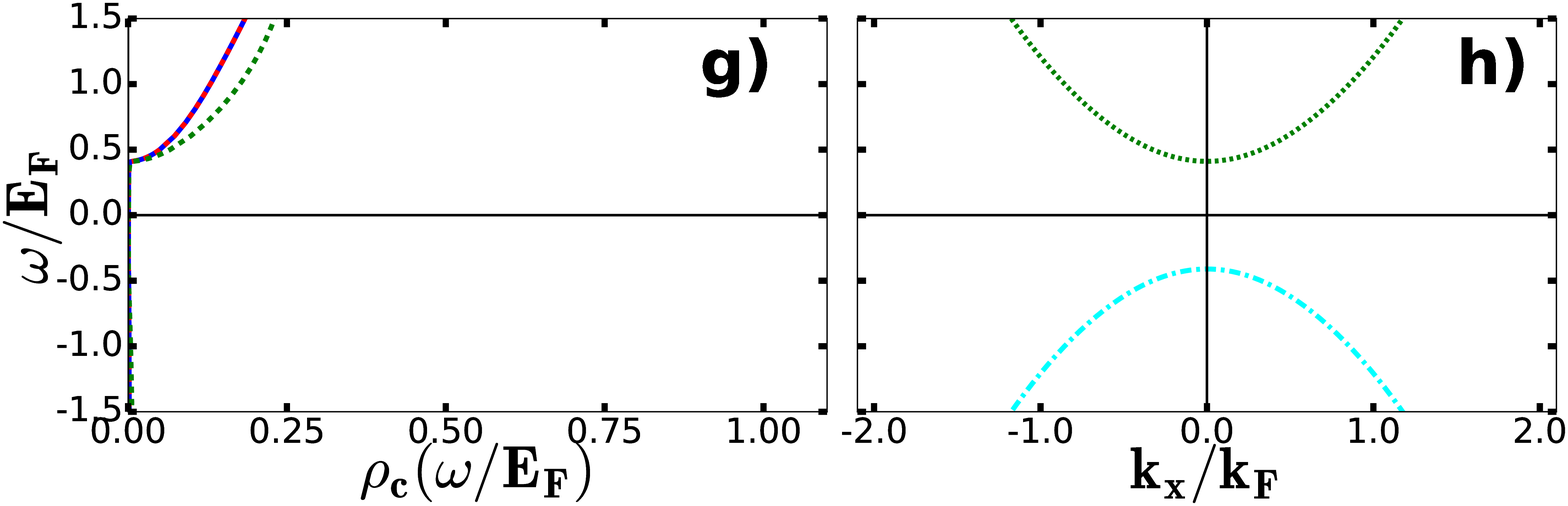,width=0.98 \linewidth}
\caption{ 
\label{fig:eleven}
(Color online)
Frequency $\omega/E_F$ versus color density of states $\rho_c (\omega/E_F)$ 
in the original $c = \{ {\rm R, G, B} \}$ basis,
and the corresponding excitation spectrum $E_j ({\bf k})$ along the $k_x$ direction. 
The parameters used are $\Omega = 0.29E_F$, $k_T = 0.35 k_F$,
$b_z = k_T^2/(2m)$ at temperature $T = 0.01 E_F$, and various interactions.
In the left panels, the red dotted line represents the Red fermions, 
the dashed green line represents the Green fermions, and 
the blue solid line represents the Blue fermions. On the right panels, 
the style-color code for energies $E_j ({\bf k})$ is: 
$E_1 ({\bf k})$ is solid blue, 
$E_2 ({\bf k})$ is dashed red, 
$E_3 ({\bf k})$ is dotted green,
$E_4 ({\bf k})$ is dash-dotted cyan,
$E_5 ({\bf k})$ is dashed-double-dotted brown,
and $E_6 ({\bf k})$ is double-dashed-dotted magenta. 
The panels a) and b) correspond to the normal phase $N3$ with 
$1/(k_F a_s) = -1.8$ $(\mu/E_F = 0.97, \vert \Delta \vert/E_F = 0)$.
The panels c) and d) correspond to the superfluid phase $R3$ with 
$1/(k_F a_s) = -0.069$ $(\mu/E_F = 0.81, \vert \Delta \vert/E_F = 0.31)$.
The panels e) and f) correspond to the superfluid phase $R1$ with 
$1/(k_F a_s) = 0.62$ $(\mu/E_F = 0.19, \vert \Delta \vert/E_F = 0.73)$.
The panels g) and h) correspond to the superfluid phase $FG$ with 
$1/(k_F a_s) = 1.1$ $(\mu/E_F = -0.73, \vert \Delta \vert/E_F = 0.99)$.
}
\end{figure}

In Fig.~\ref{fig:eleven}, we show plots of frequency $\omega/E_F$ versus
density of the states $\rho_c (\omega/E_F)$ and the corresponding excitation spectrum 
$E_j ({\bf k})$ along the $k_x$ direction, 
for parameters $\Omega = 0.29E_F$, $k_T = 0.35 k_F$,
$b_z = k_T^2/(2m)$ at temperature $T = 0.01 E_F$. 
The panels a) and b) correspond to the normal phase $N3$ with parameters 
$1/(k_F a_s) = -1.8$ $(\mu/E_F = 0.97, \vert \Delta \vert / E_F = 0)$.
The panels c) and d) correspond to the superfluid phase $R3$ with parameters 
$1/(k_F a_s) = -0.069$ $(\mu/E_F = 0.81, \vert \Delta \vert /E_F = 0.31)$.
The panels e) and f) correspond to the superfluid phase $R1$ with parameters 
$1/(k_F a_s) = 0.62$ $(\mu/E_F = 0.19, \vert \Delta \vert/E_F = 0.73)$.
The panels g) and h) correspond to the superfluid phase $FG$ with parameters 
$1/(k_F a_s) = 1.1$ $(\mu/E_F = -0.73, \vert \Delta \vert/E_F = 0.99)$.
A general feature of all the panels in Fig.~\ref{fig:eleven} is that
the density of states of the Blue and Red colors is identical, that is,
$\rho_B (\omega/E_F) = \rho_R (\omega/E_F)$. 
This symmetry is ultimately connected to the shifts in the color dispersions 
$\varepsilon_R ({\bf k}) = \varepsilon ({\bf k} - {\bf k_T})$ to positive momenta,
and 
$\varepsilon_B ({\bf k}) = \varepsilon ({\bf k} + {\bf k_T})$ to negative momenta,
which together with the eveness of $\varepsilon ({\bf k})$ 
leads to the relationship
$\varepsilon_{R} ({\bf \mp k}) = \varepsilon_{B} ({\bf \pm k})$. 
This color-parity
symmetry leads to identical density of states for the Red and Blue 
colored fermions,
since it is no longer possible to distinguish between Red and Blue 
fermions after integration over momentum states.
This last observation is an explicit consequence of the 
symmetry relations for the matrix element 
$M_{j,R} (\pm {\bf k}) = M_{j, B} (\mp{\bf k})$ 
and for the quasiparticle and quasihole energies
$E_j ({\bf k}) = E_j (-{\bf k})$, which follow from the fact that the Red (R) 
and Blue (B) states get momentum kicks $k_T$ in opposite directions.  
Another overall feature of the plots is that cusps and peaks in the density
of states $\rho_{c} (\omega/E_F)$ are associated with maxima, minima and flat
regions in the energy dispersions $E_j ({\bf k})$.

%

In Fig.~\ref{fig:eleven}a, the color density of states $\rho_c (\omega/E_F)$ 
of the normal phase $N3$ with non-zero $\Omega/E_F$ and $k_T/k_F$ has similar 
features to those of the normal state when $\Omega/E_F$ and $k_T/k_F$ 
are zero and the Red, Green and Blue color states are degenerate. 
However, non-zero $\Omega/E_F$ and $k_T/k_F$ mix the original $R, G, B$ states 
and lift degeneracies making the density of states of the Green fermions
different from that of the Red and Blue fermions, however the density of states 
of Red and Blue fermions remain the same because of the color-parity symmetry
$\varepsilon_{R} ({\bf \mp k}) = \varepsilon_{B} ({\bf \pm k})$. 
The corresponding quasiparticle/quasihole spectrum is shown in 
Fig.~\ref{fig:eleven}b.

In Fig.~\ref{fig:eleven}c, the color density of states for the $R3$ 
superfluid phase is illustrated. 
At low frequencies $\vert \omega \vert/E_F \ll \vert \Delta \vert/E_F$ 
the density of states grows linearly with frequency $\omega$,
that is $\rho_c (\omega/E_F) = \gamma_c \omega/E_F$, because 
of the three nodal lines (rings) in the excitation spectrum. 
The coefficient $\gamma_c$ depends on color. In the present case
$\gamma_R = \gamma_B \ne \gamma_G$.
Peaks in the color density of states appear at maxima and minima of 
the quasiparticle/quasihole excitation 
spectrum, as can be seen from the plots in Fig.~\ref{fig:eleven}c and d.

In Fig.~\ref{fig:eleven}e, the color density of states for the $R1$ superfluid
phase is shown. Again, the color density of states is 
$\rho_c (\omega/E_F) = \gamma_c \omega/E_F$ for low frequencies, because 
there is a nodal line (ring) in the excitation energies $E_j ({\bf k})$. 
In the $R1$ phase, the coefficient $\gamma_c$ is 
color-dependent with $\gamma_R = \gamma_B \ne \gamma_G$
similarly to the $R3$ phase. Again, peaks in $\rho_c (\omega/E_F)$ appear 
at maxima and minima of the quasiparticle/quasihole excitation spectrum as 
seen in Fig.~\ref{fig:eleven}e and f.

In Fig.~\ref{fig:eleven}g, the color density of states for the $FG$ superfluid
phase is shown. There is now a clear gap $E_g$ in the color density of states  
$\rho_c (\omega/E_F)$ as seen in the excitation spectrum $E_j ({\bf k})$ shown
in Fig.~\ref{fig:eleven}h. Deep in this phase, where the scattering
parameter $1/(k_F a_s)$ is large, the quasihole energies carry essentially no 
spectral weight. The physical reason for the very small quasihole spectral weight
is that interactions are sufficiently strong $1/(k_F a_s) \sim {\cal O} (1)$, 
and colored fermions are sufficiently non-degenerate $(\mu/E_F \ll -1)$ 
that two-body bound states (colored pairs) are well established. 
Therefore, the creation of elementary (single fermion) excitations requires the 
breaking of two-body bound states, and thus only positive energy states 
are accessible.

It is important to mention that the color density of states can also be measured 
using the photoemission spectroscopy technique developed for cold atoms, which was
used to probe the density of states of strongly interacting Fermi gas 
of $^{40}$K atoms with two internal states throughout the evolution 
from the BCS to the BEC limits~\cite{jin-2008}. Now that we have 
finalized our discussion of spectroscopic quantities, we would like 
make some final remarks, before we summarize our conclusions.

\section{Final Remarks}
\label{sec:final-remarks}

We would like to make some final remarks on a few important issues
that we have not discussed so far, such as 
Efimov states, non-uniform superfluidity, the critical temperature
of color superfluids and trap effects. It is important
to understand how these topics are affected by color-orbit and 
color-flip fields. However, this analysis lie beyond the
regions of applicability of our current work, but can be included 
in refined generalizations as discussed below.

We begin our discussion by pointing out that Efimov trimers 
in ultra-cold fermions with three internal states $\{c = R, G, B \}$
can be formed in an extremely dilute cloud of atoms. 
The densities for which these $RGB$ triatomic molecules may 
form and remain stable is typically less than $10^{12}$ atoms 
per cubic centimeter, based on a extensive description of Efimov states 
in a variety of systems~\cite{pascal-2017}. Below the upper bound 
density for $10^{12}/{\rm cm^3}$ it may be sufficient to regard the 
trimers as isolated, however with increasing atom densities in the 
range of $10^{12}/{\rm cm^3}$ to $10^{14}/{\rm cm^3}$ the surrounding 
medium largely limits the formation of stable triatomic molecules. 
In the particular case of fermions, the Pauli exclusion principle 
plays an important role as the density increases eventually preventing
the formation of Efimov trimers. Indeed the existence of these states is
observed experimentaly via an increasse in the particle loss rate mediated
by the triatomic molecules~\cite{jochim-2008, ohara-2009}. Since the
two-body collision rates for $RB, RG, BG$ are proportional to the 
product of the fermion densities $n_{R}n_{B}, n_{R}n_{G}, n_{B}n_{G}$,
respectively and the three-body collision rate for $RGB$ is proportional
to product $n_{R}n_{G}n_{B}$, there is a regime of densities within
the range $10^{12}/{\rm cm^3}$ to $10^{14}/{\rm cm^3}$, where it
is sufficient to consider two-body physics.

A conjectured zero-temperature phase diagram describing Efimov trimers 
have been sketched for a three-component (color) Fermi mixture in 
the regime of narrow Feshbach resonances~\cite{nishida-2012}, but without 
color-orbit or color-flip fields. The phase space parameters used were 
the resonance strength parameter $k_F R^*$ and the interaction parameter
$1/(k_F a_s)$, where $k_F$ is the Fermi momentum for the 
three-component system, $R^*$ is the scattering strength length 
and $a_s$ is the s-wave scattering length. Indeed at very low 
densities Efimov trimers are present, but they dissociate and disappear
as density is increased~\cite{pascal-2017,nishida-2012}. 
In the broad resonance 
regime that we are considering, the three-body parameter 
$\kappa_*^{(0)}$ plays the same role as the inverse of the length
associated with the strength of the resonance~\cite{pascal-2017}, 
that is, $(R^*)^{-1}$. Therefore, a similar qualitative phase diagram
of $k_F/\kappa_*^{(0)}$ versus $1/(k_F a_s)$ is expected in the absence of
color-orbit and color-flip fields~\cite{pascal-2017, sademelo-2017}. 
However, the inclusion of color-flip fields creates internal population
imbalances, which exacerbate the effects of the Pauli exclusion principle,
thus making the formation of Efimov states more difficult. A quantitative
analysis of Efimov states for colored fermions with three internal states 
in the presence of color-orbit and color-flip fields is currently underway
and it is the subject of a future publication~\cite{sademelo-2017} meant
to describe the low density regime $n \lesssim 10^{12}/{\rm cm}^3$ 
of the present problem.

Another important point to mention is that non-uniform color superfluidity 
may exist over a narrow area in the BCS regime of the phase diagram shown 
in Fig.~\ref{fig:three}, that is, in 
the range of parameters $\Omega/E_F \lesssim 0.2$ and 
$1/(k_F a_s) \lesssim -1$. For continuum problems in three spatial dimensions,
non-uniform superfluids occur only in the BCS regime and at very 
low temperatures  $T/T_F << 10^{-3}$, as its formation requires a large 
degree of particle-particle nesting which parabolic bands do not 
provide~\cite{larkin-1964, fulde-1964}. 
Therefore, in three-dimensional continuum systems, 
non-uniform superfluid states of the Larkin-Ovchinikov~\cite{larkin-1964} 
or Fulde-Ferrel~\cite{fulde-1964} type may exist only in a 
very limited region of phase space, which is confined to 
the BCS regime and very low temperatures. The situation is similar 
for color superfluids with color-flip and color-orbit fields.

In addition, we would like to comment on the effects of fluctuations and
on its importance in obtaining the critical temperature of the 
system away from the BCS regime. In the case of zero olor-orbit and
color-flip fields, the critical temperature of color superfluids has been
investigated as a function of the interaction parameter $1/(k_F a_s)$
using the T-matrix approach~\cite{baym-2010} and lead to 
qualitatively similar results to those of spin-$1/2$ Fermi 
superfluids~\cite{sademelo-1993}. Without color-orbit and color-flip fields
the normal state of the system evolves from a color $(RGB)$ Fermi liquid to 
a color Bose liquid with $RG$, $GB$ and $RB$ molecules of masses $M_B$ equal
to twice the mass $m$ of their consituent fermions. 
The critical temperature in the Bose-Einstein condensation regime 
is that of non-interacting
bosons of mass $M_B = 2m$ and density $n_B = n/6$, where $n$ is the 
density of fermions. Therefore, 
$
T_{BEC} 
= 
\left(
2\pi
/
M_B
\right)
\left[ 
n_B/\zeta(3/2)
\right]^{2/3}
= 0.137 E_F
$ 
is proportional to Fermi energy.

The critical temperature of spin-$1/2$ ultra-cold fermions 
was recently investigated in the presence of
Zeeman fields and of the experimentally relevant equal-Rashba-Dresselhaus 
(ERD) spin-orbit coupling~\cite{powell-2017}. 
That analysis revealed that spin-orbit coupling 
and Zeeman fields modify the masses of the Bose molecules and lead to 
an enhancement of the critical temperature in the BEC regime. Furthermore,
in the BCS region, the critical temperature lies always below that of a
system without Zeeman fields and spin-orbit coupling. 
A detailed analysis of the fluctuation effects and the critical temperature 
of color superfluids in the presence of color-orbit and color-flip fields 
is an important issue and will be carried out at a later publication following
the works on spin-$1/2$ fermions~\cite{powell-2017} and on color
superfluids~\cite{baym-2010}. 

Lastly, we would like to comment on the effects of traping potentials. 
Historically, harmonic confining potentials have been consistently used to 
trap atomic Fermi gases and the local density approximation (LDA) 
has been widely used to describe the resulting inhomogeneous states. 
For harmonic traps $V_{\rm trap} ({\bf r})$, the chemical potential $\mu$ 
of our system is mapped into a local chemical potential 
${\widetilde \mu} ({\bf r})= \mu - V_{\rm trap} ({\bf r})$ within LDA. 
Thus, several of the phases described in our phase diagram, 
shown in Fig.~\ref{fig:three}a, may coexist in a harmonic trap. 
For example, at unitarity, the single ring and the three ring color
superfluid phases can coexist in harmonic traps and the detailed 
inhomogeneous spatial structure of color superfluids needs to be mapped
for fixed color-orbit and color-flip fields.
However, the current trend in experimental work is to create different
types of trapping potentials including that of a box-type variety 
using digital micromirror devices (DMD)~\cite{neely-2016},
which can produce homogeneous states. This trend to study 
experimentally homogeneous systems is reflected in recent work 
covering both Bose~\cite{hadzibabic-2017} and 
Fermi~\cite{zwierlein-2017} atomic superfluids. 
Thus, we expect that homogeneous color superfluids with and without 
color-orbit and color-flip fields will be studied using 
box potentials in the near future, such that a direct comparison to our
work can be made.  

With our final remarks completed, 
we are ready to state our conclusions next.

\section{Conclusions}
\label{sec:conclusions}

We studied the quantum phases of interacting colored fermions
in the presence of color-orbit coupling and color-flip fields. 
Experimental candidates for the observation of such phases
include $^6$Li, $^{40}$K, and $^{173}$Yb, 
which possess at least three internal states, that can be labeled
Red $(R)$, Green $(G)$ and Blue $(B)$, and therefore can be used to 
simulate exotic phases related to quantum chromodynamic (QCD) systems
in table top experiments.
 
Among many possibilities of analogous exotic QCD phases, we focused 
on the emergence of color superfluidity, where the presence of
color-orbit coupling and color-flip fields induce Lifshitz-type topological 
phase transitions. In such transitions, the symmetry of the order parameter 
tensor does not change, but the structures of the ground state wavefunctions 
and of the energy spectrum of elementary excitations (quasiparticles
and quasiholes) do.

We constructed the low temperature phase diagram of color-flip field versus
scattering parameter (interactions) and classified the emerging color 
superfluid phases in terms of the {\it loci} of zeros of the quasiparticle 
excitation spectrum in momentum space. For fixed color-orbit coupling 
and quadratic color-shift field, we identified five gapless
phases with one, two, three, four or five rings of nodes in the excitation 
spectrum, and one fully gapped phase. In addition, we found
that a very rare quintuple point exists where five gapless superfluid phases 
with line nodes converge. Given that the phase transitions from one nodal 
superfluid to another is continuous (second order) the quintuple point 
is also pentacritical. Furthermore, in the limit of zero color-flip fields,
but finite color-orbit coupling, phase transitions from the normal state 
to a nodal superfluid, and from a nodal to fully gapped superfluids occur. 

We constrasted the phase diagram of non-zero color-orbit coupling with the
simpler case of zero color-orbit coupling, where the color-flip field versus
scattering parameter phase diagram has only one gapless and one fully 
gapped superfluid phase. In the limit of zero color-flip field, 
the gapless phase is described by an inert degenerate mixed-color fermion band 
and two fully gapped bands of quasiparticle excitations, while the fully gapped 
phase is described by an inert non-degenerate mixed-color fermion band
and two fully gapped bands of quasiparticle excitations. In this case, only a 
crossover between BCS and BEC superfluids occurs.

We used the connectivity of the nodal regions in momentum space to classify 
the topology of the superfluid phases, and analysed the order parameter tensor
structure in a mixed color representation, as well as, in a pseudo-spin 
representation exploring the singlet, triplet and quintet sectors. We found
that the nodal structure of the order parameter tensor does not coincide with
the nodes in the quasiparticle/quasihole excitation spectrum, which is the case for the 
more familiar example of spin-1/2 fermions.

In addition to topological aspects, we investigated in detail spectroscopic
properties of colored fermions in their normal and superfluid phases. 
We analysed the quasiparticle/quasihole excitation spectrum, as well as 
the momentum distribution and density of states of colored fermions and 
concluded that these properties can be used to help distinguish between 
different topological nodal phases of color superfluids, 
and can be measured using current experimental techniques.

\acknowledgments{One of us (C. A. R. SdM) acknowledges the support of 
a) the Joint Quantum Institute at the University of Maryland and 
the National Institute of Standards and Technology, where part of this work 
was completed during a Sabbatical visit, 
b) the Galileo Galilei Institute for Theoretical Physics 
via a Simons Fellowship and 
c) the Aspen Center for Physics via NSF grant PHY1607611.}

\end{document}